\documentclass[12pt]{article}
\pdfoutput=1

\newlength{\abstractwidth}
\usepackage{epsf}
\usepackage{color}
\usepackage{graphicx}
\usepackage{hyperref}

\usepackage{amsmath}
\usepackage{amssymb}
\usepackage{latexsym}

\usepackage{tikz, pgf}
\usepackage{tkz-fct}
\usepackage{pgfplots}
\usetikzlibrary{shapes.misc}
\usetikzlibrary{shapes,snakes}
\usetikzlibrary{decorations.pathmorphing}	
\usetikzlibrary{decorations.markings}

\newcommand{\midarrow}{\tikz \draw[-triangle 90] (0,0) -- +(.1,0);}

\renewcommand{\thefootnote}{\fnsymbol{footnote}}
\renewcommand{\thanks}[1]{\footnote{#1}}
\newcommand{\starttext}{
\setcounter{footnote}{0}
\renewcommand{\thefootnote}{\arabic{footnote}}}

\newcommand{\bea}{\begin{eqnarray}}
\newcommand{\eea}{\end{eqnarray}}
\newcommand{\be}{\begin{eqnarray}}
\newcommand{\ee}{\end{eqnarray}}

\def\cA{{\cal A}}
\def\cB{{\cal B}}
\def\cC{{\cal C}}

\def\cF{{\cal F}}
\def\cG{{\cal G}}
\def\cH{{\cal H}}

\def\cK{{\cal K}}

\def\cM{{\cal M}}

\def\cO{{\cal O}}
\def\cP{{\cal P}}

\def\cR{{\cal R}}
\def\cS{{\cal S}}
\def\cT{{\cal T}}

\def\cY{{\cal Y}}
\def\cZ{{\cal Z}}

\def\bp{{\bf p}}

\def\mA{\mathfrak{A}}
\def\mB{\mathfrak{B}}
\def\mC{\mathfrak{C}}
\def\mD{\mathfrak{D}}

\def\mG{\mathfrak{G}}

\def\mI{\mathfrak{I}}
\def\mJ{\mathfrak{J}}

\def\mS{\mathfrak{S}}

\def\mg{\mathfrak{g}}

\def\mj{\mathfrak{j}}

\def\ms{\mathfrak{s}}

\def\mz{\mathfrak{z}}

\def\ZZ{{\mathbb Z}}
\def\RR{{\mathbb R}}

\def\CC{{\mathbb C}}

\def\Re{{\rm Re \,}}
\def\Im{{\rm Im \,}}

\def\det{{\rm det \,}}

\def\half{{1\over 2}}
\def\p{\partial}

\def\f{\varphi}
\def\tet{\vartheta}
\def\ep{\varepsilon}
\def\om{\omega}

\def\Sep{{\Sigma_{ab}}} 

\def\pbx{\p _{\bar x}}

\def\pbw{\p _{\bar w}}
\def\pbz{\p _{\bar z}}

\def\GA{\cG}
\def\kap{\kappa}
\def\oom{\overline{\om}}

\def\SM{{\Sigma}}
\def\cBZ{{\cal Z}}
\def\BZ{{Z}}
\def\ccF{\cF}

\def\no{\nonumber}
\def\sm{\smallskip}

\def\bp#1{{#1}}

\newcommand{\gone}{g}
\newcommand{\de}{{\rm d}}
\newcommand{\I}{{\rm i}}
\newcommand{\nn}{\nonumber}
\newcommand{\IZ}{\mathbb{Z}}

\newcommand{\sugra}{(sg)}
\newcommand{\Bsugra}{\BZ^{\sugra}}
\newcommand{\cBsugra}{\cBZ^{\sugra}}
\newcommand{\Gsugra}{G^{\sugra}}
\newcommand{\GAsugra}{\cG^{\sugra}}
\newcommand{\ksugra}{\kappa^{\sugra}}
\newcommand{\gsugra}{\gamma^{\sugra}}
\newcommand{\ggsugra}{\gamma_1^{\sugra}}
\newcommand{\gggsugra}{\gamma_2^{\sugra}}
\newcommand{\ggggsugra}{\gamma_3^{\sugra}}
\newcommand{\Deltasugra}{\Delta^{\sugra}}
\newcommand{\Ysugra}{Y^{\sugra}}

\newcommand{\ttau}{S}

\newcommand{\rrho}{\tau}

\begin{document}
\starttext
\setcounter{footnote}{0}

\begin{flushright}
arXiv:1806.02691v3\\
QMUL-PH-18-10 \\
DAMTP-2018-24  
\end{flushright}

\vskip 0.3in

\begin{center}

{\Large \bf Asymptotics of the $D^8 \cR^4$ genus-two string invariant}

\vskip 0.2in

{\large Eric D'Hoker$^{(a)}$, Michael B. Green$^{(b)}$ and Boris Pioline$^{(c)}$}

\vskip 0.15in

{ \sl  (a) Mani L. Bhaumik Institute for Theoretical Physics}\\
{\sl  Department of Physics and Astronomy}\\
{\sl University of California, Los Angeles, CA 90095, USA}

\vskip 0.1in

{ \sl (b) Department of Applied Mathematics and Theoretical Physics }\\
{\sl Wilberforce Road, Cambridge CB3 0WA, UK}, and \\
{\sl Centre for Research in String Theory, School of Physics, }\\
{\sl Queen Mary University of London, Mile End Road, London, E1 4NS, England}

\vskip 0.1in

{\sl (c) 
Laboratoire de Physique Th\'eorique et Hautes \'Energies (LPTHE), \\
Sorbonne Universit\'e and CNRS,  4 place Jussieu, F-75005 Paris, France
}

\vskip 0.15in

{\tt \small dhoker@physics.ucla.edu, M.B.Green@damtp.cam.ac.uk,  pioline@lpthe.jussieu.fr}

\vskip 0.2in

\begin{abstract}
\vskip 0.1in

We continue our investigation of the modular graph functions and string invariants that arise at genus-two as coefficients of  low energy effective interactions in Type II superstring theory. In previous work, the non-separating degeneration of a genus-two modular graph function of weight $w$  was shown to be given by a Laurent polynomial  in the degeneration  parameter $t$ of degree $(w,w)$.  The coefficients of this polynomial generalize genus-one modular graph functions, up to terms which are exponentially suppressed in $t$ as $t \to \infty$. In this paper, we evaluate this expansion explicitly for the modular graph functions associated with the $D^8 \cR^4$ effective interaction for which the Laurent polynomial has degree $(2,2)$.  We also prove that the separating degeneration is given by a polynomial  in the degeneration parameter $\ln (|v|)$ up to contributions which are power-behaved in  $v$ as $v \to 0$. We further extract the complete, or tropical,  degeneration and compare it with the independent calculation of the integrand of the sum of Feynman diagrams that contributes to two-loop type II supergravity expanded to the same order in the low energy expansion.  We find that the tropical limit of the string theory integrand reproduces the supergravity integrand as its leading term, but also includes sub-leading terms  proportional to odd zeta values that are absent in supergravity and can be ascribed to higher-derivative stringy interactions.
\end{abstract}
\end{center}

\newpage
 
\setcounter{tocdepth}{2} 
\tableofcontents
\newpage

\baselineskip=15pt
\setcounter{equation}{0}
\setcounter{footnote}{0}

\section{Introduction}
\setcounter{equation}{0}
\label{sec:1}

The low energy dynamics of string theory may be described in terms of an effective action which encodes the influence of massive string states upon the massless sector. The effective action admits an expansion in powers of space-time derivatives or, equivalently, in powers of the momenta of the massless states. The expansion contains  the supersymmetrized Einstein--Hilbert action  as its leading term, plus an infinite series of higher derivative effective interactions.  The coefficients of the effective interactions are functions of the scalar fields that are associated with geometrical data of the target-space and are referred to as target-space moduli. These coefficients exhibit a rich mathematical structure.  While relatively little is known about their exact dependence  on the target-space moduli, precise statements can be made order by order in a variety of further expansions at asymptotic values of the target-space moduli.  String perturbation theory uses an expansion in powers of the string coupling $g_s$, which is related to the constant value of the dilaton field. 

\sm

The low energy effective action may be extracted  from superstring scattering amplitudes.  In closed superstring perturbation theory an amplitude is given by an infinite power series in $g_s$ where the coefficient of $g_s^{-2+2h}$ for $h \geq 0$ is  an integral over the moduli space of compact super-Riemann surfaces of genus $h$.    Many features of superstring amplitudes have been established to all orders in $g_s$,  most notably the absence of the ultraviolet divergence and anomaly problems that plague perturbative quantum field theories containing gravity.  However, explicit formulas for the  amplitudes  have been obtained so far only at  low orders in the $g_s$ expansion.  Our interest in this paper will be in the simplest non-trivial Type II amplitude, namely for the scattering of four gravitons, whose explicit form is known only for $h \leq 2$. The low energy expansion of the four-graviton amplitude is given by a sum over $k \geq 0$ of effective interactions which are schematically  of the form  $D^{2k} \cR^4$, where $D$ and $\cR$ respectively stand for the covariant derivative and the Riemann tensor of the target-space, suitably contracted. The coefficients of the effective interactions will be described next. 

\sm

The  genus-zero ($h=0$) term is the tree-level contribution. Its leading low energy expansion reproduces the amplitudes arising in classical Einstein gravity.  The coefficients of the higher order effective interactions 
 in the low energy expansion of the four-graviton amplitude are polynomials in odd Riemann zeta values with rational coefficients. More generally, the coefficients in the expansion of the tree-level amplitude with more than four gravitons  are single-valued multiple zeta values  \cite{Stieberger:2013wea,Brown:2013gia}.  

\sm

The genus-one ($h=1$) four-graviton amplitude involves an integral over the moduli space $\cH_1/SL(2,\ZZ)$ of complex tori $\Sigma $ (where $\cH_1$ is the Poincar\'e upper half plane) and an integral over four points on $\Sigma $, corresponding to the four gravitons \cite{Green:1981yb}. The coefficients of the effective interactions  in the low-energy expansion of the integral over the points only, without  integrating over $\cH_1/SL(2,\ZZ)$, are $SL(2,\ZZ)$-invariant {\sl modular graph functions} on $\cH_1$ that were recently studied in some detail  in \cite{Green:1999pv,Green:2008uj,D'Hoker:2015foa,DHoker:2015wxz}.  Although their structure still remains to be fully elucidated, it is clear that modular graph functions (and their generalizations to modular forms) generalize the multiple zeta values  that arise in the tree-level expansion, and satisfy algebraic identities that generalize  those satisfied by multiple zeta values \cite{DHoker:2015sve,DHoker:2016mwo,DHoker:2016quv} (see also \cite{Basu:2016kli,Brown:2017qwo,Brown2,Kleinschmidt:2017ege,Broedel:2018izr} for further studies). 

\sm

Much less is known about the coefficients of the low-energy expansion of higher-genus ($h\geq 2$)  amplitudes in superstring perturbation theory. The genus-two four-graviton amplitude was evaluated explicitly in \cite{DHoker:2001kkt,DHoker:2001qqx,D'Hoker:2002gw, D'Hoker:2005jc} for Type II and Heterotic strings by projecting the moduli space of super Riemann surfaces to that of Riemann surfaces. The Type II amplitude was reproduced in the pure spinor formulation  and extended to include fermions in \cite{Berkovits:2005df,Berkovits:2005ng}. The structure of the genus-two four-graviton amplitude  generalizes that of its genus-one counterpart. It is given by an integral over the moduli space $\cM_2 \approx \cH_2/Sp(4,\ZZ)$ of genus two compact Riemann surfaces $\Sigma $ (where $\cH_2$ is the Siegel upper half space of rank two, parametrized by the period matrix $\Omega$) of an integral over four points on $\Sigma $, again corresponding to the four gravitons. The low-energy expansion of the integral over the points only, without integrating over $\cM_2$, now gives rise to $Sp(4,\ZZ)$-invariant functions on $\cH_2$ which, by analogy with the genus-one case, are referred to as {\sl genus-two modular graph functions} \cite{DHoker:2017pvk}.   

\sm

In the low energy expansion of the genus-two  four-graviton amplitude in Type II superstrings, the effective interactions $\cR^4$ and $D^2\cR^4$ have vanishing coefficients. The first non-zero term is $D^4\cR^4$, whose coefficient is constant on $\cH_2$ and matches the predictions of S-duality in Type~IIB string theory \cite{D'Hoker:2005ht}. The next order term is the effective interaction $D^6 \cR^4$. Its coefficient is a non-trivial $Sp(4,\ZZ)$-invariant function $\f$ on $\cH_2$, which was shown in \cite{D'Hoker:2013eea} to be proportional to the genus-two Kawazumi-Zhang invariant  \cite{Kawazumi,Zhang} (see also \cite{Kawa1,Kawa2}). The invariant $\f$ satisfies a  Laplace eigenvalue equation on $\cH_2$  \cite{DHoker:2014oxd},  which was later used  in combination with known leading asymptotics \cite{MR1105425,zbMATH06355718} to establish its representation as a generalized Borcherds-type theta-lift \cite{Pioline:2015qha}. The latter provides the full asymptotic expansion near the boundary of moduli space $\cM_2$, including all exponentially suppressed terms.
The integral of $\f(\Omega)$ over $\cM_2$ can be computed using the eigenvalue equation and also matches the  S-duality prediction \cite{DHoker:2014oxd}.
 
\sm

The genus-two contribution to higher order effective interactions, schematically of the form $D^{2k} \cR^4$ for $k \geq 4$, may also be derived from the four-graviton amplitude and, as was pointed out in \cite{D'Hoker:2013eea}, produces further and novel genus-two $Sp(4,\ZZ)$-invariant  modular graph functions on $\cH_2$. The goal of previous work in \cite{DHoker:2017pvk}, of this paper, and of future work, is to gain understanding of these novel invariants, and of any algebraic and differential relations they may satisfy, at a level comparable to the one that has been achieved for the  Kawazumi-Zhang invariant or for the genus-one case. One important step in this direction, which has proven to be invaluable also at genus one, is to obtain the behavior of the novel  invariants under degenerations of the genus-two Riemann surface.

\sm

Powerful techniques were developed in \cite{DHoker:2017pvk} to analyze the behavior of general classes of modular graph functions at arbitrary genus near the non-separating degeneration of the Riemann  surface.  The non-separating degeneration of a genus-two surface $\Sigma $ corresponds to letting a  non-trivial homology cycle  become infinitely long while keeping independent cycles finite so that a genus-two surface $\Sigma $ degenerates to a torus $\Sigma_1$ with two punctures. 
 
  \subsection{Summary of results}
 \label{subsec:1.1}

\begin{figure}[t]
\begin{center}
\begin{tikzpicture}[scale=.27]
\begin{scope}[shift={(+22,0)}]
\filldraw [black]  (4,0) ellipse (0.7 and 0.7);
\filldraw [black]  (-4,0) ellipse (0.7 and 0.7);
\draw [ultra thick] (0,0) ellipse (4 and 3);
\draw [ultra thick] (0,3) -- (0,-3);
\draw [ultra thick] (4,0) -- (7,-4);
\draw [ultra thick] (4,0) -- (7,4);
\draw [ultra thick] (-4,0) -- (-7,-4);
\draw [ultra thick] (-4,0) -- (-7,4);
\draw (0, -7.0) node{$(c)$};
\end{scope}
\begin{scope}[shift={(-22,0)}]
\filldraw [black]  (4,0) ellipse (0.7 and 0.7);
\draw [ultra thick] (0,0) ellipse (4 and 3);
\draw [ultra thick] (-2.6,-2.4) -- (-7,-4);
\draw [ultra thick] (-2.6,2.4) -- (-7,4);
\begin{scope}[shift={(0,0)}]
\draw [ultra thick] (4,0) -- (7,-4);
\draw [ultra thick] (4,0) -- (7,4);
\draw [ultra thick] (0,3) -- (0,-3);
\draw (0, -7.0) node{$(a)$};
\end{scope}
\end{scope}
\begin{scope}[shift={(0,0)}]
\filldraw [black]  (0,0) ellipse (0.7 and 0.7);
\begin{scope}[rotate=180]   
\draw [ultra thick] (0,0) .. 
controls (-8,9) and (-8,-9) .. 
(0,0);
\end{scope}
\draw [ultra thick] (5,-2.4) -- (7,-4);
\draw [ultra thick] (5,2.4) -- (7,4);
\end{scope}
\begin{scope}[rotate=0]   
\draw [ultra thick] (0,0) .. 
controls (-8,9) and (-8,-9) .. 
(0,0);
\draw [ultra thick] (-5,-2.4) -- (-7,-4);
\draw [ultra thick] (-5,2.4) -- (-7,4);
\draw (0, -7.0) node{$(b)$};
\end{scope}
\end{tikzpicture}
\end{center}
\caption{Examples of  modifications of  two-loop supergravity diagrams in which the black nodes indicate higher-derivative local interaction vertices, which arise in the tropical limit. \label{fig:R4degen}}
\label{fig:higher}
\end{figure}
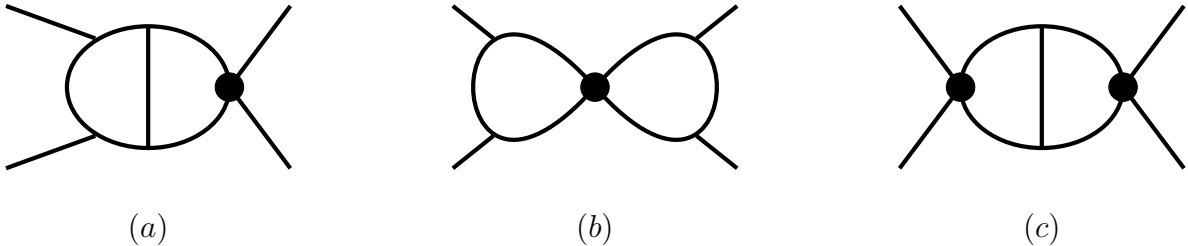

In this paper we will extend the techniques  and results of \cite{DHoker:2017pvk} 
 to obtain the expansions of the genus-two modular graph function $\cB_{(2,0)}(\Omega)$  associated with the $D^8\cR^4$ effective interaction around both the non-separating  and the separating degeneration limits.  We will also consider the further degeneration, known as the ``tropical'' limit, in which the two-dimensional
surface reduces to a two-loop irreducible graph shown in Figure \ref{fig:skeleton} on page \pageref{fig:skeleton}.  In the non-separating case this will be compared with the expression obtained from low energy expansion of two-loop supergravity.  Our  results may be summarized as follows:
\begin{enumerate}
\itemsep=0.15in
\item 
With the help of the genus-two Arakelov Green function, the string invariant $\cB_{(2,0)}(\Omega)$ may be decomposed into a sum of three non-trivial $Sp(4,\ZZ)$-invariant genus-two modular graph functions $\cBZ_i (\Omega)$ defined in \eqref{defcB1},
\bea
\cB_{(2,0)} (\Omega) = \textstyle{\half} \cBZ_1(\Omega) - \cBZ_2(\Omega) + \textstyle{\half }\cBZ_3(\Omega) 
\eea 
Throughout, $\Omega$ will denote a genus-two period matrix, and $Y$ will denote its imaginary part,  whose components are given by,
\bea
\label{OmY}
\Omega = \left ( \begin{matrix} \tau & v \cr v & \sigma \cr \end{matrix}  \right )
\hskip 1in 
Y = \Im \Omega = \left ( \begin{matrix} \tau_2 & v_2 \cr v_2 & \sigma_2 \cr \end{matrix}  \right )
\eea
with $\tau=\tau_1+i \tau_2, v=v_1+iv_2 , \sigma=\sigma _1 + i \sigma _2$, and $\tau_1,\tau_2, v_1, v_2, \sigma _1, \sigma _2 \in \RR$.  The matrix $\Omega$ takes values in the Siegel upper-half plane $\cH_2$ so that the matrix $Y $ is positive definite.
\item 
Near the {\sl non-separating degeneration} $t \to \infty$, 
Theorem 3  of \cite{DHoker:2017pvk}  states that each modular graph function $\cBZ_i(\Omega) $  is given by a Laurent polynomial in the degeneration parameter~$t$, of degree $(w,w)$ for $w=2$,  with exponentially small corrections, 
\bea
\label{LaurentBpq}
\cBZ_i  (\Omega) = \sum_{n=-w}^{w} (\pi t)^n \, \mz_i^{(n)}(v|\tau)  + \cO(e^{-2\pi t})
\eea
The subgroup of the modular group $Sp(4,\ZZ)$ which leaves the degeneration invariant is isomorphic to the Fourier-Jacobi group $SL(2,\ZZ)\ltimes (\ZZ^2 \ltimes \ZZ)$. It was shown in \cite{DHoker:2017pvk} that the non-separating degeneration takes a strikingly simple form in terms of a special combination $t$ of the moduli given by $t = \det (\Im \Omega)/ \Im (\tau)$. The  parameter $t$ and the coefficients $\mz_i^{(n)}(v|\tau)$ are invariant under the modular subgroup $SL(2,\ZZ)\ltimes \ZZ^2 $  acting on $v \in \Sigma _1$ and $\tau \in \cH_1$. The coefficients $\mz_i^{(n)}(v|\tau)$ may be thought of equivalently as non-holomorphic elliptic functions,  non-holomorphic Jacobi forms, or modular graph functions with $(\RR/\ZZ)^2$-character, and are evaluated explicitly in this paper, see Eq. \eqref{bark3}. 

\item 
Near the {\sl separating degeneration} $v\to 0$, we show that each modular graph function $\cBZ_i (\Omega)$ is given by a polynomial in $(-\ln |\hat v|)$ of degree $w=2$, up to corrections that are power behaved in $|v|^\half$ (see Eq. \eqref{Zisep}),
\bea
\cBZ_i (\Omega) = \sum _{n=0}^w (- \ln |\hat v|)^n \, \ms^{(n)} _i (\tau, \sigma) + \cO(|\hat v|^\half )
\hskip 0.7in \hat v = 2 \pi v \, \eta (\tau)^2 \eta (\sigma )^2
\label{sepcoeff}
\eea
where $\eta$ is the Dedekind eta-function. The degeneration parameter $\hat v$ and the coefficients $\ms^{(n)}_i (\tau, \sigma)$ are invariant under the residual modular group $SL(2,\ZZ) \times SL(2,\ZZ)'$ of the separating degeneration as $v \to 0$ while keeping $\tau$ and $\sigma$ fixed.
This is in fact a special case of a result valid for a general class of genus two modular
graph functions of degree $w$, as we show in Section \ref{sec_gensep}.

\item
Near the {\sl tropical degeneration}, the matrix $Y$ is uniformly scaled to $\infty$ keeping the ratios of its entries  fixed,  so that the parameter $V$ defined by $V= (\det Y)^{-1/2}$ tends to zero.  In this limit, each modular graph function $\cBZ_i (\Omega) $ is given by a Laurent polynomial with exponentially small corrections (see Eq. \eqref{Z2trop}, \eqref{Z3trop}, \eqref{Z1trop}),
\bea
\cBZ_i (\Omega) = \sum _{n=-w}^{2w} V^n \, \mj^{(n)} _i (S) + \cO(e^{-1/V} )
\hskip 0.7in S={v_2 \over \tau_2} + i \sqrt{{t \over \tau_2}}
\label{tropcoeff}
\eea 
The coefficients $\mj^{(n)}_i (S)$ are {\sl modular local Laurent polynomials}, which belong to a class of non-holomorphic modular functions
first encountered in the study of two-loop supergravity amplitudes \cite{Green:2008bf} and
further developed in the mathematics literature \cite{ZagierPC,zbMATH06251204,bringmann2014modular}. In the vicinity of the cusp $S\to i\infty$,
this degeneration is obtained by extracting the behavior of the genus-one functions $\mz_i^{(n)}(v|\tau)$ in the limit $\tau_2\to\infty$ keeping $v_2/\tau_2$ fixed in the range $0< v_2/\tau_2 <1$. The tropical  degeneration near the separating degeneration is obtained by extracting the behavior of the coefficients $\ms^{(n)}_i (\tau, \sigma)$ as both $\tau_2, \sigma_2 \to \infty$, keeping their ratio fixed.

\item 
We compare the tropical limit of $\cB_{(2,0)}$ with the coefficient,  $\cB^{(sg)}_{(2,0)}$,  of the $D^8 \cR^4$ interaction in the low energy expansion of the two-loop contribution to Type II supergravity \cite{Bern:1998ug}, which  can be expressed as a sum of  scalar field theory  diagrams as shown in Figure~\ref{fig:skeleton}.  In order to make this comparison  we  use a world-line formalism that mimics the string theory world-sheet formalism and expresses  $\cB^{(sg)}_{(2,0)}$ as a linear sum of three contributions, which are built out of the world-line  Arakelov Green function and are field theory analogues of $\cZ_i(\Omega)$.  The tropical degeneration $V\to 0$ is found to reproduce the known supergravity integrand at order $\cO(1/V^2)$, but includes additional sub-leading terms proportional to odd zeta values such as $\zeta(3) V$,  $\zeta(5) V^3$ and $\zeta(3)^2 V^4$. 

\sm

The same phenomenon holds for the Kawazumi-Zhang invariant, whose tropical limit  includes a single subleading term proportional to $\zeta(3) V^2$   \cite{Pioline:2015qha}. In field theory language each subleading term can be interpreted as a two-loop Feynman integrand where one of the supergravity interaction vertices is replaced by a higher derivative tree-level effective interaction, such as $\cR^4$, $D^4\cR^4$ or $D^6\cR^4$ as is indicated in Figure \ref{fig:R4degen}. 

\sm

The effect of such higher derivative interactions implies a particular pattern of logarithmic divergences when the amplitude is considered in lower dimensions by compactification on a torus.
 We leave a detailed analysis of this phenomenon to a forthcoming 
 publication~\cite{DGP2018}.

\end{enumerate}

\sm

These  results should be important  for elucidating further properties of higher genus modular graph functions.  Among many outstanding issues still to be understood are algebraic and differential identities between the genus-two modular graph function and the  possibility of generalised theta-lift representation for these functions analogous to the representation of the Kawazumi--Zhang invariant found in \cite{Pioline:2015qha}.
Moreover, these results are important in determining the genus-two contribution to the  coefficient of the $D^8\cR^4$  effective interaction, which is given by integration of $\cB_{(2,0)}$  over the moduli space of genus-two Riemann surfaces, or equivalently over a fundamental domain $\cH_2/Sp(4,\IZ)$ in the Siegel upper-half plane.

\newpage

\subsection{Organization}

The remainder of this paper is organized as follows.  In Section~\ref{sec:2}, we review the low energy expansion of the genus-two contribution to the four graviton amplitude.   Appendix \ref{sec:A}  reviews related material concerning some basic features of genus-one surfaces, including details of how various genus-two integrals  reduce to integrals over genus-one surfaces with punctures.
 In Section~\ref{sec:3} we use the methods developed in \cite{DHoker:2017pvk} to give a detailed evaluation of the expansion of  $\cB_{(2,0)}(\Omega)$, the $Sp(4,\IZ)$ invariant coefficient of $D^8 \cR^4$, around the non-separating degeneration. The determination of the expansion coefficients $\mz_i^{(n)}(v|\tau)$ in \eqref{LaurentBpq} involves a large number of steps, which are detailed  in Appendix~\ref{sec:B}.    In Section~\ref{sec_sep} we obtain a general result on the separating degeneration of genus-two modular graph functions, and apply this to  obtain the explicit coefficients, $ \ms^{(n)} _i $ (defined in \eqref{sepcoeff}) of the expansion of $\cB_{(2,0)}(\Omega)$. In Section~\ref{sec_trop} we take the further limit that gives the  complete (or tropical) degeneration and evaluate the coefficients, $ \mj^{(n)} _i $ (defined in \ref{tropcoeff}), relevant to this degeneration.  Technical details needed in deriving the coefficients in the tropical limit are given in Appendix \ref{sec:D}.  In  Section~\ref{sec:4}, we review and extend the computation of the coefficient of $D^8 \cR^4$ in the expansion of the  two-loop  supergravity amplitude  using world-line techniques, and compare this with the tropical limit of the non-separating degeneration of the string amplitude.

\subsection{Acknowledgments}
MBG and BP are grateful to the Mani L. Baumik Institute at UCLA for hospitality  during the early part of this project. All authors are grateful to the organizers of the workshops {\sl Automorphic forms, Mock modular forms and String theory} at the Banff International Research Station, and {\sl 50 years
of the Veneziano Model} at the Galileo Galilei Institute in Florence, where part of this work was completed. ED would like to thank DAMTP in Cambridge, LPTHE at Jussieu, and LPTENS at the Ecole Normale Sup\'erieure in Paris for their hospitality during the final part of this work.  MBG would like to thank Pierre Vanhove and Arnab Rudra for conversations related to this work.

\sm

The research of ED  is supported in part by the National Science Foundation under research grant PHY-16-19926, and by  a Fellowship from the Simons Foundation.  MBG has been partially supported by STFC consolidated grant ST/L000385/1 and by a Leverhulme Fellowship. The research of BP
is supported in part by French state funds managed by the Agence Nationale de la Recherche (ANR) in the context of the LABEX ILP (ANR-11-IDEX-0004-02, ANR-10- LABX-63).

\newpage

\section{Structure of genus-two string invariants}
\setcounter{equation}{0}
\label{sec:2}

The genus-two contribution to the four-graviton scattering amplitude $\cA^{(2)} (\epsilon_i, k_i)$ is proportional to an  integral of a scalar function $\cB^{(2)} (s_{ij} | \Omega)$ over the moduli space $\cM_2$ of genus-two compact Riemann surfaces \cite{D'Hoker:2005jc,D'Hoker:2005ht}, 
\bea
\label{gen2amp}
\cA^{(2)} (\epsilon_i, k_i)
 = 
\frac{ \pi}{4} \kappa _{10}^2 \, g_s^2 \, \cR^4 
\int_{\cM_2} \frac{ |d^3 \Omega|^2}{ (\det Y)^3} \,
\cB^{(2)} (s_{ij} | \Omega)
\eea
The polarization tensors and momenta of the four gravitons are respectively denoted by $\epsilon_i$ and $k_i$ with $i=1,2,3,4$. The  kinematic invariants are defined by $s_{ij}=-\alpha' k_i\cdot k_j/2$ and satisfy the relations $s_{12}=s_{34}$, $s_{13}=s_{24}$, $s_{14}=s_{23}$, and $s_{12}+s_{13}+s_{14}=0$ due to momentum conservation. The  gravitational constant in ten-dimensional space-time is denoted by  $\kappa_{10}^2$, while $\alpha'$ is the string scale.  The quantity $\cR^4$ represents a particular scalar contraction of four powers of the linearized Riemann curvature tensor whose detailed form {can be found in \cite{Green:1981yb} and} is dictated by supersymmetry. Its explicit expression will not be needed  here.  
 
 \sm
 
The period matrix $\Omega \in \cH_2$ parametrizes  the complex structure of a genus-two  Riemann surface $\SM$. Given a choice of canonical homology cycles $\mA_I,\mB_I$ for $I=1,2$, and dual holomorphic one-forms $\omega_I$  on $\SM$, the period matrix is defined by, 
\bea
\label{AB}
\oint _{\mA_I} \om_J = \delta _{IJ} 
\hskip 1in 
\oint _{\mB_I} \om_J = \Omega _{IJ} 
\eea
The Riemann bilinear relations imply that $\Omega$ is a symmetric matrix, and that its imaginary part $Y$ is positive definite, along with the following integral relation,\footnote{Throughout, we shall use the Einstein summation convention for the indices $I,J=1,2$ which implies summation over any repeated lower and upper index of the same name. When no confusion is expected to arise, we shall not exhibit the dependence on the moduli and the coordinates  in differential forms, but we shall  exhibit these dependences for functions.}
\bea
\label{2a2}
{ i \over 2} \int _{\SM} \om_I \wedge \oom ^J = \delta _I {}^J 
\hskip 1in 
\oom ^J=  (Y^{-1} )^{JK} \, \overline{\om _K} 
\eea
where $Y^{-1}$ denotes the inverse of the matrix $Y$.
Further properties of the period matrix, including its behavior under modular transformations $Sp(4,\ZZ)$,   are well-known and are reviewed, for example, in subsection 2.2 of \cite{DHoker:2017pvk}. The measure factor in the integrand of (\ref{gen2amp}) is the $Sp(4,\ZZ)$-invariant  volume form for the Siegel metric on $\cH_2$, which will not be needed  in the sequel. By construction, the function $\cB^{(2)} (s_{ij} | \Omega)$ will be invariant under $Sp(4,\ZZ)$ transformations of $\Omega$, for arbitrary values of $s_{ij}$.  Therefore, the integral in (\ref{gen2amp}) is intrinsically defined, and we may represent $\cM_2$ by a fundamental domain $\cH_2/Sp(4,\ZZ)$ for the action of the modular group $Sp(4,\ZZ)$ on the Siegel upper half space $\cH_2$.

\subsection{Genus-two string invariants}
 
The function  $\cB^{(2)}(s_{ij} | \Omega)$ is the starting point of our study, and may be viewed as a generating function in the parameters $s_{ij}$ for genus-two modular graph functions derived from string theory.  It is given by an integral over four points $z_i$, corresponding to the four gravitons,   on a genus-two Riemann surface $\SM$ with period matrix $\Omega$,\footnote{In the earlier paper \cite{DHoker:2017pvk}, a Riemann surface of genus $h$ was denoted $\Sigma _h$ and we used the notations $\kappa_h$ for the  canonical K\"ahler form, $\cG_h$ for the Arakelov Green function, and $G_h$ for the string Green function. Since the present paper deals exclusively with genus-two surfaces and their degenerations, we shall drop the subscript ``2" throughout, and use the notation $g=\cG_1$ for the standard Green function on the torus.}
\bea
\label{B2g}
\cB ^{(2)} (s_{ij} |\Omega) 
=  { 1 \over 16} \int _{\SM ^4} { \cY \wedge \bar \cY \over (\det Y)^2}
\exp \left \{ \sum _{1 \leq i < j \leq 4} s_{ij} \, \GA(z_i, z_j |\Omega) \right \}
\eea
Here, $\cY$ is a holomorphic  $(1,0)$-form in each point $z_i \in \SM$ and is linear in the  $s_{ij}$.  It was constructed in  \cite{D'Hoker:2005jc}, and is given for example  in eq. (2.29) of \cite{DHoker:2017pvk}, but its explicit form will be needed only for a special arrangement of the parameters $s_{ij}$ which will be given below. 

\sm

The Green function  $\cG(z_i,z_j|\Omega)$ is formally the inverse of the Laplace operator on $\SM$. Due to the zero mode of the Laplace operator, the inverse is not unique, and generally fails to be conformal invariant. 
However, momentum conservation relations between the parameters~$s_{ij}$ imply that the exponential in (\ref{B2g}) is invariant under the following shift of $\cG$, 
\bea
\label{shift}
\cG(z_i,z_j|\Omega) \to \cG(z_i,z_j|\Omega) + c(z_i) + c (z_j)
\eea 
for an arbitrary function $c$. Since a conformal transformation on $\cG$ produces a shift in $\cG$ which is precisely of the form (\ref{shift}) (see for example \cite{D'Hoker:1988ta}), the exponential in (\ref{B2g}) and thus  $\cB^{(2)}(s_{ij}|\Omega)$ are  conformal invariant, as is essential in a consistent string theory amplitude. Two convenient choices will be used below for $\cG$, the first being the familiar string Green function  \cite{D'Hoker:1988ta}, the other being the Arakelov Green function (see for example \cite{AlvarezGaume:1987vm}), both of which will be discussed in subsection \ref{Arak}. We note that the behavior as $z_j\to z_i$ of either of these scalar Green functions is $\cG(z_i,z_j|\Omega) \approx - \ln |z_i-z_j|^2$.
 
 \sm
 
The integrals  in (\ref{B2g}) are absolutely convergent for $\Re(s_{ij}) <1$, and admit a Taylor series in powers of $s_{ij}$ with unit radius of convergence in $s_{12}, s_{13}$ and $s_{14}$. Expanding the exponential in powers of $s_{ij}$, using the relation $s_{12}+s_{13}+s_{14}=0$ and the fact that $\cB^{(2)} (s_{ij} |\Omega)$  is symmetric in the variables $s_{ij}$,  leads to an infinite series which may be organized as follows  \cite{Green:1999pv},
\bea
\label{Bpq}
\cB^{(2)} (s_{ij} |\Omega) 
 = \sum_{p,q=0}^\infty  \cB_{(p,q)}(\Omega) \, { \sigma _2^p \, \sigma _3^q \over p! \, q!}
\eea
where $\sigma_n$ are symmetric polynomials in $s_{ij}$ defined by $\sigma _n= (s_{12})^n + (s_{13})^n +(s_{14})^n$. Since $\cB^{(2)} (s_{ij}|\Omega)$ is $Sp(4,\ZZ)$-invariant for any value of $s_{ij}$, each coefficient $\cB_{(p,q)}  (\Omega)$  is itself $Sp(4,\ZZ)$-invariant and defines a genus-two modular graph function, in the sense of \cite{DHoker:2017pvk}. To identify the structure of the effective interaction to which each coefficient corresponds, it will be convenient to introduce the notion of {\sl weight}, defined as the number of Green function $\cG$ factors in the Taylor series expansion in powers of $s_{ij}$, which is  given by, 
\bea
w = 2p+3q-2
\label{wdef}
\eea
The linearity of $\cY$ in $s_{ij}$ implies that a modular graph function $\cB_{(p,q)}(\Omega)$ of weight $w$ corresponds to an effective interaction of the form $D^{2w+4} \cR^4$. For low weights, $w \leq 3$, there is a unique  effective interaction for each $w$, but  for $w\geq 4$ several independent effective interactions may correspond to the same weight.

\subsection{Convergence of the integrals over $\cM_2$}

The integrals in (\ref{B2g}) over the points $z_i \in \SM$ are absolutely convergent for any point $\Omega$ in the interior of moduli space as long as $\Re(s_{ij})<1$. However,  this convergence is non-uniform as~$\Omega$ moves to the boundary of $\cM_2$ so that the summation in (\ref{Bpq}) and the integration in (\ref{gen2amp}) cannot be legitimately  interchanged. Mathematically, this is due to the fact that $\cG$ grows linearly with $Y$, so that the domain of absolute convergence of the integral over $\cM_2$ is restricted to $\Re(s_{ij})=0$. Away from this set, analytic continuation in~$s_{ij}$ is required, and may be carried out along similar lines as the construction of the genus-one amplitude in \cite{DHoker:1994gnm}. Physically, the divergences arise because non-analyticities, such as logarithmic branch cuts, in the variables $s_{ij}$ are produced by this analytic continuation, and these functions  cannot be expanded in  a convergent Taylor series at $s_{ij}=0$.

\sm

Even when the integrals of $\cB_{(p,q)}(\Omega)$ over $\cM_2$ are not absolutely convergent, due to the appearance of non-analyticities in $s_{ij}$ as explained above, it is still possible to extract the strength of the corresponding effective interactions. However, this requires isolating the contribution  to the  integral from the boundary of $\cM_2$ first, carrying out its analytic continuation, and then identifying the low energy expansion of the analytic remainder of the amplitude. Carrying out this procedure will be the subject of future work \cite{DGP2018}.

\subsection{Low weights: the  Kawazumi-Zhang invariant}

In this subsection, we briefly review the results for the string invariants for weights $w\leq 1$.
Since $\cY$ is linear in $s_{ij}$, we have $\cB_{(0,0)}(\Omega)=0$, reflecting the absence of genus-two corrections to the effective interaction $\cR^4$.  Weight zero corresponds to the effective interaction $D^4\cR^4$ whose coefficient $\cB_{(1,0)}(\Omega)$ is constant on $\cM_2$ (equal to $-2$ in our conventions) and, upon integration over $\cM_2$, provides an important consistency check with the implications of S-duality in Type~IIB superstrings \cite{D'Hoker:2005ht}. 

\sm

The weight $w=1$ coefficient $\cB_{(0,1)}(\Omega)$ of the effective interaction $D^6\cR^4$ is proportional to the Kawazumi-Zhang invariant $\f (\Omega) $ for genus two, specifically $B_{(0,1)}(\Omega) =4 \, \f(\Omega) $. The contribution from the exponential in (\ref{B2g}) is linear in $\cG$ and therefore one may integrate explicitly over two of the four points in (\ref{B2g}), using (\ref{2a2}), giving the following formula, 
\bea
\f (\Omega) = - { 1 \over 8} \Big ( 2 \, \delta _{J_1}{}^{I_2} \delta _{J_2}{}^{I_1} - \delta _{J_1}{}^{I_1} \delta _{J_2}{}^{I_2} \Big ) 
\int _{\SM^2} \om_{I_1} (z_1)\, \oom ^{J_1}(z_1)\, \om_{I_2} (z_2) \, \oom^{J_2}(z_2)\, \cG(z_1, z_2|\Omega)
\eea
The complete asymptotics of $\f (\Omega)$ is known thanks to the theta-lift representation established
in \cite{Pioline:2015qha}, based on the Laplace-Beltrami eigenvalue equation derived in 
\cite{DHoker:2014oxd} and on known leading asymptotics \cite{MR1105425,zbMATH06355718}.
Its integral over $\cM_2$ can be computed using the eigenvalue equation and is also in 
agreement with S-duality \cite{DHoker:2014oxd}.

\subsection{The string invariant $\cB_{(2,0)}$}

At weight $w=2$, there is a single effective interaction, of the form $D^8 \cR^4$ corresponding to $p=2$ and $ q=0$ in (\ref{Bpq}) for a single kinematic invariant $(\sigma _2)^2$. The explicit form of the corresponding string invariant was given in \cite{D'Hoker:2013eea,DHoker:2017pvk}. It may be obtained  by expanding (\ref{B2g}) to second order in $s_{ij}$ (equivalently to second order in $\cG$), and  setting $s_{13}=0$ so that the expression for $\cY$ simplifies. The function $ \cB_{(2,0)} (\Omega) $ is given as follows,
\be
\label{defcB}
\cB_{(2,0)} (\Omega) =\int _{\SM ^4} \frac{|\Delta(z_1,z_3)\Delta(z_2,z_4)|^2}{ 64 \,  (\det Y)^2}
\Big ( \cG(z_1,z_2|\Omega) + \cG(z_3,z_4|\Omega) - (z_1 \leftrightarrow z_3)  \Big ) ^2
\ee
where 
$ \Delta(z_i,z_j)$ is the holomorphic two-form on $\SM^2$ defined by,
\be
\label{defDelta}
\Delta(x,y) = \omega_1(x)\wedge \omega_2(y)-\omega_2(x)\wedge \omega_1(y)
\ee
The string invariant $\cB_{(2,0)} (\Omega)$ will be the central object of our study in this paper. 

\sm

\begin{figure}[h]
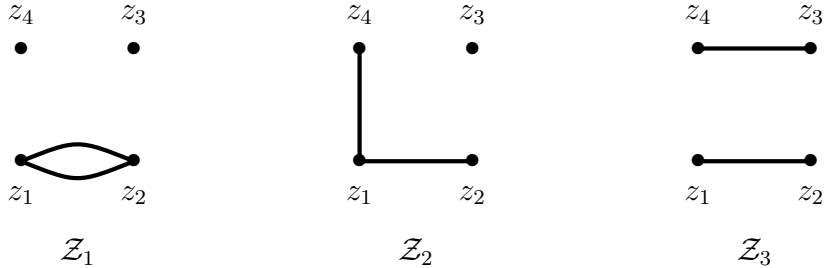

\begin{center}
\tikzpicture[scale=1.5]
\scope[xshift=-5cm,yshift=0cm]
\draw (0,0) node{$\bullet$};
\draw (0,1) node{$\bullet$};
\draw (1,0) node{$\bullet$};
\draw (1,1) node{$\bullet$};
\draw (0,-0.3) node{$z_1$};
\draw (1,-0.3) node{$z_2$};
\draw (0,1.3) node{$z_4$};
\draw (1,1.3) node{$z_3$};
\draw (0.5, -0.8) node{$\cBZ_1$};
\draw[ultra thick] (0,0) .. controls (0.5, -0.2) .. (1,0);
\draw[ultra thick] (0,0) .. controls (0.5, 0.2) .. (1,0);
\draw (3,0) node{$\bullet$};
\draw (3,1) node{$\bullet$};
\draw (4,0) node{$\bullet$};
\draw (4,1) node{$\bullet$};
\draw (3,-0.3) node{$z_1$};
\draw (4,-0.3) node{$z_2$};
\draw (3,1.3) node{$z_4$};
\draw (4,1.3) node{$z_3$};
\draw (3.5, -0.8) node{$\cBZ_2$};
\draw[ultra thick] (3,0) -- (4,0);
\draw[ultra thick] (3,0) --  (3,1);
\draw (6,0) node{$\bullet$};
\draw (6,1) node{$\bullet$};
\draw (7,0) node{$\bullet$};
\draw (7,1) node{$\bullet$};
\draw (6,-0.3) node{$z_1$};
\draw (7,-0.3) node{$z_2$};
\draw (6,1.3) node{$z_4$};
\draw (7,1.3) node{$z_3$};
\draw (6.5, -0.8) node{$\cBZ_3$};
\draw[ultra thick] (6,0) -- (7,0);
\draw[ultra thick] (6,1) --  (7,1);
\endscope
\endtikzpicture
\caption{Graphs representing the three distinct contributions $\cBZ_i$ to $\cB_{(2,0)}$ in (\ref{totalcB}), where each line represents a factor of the Green function $\cG$.} \label{fig:1}
\end{center}
\end{figure}

In the sequel, we shall find it useful to  decompose $\cB_{(2,0)}(\Omega)$ into a sum of three terms by expanding the integrand in (\ref{defcB}),
\bea
\cB_{(2,0)} (\Omega)= \half \cBZ_1(\Omega) - \cBZ_2(\Omega) + \half \cBZ_3(\Omega)
\label{totalcB}
\eea
where the 8-fold symmetry group generated by the permutations $(3214)$, $(1432)$ and $(3412)$ 
allows us to reduce the terms bilinear in the Green functions as follows,
\bea
\label{defcB1}
\cBZ_1 (\Omega) &=& \int _{\SM ^4} \frac{|\Delta(z_1,z_3)\Delta(z_2,z_4)|^2}{ 8\, (\det Y)^2} \, 
  \cG(z_1,z_2|\Omega) ^2 
\no \\
\cBZ_2 (\Omega) &=& \int _{\SM ^4} \frac{|\Delta(z_1,z_3)\Delta(z_2,z_4)|^2}{8\, (\det Y)^2} \, 
\cG(z_1,z_2|\Omega ) \, \cG(z_1,z_4|\Omega) 
\no \\
\cBZ_3 (\Omega) &=& \int _{\SM ^4} \frac{|\Delta(z_1,z_3)\Delta(z_2,z_4)|^2}{8\, (\det Y)^2} \,
\cG(z_1,z_2|\Omega) \, \cG(z_3,z_4|\Omega)  
\eea
The modular graph function $\cB_{(2,0)}(\Omega)$ is invariant under shifts of  $\cG$ given in (\ref{shift}), just as the generating function $\cB^{(2)}(s_{ij}|\Omega)$ was from which $\cB_{(2,0)}(\Omega)$ is derived. Thus, $\cB_{(2,0)}(\Omega)$ is independent of the type of  Green function chosen to represent it, and is conformal invariant. However, once we split $\cB_{(2,0)}(\Omega)$ into a sum of three terms, as in (\ref{totalcB}), each individual term $\cBZ_i(\Omega)$  will generally fail to be invariant under the shifts (\ref{shift}), and  fail to be conformal invariant. This shortcoming may be remedied  by using  the conformal invariant Arakelov Green function in each term $\cBZ_i $. As a result, each $\cBZ_i$ will be a genuine  genus-two modular graph function in the sense of \cite{DHoker:2017pvk}, and may be represented graphically as in  Figure~\ref{fig:1}.    

The expressions for the modular graph functions $\cBZ_1$ and $\cBZ_2$ may be simplified by integrating over the points $z_3$ and $z_4$ in $\cBZ_1$ and the point $z_3$ in $\cBZ_2$ using (\ref{2a2}) and (\ref{defDelta}). 
The resulting expressions are as follows,
\bea
\label{defB2}
\cBZ_1(\Omega) &=& 8 \int _{\SM ^2} \kappa (z_1) \kappa(z_2) \, \cG(z_1,z_2|\Omega) ^2 
\no \\
\cBZ_2(\Omega) &=&\int _{\SM ^3} \kappa (z_1) \frac{| \Delta(z_2,z_4)|^2}{ \det Y} \,
\cG(z_1,z_2|\Omega) \, \cG(z_1,z_4|\Omega) 
\eea
{where $\kappa(z)$ is the canonical K\"ahler form defined in \eqref{defkap}.}

We close this subsection by noting two further motivations for splitting $\cB_{(2,0)}$ into three individual modular graph functions $\cBZ_i$. One motivation will be to establish detailed, graph by graph  agreement with the supergravity calculations -- even though the original supergravity calculation for the complete  $\cB_{(2,0)}$ integrand \cite{Green:2008bf} is much  simpler 
than the one required for each of the  three terms separately.    A second motivation stems from the ultimate goal to obtain algebraic and differential equations satisfied by genus-two modular graph functions of weight $w \geq 1$. Experience with the corresponding problem at genus one has revealed that, for high enough weight, one has to deal with a system of equations involving several modular graph functions rather than a single equation for a single function \cite{D'Hoker:2015foa}. Therefore, it may be useful to build a ``library" of  functions such as the  individual modular graph functions $\cBZ_i$.

\subsection{The Arakelov Green function}
\label{Arak}

As discussed in the preceding subsection, the use of the Arakelov Green function is crucial for obtaining a decomposition of the integrand  (\ref{totalcB}) into a sum of well-defined conformal invariant modular graph functions $\cBZ_i$. In this subsection, we review the salient features of the Arakelov Green function on a genus-two Riemann surface $\SM$, and evaluate it concretely. The starting point is the canonical K\"ahler form $\kappa$ normalized to unit integral,
\bea
\label{defkap}
\kap = \frac{i}{4} \, \omega_I \wedge \oom^I = \frac{i}{4} (Y^{-1})^{IJ} \om_I \wedge \oom _J
\hskip 1in
\int _{\SM} \kappa =1
\eea
The canonical K\"ahler form  depends only on the holomorphic one-forms $\om_I$ and their periods. It is conformal and modular invariant, and uniquely determined by its  integral over $\SM$.

\sm

The Arakelov Green function $\GA(z,y|\Omega)$ on a genus-two Riemann surface $\SM$ is a real-valued symmetric function  on $\SM \times \SM \times \cH_2$, which provides an inverse to the scalar Laplace operator on $\SM$ equipped with the canonical K\"ahler form $\kappa$. Expressing $\kappa$ in local complex coordinates $ \kap = {i \over 2}  \kap _{z\bar z} (z)\, dz \wedge d\bar z$,  the Arakelov Green function is defined by the following equations,\footnote{Throughout, the ``coordinate" Dirac $\delta$-function is   normalized by ${ i \over 2} \int _\Sigma  dz \wedge d \bar z \, \delta ^{(2)} (z,y)=1$. }
\bea
\label{AraG}
 \pbz  \, \p_z \, \GA(z,y|\Omega )  =  - \pi \, \delta^{(2)} (z,y) +  \pi \, \kap _{z \bar z}  (z)  
 \hskip 0.8in 
 \int _{\SM} \kap (z) \, \GA(z,y|\Omega) = 0
\eea
An explicit expression for $\GA$ may be obtained by relating it to another Green function $G$ which is often used in string theory \cite{D'Hoker:1988ta}, and defined by, 
\bea
\label{G}
G(x,y|\Omega) = - \ln |E(x,y|\Omega)|^2 + 2 \pi \, \Im \left ( \int _y ^x \omega _I \right ) (Y^{-1})^{IJ} \, \Im \left ( \int _y ^x \omega _J \right )
\eea
where $E(x,y|\Omega)$ is the prime form \cite{fay73}, which is a holomorphic form of weight $(-1/2,0)$ in each variable $x,y$ on the covering space of $\SM^2$. As a result, the Green function $G(x,y|\Omega)$ is not a conformal scalar.  Therefore, one should be careful to calculate with $G(x,y|\Omega)$ on a simply connected fundamental domain for $\SM$ obtained by cutting the surface along suitably chosen curves $\mA_I, \mB_I$. The Green functions $G$ and $\cG$ are related as follows,
\bea
\label{GAG}
\GA(x,y|\Omega) = G(x,y|\Omega) - \gamma (x|\Omega) - \gamma (y|\Omega) + \gamma_1(\Omega)
\eea
where
\bea
\label{gamma}
\gamma (x|\Omega) =  \int _{\SM} \kap (z) G(x,z|\Omega) 
\hskip 1in 
\gamma_1(\Omega)  = \int _{\SM} \kap (z) \gamma (z|\Omega)
\eea
These relations ensure that $\cG$ integrates to zero against the canonical K\"ahler form $\kappa$, as required by (\ref{AraG}), and that  the function $\cG$ defined by (\ref{GAG}) is invariant  when $G$ is shifted as in (\ref{shift}). As a result, the Arakelov Green function is conformal invariant, and each individual function $\cBZ_i$ defined in (\ref{defcB1}) will be a conformal invariant modular graph form, as promised earlier.
Henceforth, $\cG$ will denote the Arakelov Green function, and it will understood throughout that (\ref{defcB1}) is expressed in terms of the Arakelov Green function $\cG$.


\section{The non-separating degeneration}
\setcounter{equation}{0}
\label{sec:3}

In this section we shall obtain the non-separating degeneration, in the form given in (\ref{LaurentBpq}), of the genus-two modular graph functions $\cBZ_i$, which were expressed  in terms of the Arakelov Green function $\cG$ in (\ref{defcB1}). We begin  with a review of the methods developed in \cite{DHoker:2017pvk} for parametrizing and calculating this degeneration, apply the method to reproduce the degeneration of  $\cB_{(0,1)}$ (proportional to the Kawazumi-Zhang  invariant), and  then calculate the degenerations of  the string invariants $\cBZ_i$ and $\cB_{(2,0)}$.

\subsection{Funnel construction of the non-separating degeneration}

In the non-separating degeneration, a compact genus-two surface $\SM$ degenerates to a genus-one surface $\Sigma _1 \setminus \{ p_a, p_b \}$, where $\Sigma _1$ is a compact genus-one surface and $p_a,p_b$ are the two points which are the remnants of the degenerating funnel of the surface $\SM$. Our interest is in evaluating the expansion of the form (\ref{LaurentBpq}) in a neighborhood of the non-separating node where a non-trivial homology cycle of the surface $\SM$ becomes a long and skinny, but finite, funnel.  The imaginary part of the period matrix $Y$, introduced in (\ref{OmY}), and its inverse $Y^{-1}$,  may be parametrized as follows,
\bea
\label{3a1}
Y=\begin{pmatrix} \tau_2 & \tau_2 u_2 \cr \tau_2 u_2 & t+\tau_2 u_2^2\end{pmatrix}
\hskip 1in
Y^{-1} =\begin{pmatrix} \tau_2^{-1} & 0 \cr 0 & 0  \end{pmatrix}
+ { 1 \over t} \begin{pmatrix} u_2^2 & -u_2 \cr -u_2 & 1  \end{pmatrix}
\eea
where $v_2=\tau_2 u_2$ and $t= (\det Y)/\tau_2= \sigma _2 - \tau_2 u_2^2$. The non-separating degeneration corresponds to letting $t$ become large while keeping the other independent moduli finite. We stress that the above expression for $Y^{-1}$ in terms of $t$ is exact. 

\sm

The methods developed in \cite{DHoker:2017pvk} are tailored to obtaining the expansion of (\ref{LaurentBpq}), {\sl exactly to all orders in powers of $t$} while neglecting any contributions that vanish exponentially in the large $t$ limit.  To carry out the construction, the genus-two surface $\SM$ is parametrized in terms of $t$ as well. This parametrization may be approached from two opposite directions which are intimately connected and equivalent to one another. The first approach starts from the genus-two surface, degenerates the period matrix according to (\ref{3a1}) for large but finite $t$, and infers the degenerations of other functions and forms on $\SM$, such as the canonical K\"ahler form, the string Green function, and the Arakelov Green function. The second approach constructs the genus-two surface $\SM$, near a non-separating degeneration node, in terms of a compact genus-one surface $\Sigma _1$. As was shown in \cite{DHoker:2017pvk}, the link between these two approaches is a family of Morse   functions $f(z)$ which may be constructed from either approach. 

\sm

For our purpose, it will be convenient to construct $\Sigma$ starting from a compact torus $\Sigma _1$.
We shall denote by $g(z,y|\tau)=g(z-y|\tau)$ the genus-one scalar Green function on $\Sigma _1$ which, by translation invariance, depends only on the difference $z-y$, and obeys,
\bea
\tau_2 \pbz \p_z g(z |\tau) = - \pi \tau_2 \delta^{(2)}  (z) + \pi
\hskip 1in
\int _{\Sigma _1} dz \wedge d\bar z \, g(z|\tau)=0
\eea
Its explicit expression in terms of $\tet$-functions  is given by,
\bea
\label{5c1}
\gone(z|\tau) = - \ln \left | { \tet _1 (x-y |\tau) \over \eta (\tau) } \right |^2 + { 2 \pi \over \tau_2} \left ( \Im z \right )^2
\eea
We add two punctures $p_a,p_b$ on $\Sigma_1$ to produce a punctured genus-one surface $\Sigma _1 \setminus \{ p_a,p_b\}$.  On this surface we introduce the Morse-type function $f$ defined by, 
\bea
f(z|\tau) = g(z,p_b |\tau) - g(z,p_a|\tau)
\eea
We observe that $f$ is well-defined, single-valued, and  harmonic on $\Sigma _1 \setminus \{ p_a,p_b\}$, and tends to $-\infty$ as $z \to p_a$ and to $+\infty$ as $z\to p_b$. We define level sets $\mC_a$ and $\mC_b$ by,
\bea
\mC_a & = & \{ z \in \Sigma _1 \hbox{ such that } f(z) = - 2 \pi t \}
\no \\
\mC_b & = & \{ z \in \Sigma _1 \hbox{ such that } f(z) = + 2 \pi t \}
\eea
For sufficiently large values of $t$, each level set is connected and has the topology of a circle. (As $t$ is decreased, each level set ultimately becomes disconnected and splits into two circles.) We define the genus-one surface $\Sigma _{ab}$ with a boundary  $\p \Sigma _{ab}=\mC_a \cup \mC_b$ by cutting out of $\Sigma_1$ the two discs with boundaries $\mC_a$ and $\mC_b$, or equivalently, 
\bea
\Sigma _{ab} = \{ z \in \Sigma _1 \hbox{ such that } - 2 \pi t \leq f(z) \leq + 2 \pi t \}
\eea
The genus-two surface $\SM$ is obtained from $\Sigma _{ab}$ by gluing together its boundary curves $\mC_a$ and $\mC_b$. The full moduli space of $\SM$ requires identifying $\mC_a$ and $\mC_b$ after a twist by the angle $\Re (\sigma )$, where $\sigma$ is the bottom diagonal entry of $\Omega $ in (\ref{3a1}). However, the Laurent polynomial part  in the expansion of (\ref{LaurentBpq}) is independent of $\Re(\sigma )$, since any dependence on $\Re (\sigma)$ of a modular invariant function  must be exponential in $\sigma$ in view of the periodicity $\sigma \to \sigma +1$. Thus, the twist by $  \Re(\sigma)$ is immaterial for our purposes, and  may be ignored.

\sm

\begin{figure}[h]
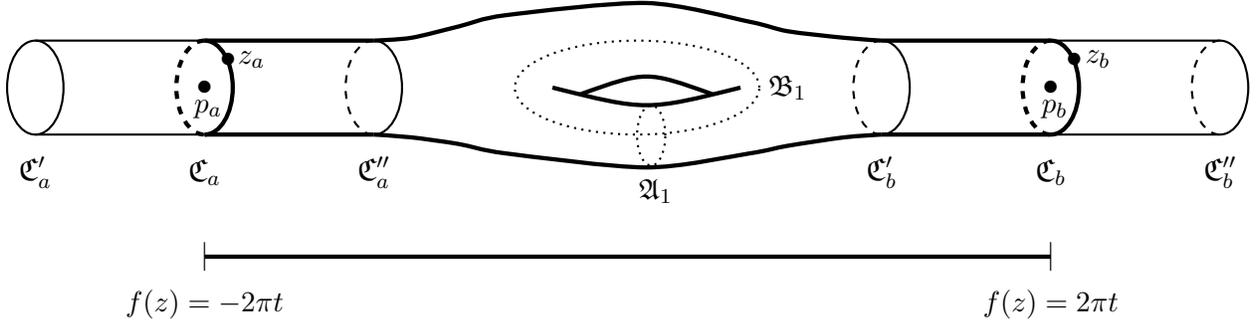

\begin{center}
\tikzpicture[scale=1.25]
\scope[xshift=-5cm,yshift=0cm]
\draw [thick] (0.3,-0.5) arc (0:360:0.3 and 0.5);
\draw [ultra thick] (1.8,-1) arc (-90:90:0.3 and 0.5);
\draw [ultra thick, dashed] (1.8,0) arc (90:270:0.3 and 0.5);
\draw [thick] (3.6,-1) arc (-90:90:0.3 and 0.5);
\draw [thick, dashed] (3.6,0) arc (90:270:0.3 and 0.5);
\draw [thick] (0,0) -- (1.8,0);
\draw [thick] (0,-1) -- (1.8,-1);
\draw [ultra thick] (1.8,0) -- (3.6,0);
\draw [ultra thick] (1.8,-1) -- (3.6,-1);
\draw[ultra thick] (5.5,-0.5) .. controls (6.5, -0.75) .. (7.5,-0.5);
\draw[ultra thick] (5.8,-0.57) .. controls (6.5, -0.32) .. (7.2,-0.57);
\draw [thick] (9,-1) arc (-90:90:0.3 and 0.5);
\draw [thick, dashed] (9,0) arc (90:270:0.3 and 0.5);
\draw [ultra thick] (10.8,-1) arc (-90:90:0.3 and 0.5);
\draw [ultra thick, dashed] (10.8,0) arc (90:270:0.3 and 0.5);
\draw [thick] (12.6,-1) arc (-90:90:0.3 and 0.5);
\draw [thick, dashed] (12.6,0) arc (90:270:0.3 and 0.5);
\draw [ultra thick] (9,0) -- (10.8,0);
\draw [ultra thick] (9,-1) -- (10.8,-1);
\draw [thick] (10.8,0) -- (12.6,0);
\draw [thick] (10.8,-1) -- (12.6,-1);
\draw[ultra thick] plot [smooth] coordinates {(3.6,0) (4,0.03) (4.5, 0.15) (5, 0.27) (6.5,0.4)  (7.6, 0.2) (8, 0.10) (8.6,0.03) (9,0)};
\draw[ultra thick] plot [smooth] coordinates {(3.6,-1) (4,-1.03) (4.5, -1.10)  (5, -1.2) (6.5,-1.35) (7.6, -1.2) (8, -1.12) (8.6,-1.03) (9,-1)};
\draw [thick, dotted] (7.7,-0.5) arc (0:360:1.3 and 0.5);
\draw [thick, dotted] (6.7,-1.01) arc (0:360:0.15 and 0.32);
\draw (0,-1.4) node{$\mC_a'$};
\draw (1.8,-1.4) node{$\mC_a$};
\draw (3.6,-1.4) node{$\mC_a''$};
\draw (9,-1.4) node{$\mC_b'$};
\draw (10.8,-1.4) node{$\mC_b$};
\draw (12.6,-1.4) node{$\mC_b''$};
\draw (1.8,-0.5) node{$\bullet$};
\draw (10.8,-0.5) node{$\bullet$};
\draw (1.84,-0.73) node{{\small $p_a$}};
\draw (10.84,-0.73) node{\small $p_b$};
\draw (2.05,-0.2) node{$\bullet$};
\draw (11.05,-0.2) node{$\bullet$};
\draw (2.3,-0.2) node{{\small $z_a$}};
\draw (11.3,-0.2) node{\small $z_b$};
\draw (8,-0.5) node{{\small $\mB_1$}};
\draw (6.6,-1.6) node{{\small $\mA_1$}};
\draw (1.8,-2.8) node{{\small $f(z)= - 2 \pi t$}};
\draw (10.8,-2.8) node{{\small $f(z)= 2 \pi t$}};
\draw [ultra thick] (1.8,-2.3) -- (10.8,-2.3);
\draw[ultra thick]  (1.8,-2.3) node{{\small $|$}};
\draw [ultra thick] (10.8,-2.3) node{{\small $|$}};
\endscope
\endtikzpicture
\caption{The funnel construction near the non-separating divisor of a genus-two surface~$\Sigma$.  The surface  $\Sep $ is obtained from the compact surface $\Sigma_1$ by removing the discs with boundaries $\mC_a$ and $\mC_b$ centered at the punctures $p_a,p_b$ respectively. The surface $\Sigma$ is obtained from $\Sep$ by pairwise  identifying the cycles  $\mC_a \approx \mC_b$, (as well as identifying $\mC_a' \approx \mC_b'$ and $\mC_a'' \approx \mC_b''$). A canonical homology basis for $\Sigma$ is obtained by choosing the cycles $\mA_1, \mB_1$  of the surface $\Sigma_1$, along with a cycle $\mA_2$ homologous to  the cycles $\mC_a, \mC_a', \mC_a'',\mC_b, \mC_b', \mC_b''$. The  cycle $\mB_{2}$ is obtained by connecting $z_a$ to $z_b$ by a curve in $\Sigma _{ab}$ and identifying the points $z_a \approx z_b$. The punctures $p_a, p_b$ lie on $\Sigma_1$, but do not belong to either $\Sep$ or $\Sigma$. The function $f(z)$ is constant on  $\mC_a$ and  $\mC_b$  and increases from $-2\pi t$ on $\mC_a$ to $2\pi t$ on  $\mC_b$. } \label{fig:2}
\end{center}
\end{figure}

To complete the construction of $\SM$, we specify a canonical homology basis $\mA_I, \mB_I$ for $H_1(\SM, \ZZ)$, and its dual basis of holomorphic one-forms $\om_I$ for $I=1,2$.  The cycles $\mA_1, \mB_1$ are chosen to be a canonical homology basis for $H_1(\Sigma _1, \ZZ)$,  the cycle $\mA_2$ is homologous to $ \mC_a \approx \mC_b$, and the cycle $\mB_2$ consists of a  curve which lies in $\Sigma _{ab}$ and which connects $\mC_a$ to $\mC_b$, as shown in Figure \ref{fig:2}. 

\sm

To specify the holomorphic one-forms on $\SM$, represented by $\Sigma _{ab}$ with identified boundary components, we represent the torus by $\Sigma _1=\CC/(\ZZ+ \tau \ZZ)$ and introduce a complex coordinate $z$ subject to the identifications $z \approx z+1$ along $\mA_1$ and $z \approx z+\tau$ along $\mB_1$.  The dual basis of  holomorphic one-forms then consists of the normalized holomorphic one-form $\omega_1$ on $\Sigma _1$  and the holomorphic one-form $\omega_2=\omega_t + u_2 \omega_1$ defined on $\Sigma _{ab}$ by,  
\be
\label{om1t}
\omega_1=d z \hskip 1in  \omega_t = \frac{i}{2\pi} \partial_z f(z)\, d z\ ,
\ee
It follows from the above construction that $\om_I$ is canonically normalized on the cycles $\mA_I$ as in (\ref{AB}), while on $\mB_I$ cycles we recover $\Omega$  of (\ref{3a1}) with $v=p_b-p_a$ and $t$ given by the construction above. Since we have not included the twist when identifying $\mC_a$ and $\mC_b$ we do not have access to the entry $\Re(\sigma)$ but, as argued earlier, we have no need for this variable here.
The non-separating degeneration corresponds to $t\to \infty$ keeping $\tau$ and $v$ fixed.  

\sm

\subsection{Degeneration of the Green functions}

The key to the striking results for the non-separating degeneration obtained in \cite{DHoker:2017pvk} is the use not of the naive modulus $\sigma_2$ but rather instead of the special parameter $t$ which is invariant under the Jacobi group, $SL(2,\mathbb{Z})\ltimes (\mathbb{Z}^2 \ltimes \mathbb{Z})$. It is in terms of $t$ that the power series expansion terminates and becomes a Laurent polynomial of finite degree. 

\sm

We begin by recalling the degeneration of the genus-two canonical K\"ahler form $\kappa$ in terms of the normalized genus-one K\"ahler form $\kappa_1$, 
\bea
\label{limkappa}
\kappa = \half  \kappa_1  +  \frac{i}{4 t} \omega_t \wedge \bar\omega_t + \cO(e^{-2\pi t})
\hskip 1in \kappa _1 = \frac{i}{2 \tau_2} \om_1 \wedge \oom_1
\eea
The degeneration of the string Green function $G$ is given by,\footnote{Henceforth, when no confusion is expected to arise, we shall often suppress the dependence on the periods $\tau$ and $\Omega$ to simplify and shorten the notations.}
\bea
\label{limG}
G(x,y) = \gone(x,y) + \frac{1}{8\pi t} \big (f(x) - f(y) \big )^2 + \cO(e^{-2\pi t}) 
\eea
while the degeneration of the Arakelov Green function $\cG(x,y)$ is given by,  
\bea
\label{limGA}
\GA(x,y) &=& 
 \frac{\pi t}{12} + \gone(x,y) -\frac14\Big ( \gone(x,p_a)+\gone(x,p_b)+ \gone(y,p_a)+\gone(y,p_b) -\gone(p_a,p_b)
\Big ) \nn \\
&& + \frac{1}{16\pi t} \Big (  f(x)^2 + f(y)^2 -4 f(x) f(y) -2 F_2(v) \Big )  + \cO(e^{-2\pi t}) 
\eea
Here and below, we find it useful to introduce the following notation, 
\bea
\label{Fn}
F_k (v)= { 1 \over k!} \int _{\Sigma _1} \kappa _1 (z) f(z)^k
\eea
Clearly, $F_k$ vanishes when $k$ is odd in view of translation and reflection symmetry of the genus-one Green function $\gone(z)$. For even $k$, we shall evaluate $F_k$ in (\ref{defFlk}) and (\ref{F4}). The combination $F_2$ may be evaluated explicitly, using its definition, and we find,
\bea
\label{F2}
F_2(v) = E_2 - g_2(v)
\eea
where $E_2$ is the genus-one non-holomorphic Eisenstein series, and $g_k$ is defined to be the genus-one Green function for $k=1$ and for higher values is defined recursively as follows,
\bea
\label{gn}
g_{k+1} (z) = \int _{\Sigma _1} \kappa _1 (x) \,  g(z,x) \, g_k(x)
\eea
The non-holomorphic Eisenstein series $E_k$ is simply related to $g_k$ by $E_k = g_k(0)$.

\sm

For the degeneration of both the Green functions $G$ and $\cG$, the asymptotic Laurent polynomial is at most of degree $(1,1)$. The functions $\gamma (x)$ and $\gamma _1$, which account for the difference between $G(x,y)$ and $\GA(x,y)$ through (\ref{GAG}), are given as follows,
\bea
\label{gamx1}
\gamma (x)   & = &   \frac{\pi t}{12}  + \frac14 \gone(x,p_a)+ \frac14 \gone(x,p_b)  + \frac{f(x)^2}{16\pi t}    + \frac{F_2(v)}{4\pi t} +\cO(e^{-2\pi t})
\no \\
\gamma _1 & =  & \frac{\pi t}{4}  + \frac14 \gone(v) + \frac{3 F_2(v)}{8\pi t} +\cO(e^{-2\pi t})
\eea
The invariants  $\cB_{(p,q)}$ that we wish to calculate involve integrals of powers of the Green function over the genus-two surface.  In view of the number of terms in \eqref{limGA} compared to \eqref{limG} it will prove convenient to first study the asymptotics of the  integrals defined using the Green function $G(x,y)$ rather than the Arakelov Green funtion.  We may then obtain the asymptotics of the individual modular graph invariants associated with the individual graphs, such as $\cBZ_i$ (in Figure~\ref{fig:1}) by a simple conversion formula involving $\gamma (x)$ and $\gamma _1$.

\subsection{Degeneration of the Kawazumi--Zhang invariant }

In terms of the Arakelov Green function $\cG$, the Kawazumi-Zhang invariant 
\bp{ is given by}
\bea
\label{KZA}
\f (\Omega) = - {1 \over 4}  \int _{\SM^2} \om_{I} (z_1) \oom ^{J}(z_1) \om_{J} (z_2)  \oom^{I}(z_2) \cG(z_1, z_2)
\eea
Its  degeneration limits were discussed in \cite{Pioline:2015qha}.   In the approach to the non-separating limit, $t\to \infty$,  it has an expansion, 
\bea
\varphi(\Omega)= \frac{1}{6} \pi t +\frac{1}{2} \gone(v)  + \frac{5 F_2(v)}{4 \pi t} + \cO(e^{-2 \pi t})
\label{nonkz}
\eea
In accord with the general Theorem on the non-separating degeneration   (\ref{LaurentBpq}), the Kawazumi-Zhang invariant indeed produces a Laurent polynomial in $t$ of degree $(1,1)$ with coefficients which are modular functions
and their generalizations to include the dependence on $v$. The precise nature of the generalization this entails will be spelled out in subsection \ref{poly}.

\subsection{Degeneration of the string invariant $\cB_{(2,0)}$}

The modular graph function $ \cB_{(2,0)}(\Omega)$ may be expressed, via the Arakelov Green function and \eqref{totalcB},  as a sum of three modular graph functions $\cBZ_i(\Omega) $, which were defined in (\ref{defcB1}) with simplified expressions given in (\ref{defB2}). To evaluate their non-separating degeneration, we observe in (\ref{limG}) and (\ref{limGA}) that the degeneration formula for the string Green function $G$ is simpler than the one for the Arakelov Green function $\cG$. Therefore,  we shall first calculate the degeneration of the analogues $Z_i$ of $\cBZ_i$ in which $G(x,y)$ is replaced by $\GA(x,y)$ in (\ref{defcB1}) and (\ref{defB2}), which leads to the following definitions,
\bea
\label{defB1}
\BZ_1(\Omega) &=& 8 \int _{\SM ^2} \kappa (z_1) \kappa(z_2) G(z_1,z_2) ^2 
\no \\
\BZ_2(\Omega) &=&\int _{\SM ^3} \kappa (z_1) \frac{\Delta(z_2,z_4)|^2}{ \det Y} \,
G(z_1,z_2) \, G(z_1,z_4) 
\no \\
\BZ_3(\Omega) &=&\int _{\SM ^4} \frac{|\Delta(z_1,z_3)\Delta(z_2,z_4)|^2}{8 \, (\det Y)^2} \,
G(z_1,z_2) \, G(z_3,z_4)  
\eea

Unlike the functions $\cBZ_i(\Omega)$, each individual function $\BZ_i(\Omega)$ fails to be conformally invariant, but the expression for $\cB_{(2,0)}(\Omega) $, which is given by,  
\bea
\label{defBx}
\cB_{(2,0)} = \half \cBZ_1 - \cBZ_2 + \half \cBZ_3= \half \BZ_1- \BZ_2+\half \BZ_3
\eea
clearly remains conformal invariant. The relations between the individual functions $\cBZ_i(\Omega) $ and $\BZ_i(\Omega)$ may be recovered in terms of three functions $\gamma _i (\Omega)$, 
\begin{align}
\cBZ_1 &= \BZ_1+8\gamma_1^2 - 16 \gamma_2 
& 
\gamma _1 & = \int _{\SM} \kappa (x) \gamma (x)
\no \\ && \no \\
\cBZ_2 &= \BZ_2 +8\gamma_1^2 - 8\gamma_2  -\gamma_3 
&
\gamma_2 & =  \int_{\SM}  \kappa(x) \gamma(x)^2 
\no \\ && \no \\
\cBZ_3 &= \BZ_3+8\gamma_1^2  -2\gamma_3
& 
\gamma_3 & =   \int_{\SM^2} \frac{ |\Delta(x,y)|^2}{\det Y} \gamma(x) \gamma(y) 
\label{BtocB}
\end{align} 
The calculations required to extract the $t$-dependence of these integrals are complicated and have been relegated to Appendix~\ref{sec:B}.  The key steps involved are as follows.
\begin{enumerate}
\itemsep=0in
\item The integrals of (\ref{defB1}) and (\ref{BtocB})  over the compact genus-two Riemann surface $\SM$ are expressed in terms of integrals over the genus-one surface $\Sep$ with boundary. The Green functions on $\SM$ are expressed in terms of $g$ and $f$ using (\ref{limG}), while the integration measure is expressed in terms of the genus-one differentials $\om_1$ and $\om_t$ using (\ref{limkappa}) and, 
\bea
\Delta (z_i, z_j)  = \om_1 (z_i) \wedge \om_t (z_j) - \om_t(z_i) \wedge \om_1(z_j)  
\eea
\item The remaining determinant factor is given by $\det Y = t \tau_2$.
\item The integrals over $\Sep$ obtained in this manner are then analyzed and recast in the form of a Laurent polynomial in $t$ with coefficients which can be expressed as convergent integrals over the compact genus-one Riemann surface $\Sigma _1$. The difficulty involved in this last step is strongly correlated with the structure of the associated Feynman graph and its renormalization properties, and will be given systematically in Appendix~\ref{sec:B}.
\end{enumerate}

The resulting expressions involve various modular functions and their generalizations which will be defined and discussed in subsection \ref{poly} and  in Appendix~\ref{sec:B}.   The functions $\BZ_i(\Omega)$ and $\gamma _i(\Omega)$ will be computed in Appendix~\ref{sec:B}, and give the following result for $\cBZ_i$,
\bea
\cBZ_1 (\Omega) 
& = &  
\frac{13\pi^2 t^2}{90}  
+ \frac{\pi t}{3} \gone(v) 
+ 4 E_2+ \half \gone(v)^2- \half F_2(v)
\no \\ &&
+ \frac{1}{ \pi t} \Bigg ( - D_3- D_3^{(1)}(v) 
 - \half \gone F_2(v)   + 2 g_3(v) + 4 \zeta (3) +{1 \over 4 \pi} \Delta _v F_4(v)  \Bigg )
 \no \\ &&
 + { 1 \over 8 \pi^2 t^2} \bigg ( 3F_2(v) ^2 + 12 F_4(v) +\cK^c(v)  \bigg ) 
 + \cO(e^{-2\pi t})
\no \\
\cBZ_2 (\Omega)
&= & 
-\frac{7\pi ^2 t^2}{90}
-\frac{\pi t}{3}  \gone(v) -E_2 - \half \gone(v) ^2 + \half F_2(v) 
\no \\ && 
+ \frac{1}{ \pi t} \Bigg ( - 2 D_3   + \half  \gone(v) F_2(v) 
 + 2 g_3(v) + 2  \zeta(3) - \frac{1}{16\pi} \Delta_v \left ( F_2(v)^2 + 2 F_4(v) \right )  \Bigg )
 \no \\ && 
 -\frac{ (\Delta_\tau + 5)F_4(v)}{4 \pi ^2 t^2}   + \cO(e^{-2\pi t})
  \no  \\
\cBZ_3 (\Omega) &=& 
\frac{ (\pi t)^2} {18}
+\frac{\pi t}{3} \gone(v) 
+\frac{1}{6}   F_2(v)+ \half    \gone(v)^2
  +  \frac{1}{ \pi t}  \left( - \half \gone(v) F_2(v) + \frac{1}{8\pi} \Delta_vF_2(v)^2 \right)\nn\\
&& + \frac{1}{ 8 \pi^2 t^2} ( \Delta_\tau +5)F_2(v)^2 + \cO(e^{-2\pi t})
\label{bark3}
\eea
The total string invariant $\cB_{(2,0)}$ is then obtained from (\ref{totalcB}) and is given by, 
\bea
\label{B20minsep}
\cB_{(2,0)} (\Omega)
&=&\frac{8 \pi^2  t^2}{45} + \frac{2\pi t}{3} \gone (v) 
+ 3 E_2 +  \gone(v)^2 - \frac{2}{3}F_2(v) 
\nn \\
&&+ \frac{1}{\pi t}   \left({3 \over 2}  D_3- \half D_3^{(1)}(v)   -  g_3(v) 
-   \gone(v) F_2(v) + \frac{1}{8\pi} 
\Delta_v \left ( F_2(v)^2 + 2 F_4(v) \right )  \right)
\nn\\
&& + \frac{1}{16 \pi^2 t^2}   \left(\Delta_\tau + 8 \right)\, \Big ( F_2(v)^2+ 4 F_4(v) \Big )   
+ {\cK^c(v) \over 16 \pi^2 t^2}  + \cO(e^{-2\pi t})
\label{barktot}
\eea
The definition of the various modular graph functions involved in these results will be  given   in the next subsections, while the corresponding  derivations are relegated to Appendix~\ref{sec:B}.

\subsection{Modular graph functions occurring in $\cBZ_i$}

\begin{figure}[h]
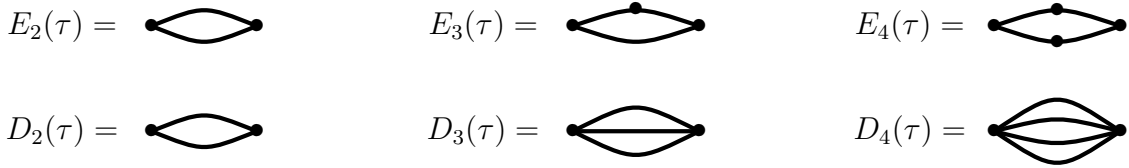

\begin{center}
\tikzpicture[scale=1.4]
\scope[xshift=0cm,yshift=0cm]
\draw (0.15,-1) node{$E_2(\tau)=$};  
\draw[ultra thick] (1,-1) .. controls (1.5, -1.2) .. (2,-1);
\draw[ultra thick] (1,-1) .. controls (1.5, -0.8) .. (2,-1);
\draw (1,-1) node{$\bullet$};
\draw (2,-1) node{$\bullet$};
\draw (4.15,-1) node{$E_3(\tau)=$};  
\draw[ultra thick] (5,-1) .. controls (5.6, -1.2) .. (6.2,-1);
\draw[ultra thick] (5,-1) .. controls (5.6, -0.8) .. (6.2,-1);
\draw (5,-1) node{$\bullet$};
\draw (6.2,-1) node{$\bullet$};
\draw (5.6,-0.84) node{$\bullet$};
\draw (8.2,-1) node{$E_4(\tau)=$};  
\draw[ultra thick] (9,-1) .. controls (9.6, -1.2) .. (10.2,-1);
\draw[ultra thick] (9,-1) .. controls (9.6, -0.8) .. (10.2,-1);
\draw (9,-1) node{$\bullet$};
\draw (10.2,-1) node{$\bullet$};
\draw (9.6,-0.85) node{$\bullet$};
\draw (9.6,-1.15) node{$\bullet$};
\draw (0.15,-2) node{$D_2(\tau)=$};  
\draw[ultra thick] (1,-2) .. controls (1.5, -2.2) .. (2,-2);
\draw[ultra thick] (1,-2) .. controls (1.5, -1.8) .. (2,-2);
\draw (1,-2) node{$\bullet$};
\draw (2,-2) node{$\bullet$};
\draw (4.15,-2) node{$D_3(\tau)=$};  
\draw[ultra thick] (5,-2) .. controls (5.6, -2.3) .. (6.2,-2);
\draw[ultra thick] (5,-2) -- (6.2,-2);
\draw[ultra thick] (5,-2) .. controls (5.6, -1.7) .. (6.2,-2);
\draw (5,-2) node{$\bullet$};
\draw (6.2,-2) node{$\bullet$};
\draw (8.2,-2) node{$D_4(\tau)=$};  
\draw[ultra thick] (9,-2) .. controls (9.6, -2.15) .. (10.2,-2);
\draw[ultra thick] (9,-2) .. controls (9.6, -1.85) .. (10.2,-2);
\draw[ultra thick] (9,-2) .. controls (9.6, -2.4) .. (10.2,-2);
\draw[ultra thick] (9,-2) .. controls (9.6, -1.6) .. (10.2,-2);
\draw (9,-2) node{$\bullet$};
\draw (10.2,-2) node{$\bullet$};
\endscope
\endtikzpicture
\caption{Modular graph functions $E_k(\tau)$ and $D_k(\tau)$.  } \label{fig:3}
\end{center}
\end{figure}

All the genus-one modular graph functions and their generalizations occurring in the non-separating degeneration of the genus-two modular graph functions $\cBZ_i$ are built from the canonical volume form $\kappa_1(x)$ and the scalar Green function $g(x,y)=g(x-y)$ on the compact genus-one surface $\Sigma_1$. The most familiar such function is the non-holomorphic Eisenstein series $E_k$ which may be defined by,\footnote{In this subsection and the next, we exhibit the dependence on moduli for added clarity.}
\bea
\label{defEk}
E_k(\tau) =  \prod _{i=1}^k  \int _{\Sigma _1}\kappa _1 (z_i) \, g(z_i-z_{i+1}|\tau) 
=\sum _{m,n \in \ZZ}^\prime { \tau_2^k \over \pi^k |m+ \tau n|^{2k}}
\eea
where $z_{k+1}=z_1$, and the prime on the sum indicates that the term $m=n=0$ is omitted. Another familiar family of modular graph functions is defined by,
\bea
\label{defDk}
D_k(\tau) =  \int _{\Sigma _1}\kappa _1 (z) \, g(z|\tau )^k 
=\sum _{m_r,n_r \in \ZZ}^\prime \delta _{m,0} \delta _{n,0}
\prod _{r=1}^k { \tau_2 \over \pi |m_r+ \tau n_r|^2}
\eea
where $m=m_1+ \cdots + m_k$ and $n = n_1 +\cdots + n_k$. Both $E_k(\tau)$ and $D_k(\tau)$ are given by convergent integrals and sums for $k \geq 2$, are invariant under $SL(2,\ZZ)$ modular transformations of $\tau$, and obey the following coincidence relations $D_2=E_2$ and $D_3 = E_3 +\zeta(3)$. Their graphical representation is well known, and illustrated in Figure \ref{fig:3}. 

\subsection{Generalized modular graph functions occurring in $\cBZ_i$}
\label{poly}

The remaining coefficients with non-trivial $\tau$-dependence in $\cBZ_i$ also depend on the punctures through the combination $v=p_b-p_a$. The simplest of these is given by the Green function $g(v)=g(v|\tau)$ itself. Closely related are the iterated Green functions, $g_k(v)=g_k(v|\tau)$, which may defined recursively by (\ref{gn}), or in terms of a Kronecker-Eisenstein sum by, 
\bea
\label{defgk}
g_k(v |\tau)  =\sum _{m,n \in \ZZ}^\prime { \tau_2^k \, 
e^{ 2 \pi i (m u_2 -n u_1)} \over \pi^k |m+ \tau n|^{2k}} 
\eea
where $u_1, u_2$ are real and defined by $v = u_1 + \tau u_2$. For $k=1$, we recover the Green function, $g_1(v|\tau)=g(v|\tau)$, while we also have $ g_k(0|\tau)=E_k(\tau)$. The functions $g_k(v|\tau)$ are invariant under $SL(2,\ZZ)$ transformations, 
\bea
\label{modgk}
g_k ( v' | \tau' ) = g_k (v|\tau)
\hskip 0.7in 
v' = { v \over c \tau +d}
\hskip 0.5in 
\tau' = { a \tau + b \over c \tau +d} 
\eea
for $a,b,c,d \in \ZZ$ and $ad-bc=1$. The transformation induced on the real variables $u_1, u_2$ 
is linear and given by $u_1 ' = a u_1 - b u_2$ and $u_2 '= - c u_1 + d u_2$. Thus, $(u_1, u_2)$ provides a $(\RR/\ZZ)^2$-valued character and the generalization of modular graph functions provided by the functions $g_k(v|\tau)$ may be viewed as the result of introducing {\sl $\RR/\ZZ$-valued characters} in the Kronecker-Eisenstein sums. Their graphical representation is given in Figure \ref{fig:4}.

\begin{figure}[h]
\begin{center}
\tikzpicture[scale=1.3]
\scope[xshift=0cm,yshift=0cm]
\draw (0.15,0) node{$g(v|\tau)=$};  
\draw[ultra thick] (1,0) -- (2,0);
\draw (1,0) [fill=white] circle(0.05cm) ;
\draw (2,0) [fill=white] circle(0.05cm) ;
\draw (1,-0.3) node{$p_a$};
\draw (2,-0.3) node{$p_b$};
\draw (4.15,0) node{$g_2(v|\tau)=$};  
\draw[ultra thick] (5,0) -- (6,0);
\draw (5,0) [fill=white] circle(0.07cm) ;
\draw (6,0) [fill=white] circle(0.07cm) ;
\draw (5.5,0) node{$\bullet$};
\draw (5,-0.3) node{$p_a$};
\draw (6,-0.3) node{$p_b$};
\draw (8.15,0) node{$g_3(v|\tau)=$};  
\draw[ultra thick] (9,0) -- (11,0);
\draw (9,0) [fill=white] circle(0.07cm) ;
\draw (11,0) [fill=white] circle(0.07cm) ;
\draw (9.66,0) node{$\bullet$};
\draw (10.33,0) node{$\bullet$};
\draw (9,-0.3) node{$p_a$};
\draw (11,-0.3) node{$p_b$};
\endscope
\endtikzpicture
\caption{Modular graph functions  $g_k(v|\tau)$.  } \label{fig:4}
\end{center}
\end{figure}

Another interpretation of the generalization is to note the relation between $g_k(v|\tau)$ and the {\sl single-valued elliptic polylogarithms} $D_{k,\ell}(v|\tau)$ introduced by Zagier \cite{zbMATH04144378}, 
\bea
D_{k,\ell} (v|\tau )  =  
{ (2 i \tau_2)^{k+\ell-1} \over 2 \pi i} \sum _{(m,n) \not= (0,0)} { e^{2 \pi i (m u_2 -nu_1 )} \over (m + n \tau)^k
(m + n \bar \tau)^\ell } 
\eea
While $g_k(v |\tau)$ are modular functions satisfying  (\ref{modgk}), $D_{k,\ell}$ 
{transforms as a modular form of weight $(1-\ell,1-k)$}. The functions $g_k(v|\tau)$ are special cases of Zagier's $D_{k,\ell}$-forms when $\ell=k$, 
\bea
\label{5e2}
D_{k,k} (v|\tau) & = & (- 4 \pi \tau_2)^{k-1} g_k(v|\tau)
\eea
Further properties and interrelations satisfied by the forms $D_{k,\ell}$ and the functions $g_k$ are provided in Appendix~\ref{sec:A}.

\sm

\begin{figure}[h]
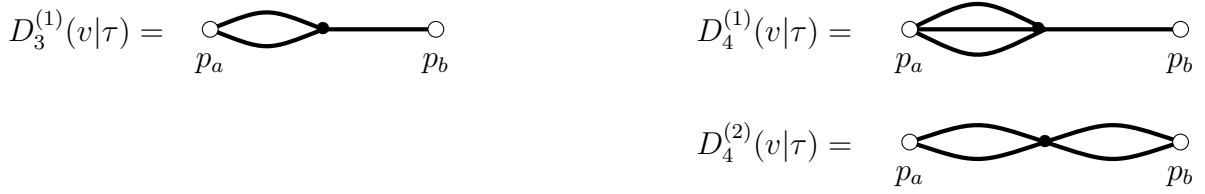

\begin{center}
\tikzpicture[scale=1.5]
\scope[xshift=0cm,yshift=0cm]
\draw (-0.1,0) node{$D_3^{(1)}(v|\tau)=$};  
\draw[ultra thick] (1,0) .. controls (1.5, -0.2) .. (2,0);
\draw[ultra thick] (1,0) .. controls (1.5, 0.2) .. (2,0);
\draw (2,0) node{$\bullet$};
\draw[ultra thick] (2,0) -- (3,0);
\draw (1,0) [fill=white] circle(0.07cm) ;
\draw (3,0) [fill=white] circle(0.07cm) ;
\draw (1,-0.3) node{$p_a$};
\draw (3,-0.3) node{$p_b$};
\draw (6,0) node{$D_4^{(1)}(v|\tau)=$};  
\draw[ultra thick] (7.2,0) .. controls (7.8, -0.3) .. (8.4,0);
\draw[ultra thick] (7.2,0) -- (8.4,0);
\draw[ultra thick] (7.2,0) .. controls (7.8, 0.3) .. (8.4,0);
\draw[ultra thick] (8.4,0) -- (9.6,0);
\draw (8.34,0) node{$\bullet$};
\draw (7.2,0) [fill=white] circle(0.07cm) ;
\draw (9.6,0) [fill=white] circle(0.07cm) ;
\draw (7.2,-0.3) node{$p_a$};
\draw (9.6,-0.3) node{$p_b$};
\draw (6,-1) node{$D_4^{(2)}(v|\tau)=$}; 
\draw[ultra thick] (7.2,-1) .. controls (7.8, -1.2) .. (8.4,-1);
\draw[ultra thick] (7.2,-1) .. controls (7.8, -0.8) .. (8.4,-1);
\draw[ultra thick] (8.4,-1) .. controls (9, -1.2) .. (9.6,-1);
\draw[ultra thick] (8.4,-1) .. controls (9, -0.8) .. (9.6,-1);
\draw (7.2,-1) [fill=white] circle(0.07cm) ;
\draw (9.6,-1) [fill=white] circle(0.07cm) ;
\draw (7.2,-1.3) node{$p_a$};
\draw (9.6,-1.3) node{$p_b$};
\draw (8.4,-1) node{$\bullet$};
\endscope
\endtikzpicture
\caption{Modular graph functions  $D_\ell ^{(k)}(v|\tau)$.  } \label{fig:5}
\end{center}
\end{figure}

The remaining coefficient functions are all generalized modular graph functions in the sense defined above, either as modular graph functions with character, or as single-valued elliptic polylogarithms. We give below the definitions of these functions, along with their graphical representations. We have the following infinite families,
\bea
\label{Dlk}
D_\ell ^{(k)} (v |\tau) = \int _{\Sigma _1} \kappa _1 (z) g(v+z|\tau)^k g(z|\tau)^{\ell-k}
\eea
for $k, \ell \geq 0$ integers. They obey the symmetry relation $D_\ell^{(k)} (v|\tau) = D_\ell ^{(\ell-k)} (v |\tau)$, and  restrict to modular graph functions by the relation $D_\ell ^{(k)} (0|\tau) = D_\ell (\tau)$,  while for $k=1$, they satisfy a simple differential relation, 
\bea
\label{delDk} 
\Delta _v D_\ell ^{(1)} (v |\tau) = 4\pi D_{\ell-1}(\tau) - 4\pi g(v|\tau)^{\ell-1}
\eea
The graphical representation of these functions is illustrated in Figure \ref{fig:5}.

The modular graph function $F_\ell(v|\tau)$, defined in (\ref{Fn}), may be expressed as a linear combination of these functions, 
\bea
\label{defFlk}
F_\ell(v|\tau) = \sum _{k=0}^\ell { (-)^{\ell-k} \over k! \, (\ell-k)!} \, D_\ell ^{(k)} (v |\tau)
\eea
For odd values of $\ell$ the sum vanishes in view of the symmetry relation of $D_\ell ^{(k)}$, for $\ell=2$ we have (\ref{F2}), while for $F_2$ and $F_4$, the formula reduces to,
\bea
\label{F4}
F_2 (v |\tau ) & = & E_2(\tau) - g_2(v|\tau )
\no \\
F_4(v|\tau ) & = & {1 \over 12} D_4(\tau)  - { 1 \over 3} D_4 ^{(1)}(v|\tau ) + { 1 \over 4 } D_4 ^{(2)}(v|\tau )
\eea
where we have made use of $D_2(\tau) =E_2(\tau) $ and $D_2^{(1)}(v|\tau ) = g_2(v|\tau )$ on the first line.

\subsection{Higher generalized modular graph functions}

\begin{figure}[hbt]
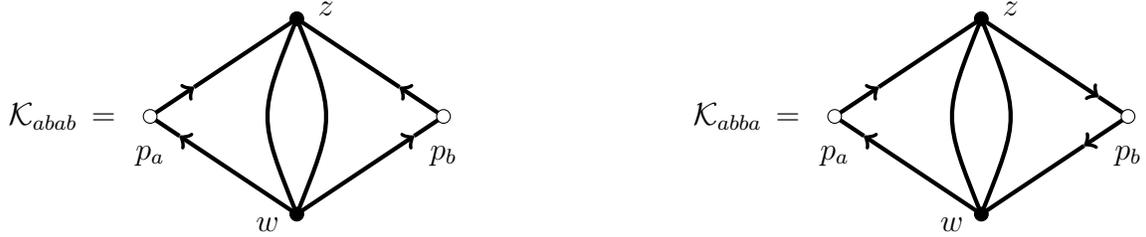

\begin{center}
\tikzpicture[scale=1.3]
\scope[xshift=0cm,yshift=0cm]
\draw (-0.9,0) node{$\cK_{abab } \, =$};  
\draw[->,ultra thick] (0,0) -- (0.45,0.3);
\draw[ultra thick] (0.45,0.3) -- (1.5,1);
\draw[ultra thick] (0,0) -- (0.3,-0.2);
\draw[->,ultra thick]  (1.5,-1) -- (0.3,-0.2);
\draw[->,ultra thick] (3,0) -- (2.55,0.3);
\draw[ultra thick] (2.55,0.3) -- (1.5,1);
\draw[ultra thick] (3,0) -- (2.7,-0.2);
\draw[->,ultra thick]  (1.5,-1) -- (2.7,-0.2);
\draw[ultra thick] (1.5,1) .. controls (1.1, 0) .. (1.5,-1);
\draw[ultra thick] (1.5,1) .. controls (1.9, 0) .. (1.5,-1);
\draw (0,0) [fill=white] circle(0.07cm) ;
\draw (3,0) [fill=white] circle(0.07cm) ;
\draw (1.5,1) [fill=black] circle(0.07cm) ;
\draw (1.5,-1) [fill=black] circle(0.07cm) ;
\draw (0,-0.4) node{$p_a$};
\draw (3,-0.4) node{$p_b$};
\draw (1.8,1.1) node{$z$};
\draw (1.2,-1.1) node{$w$};
\draw (6.1,0) node{$\cK_{abba } \, =$};  
\draw[->,ultra thick] (7,0) -- (7.45,0.3);
\draw[ultra thick] (7.45,0.3) -- (8.5,1);
\draw[ultra thick] (7,0) -- (7.3,-0.2);
\draw[->,ultra thick]  (8.5,-1) -- (7.3,-0.2);
\draw[->,ultra thick] (10,0) -- (9.55,-0.3);
\draw[ultra thick] (9.55,-0.3) -- (8.5,-1);
\draw[ultra thick] (10,0) -- (9.7,0.2);
\draw[->,ultra thick]  (8.5,1) -- (9.7,0.2);
\draw[ultra thick] (8.5,1) .. controls (8.1, 0) .. (8.5,-1);
\draw[ultra thick] (8.5,1) .. controls (8.9, 0) .. (8.5,-1);
\draw (7,0) [fill=white] circle(0.07cm) ;
\draw (10,0) [fill=white] circle(0.07cm) ;
\draw (8.5,1) [fill=black] circle(0.07cm) ;
\draw (8.5,-1) [fill=black] circle(0.07cm) ;
\draw (7,-0.4) node{$p_a$};
\draw (10,-0.4) node{$p_b$};
\draw (8.8,1.1) node{$z$};
\draw (8.2,-1.1) node{$w$};
\endscope
\endtikzpicture
\caption{Modular graph functions $\cK_{abab}$ and $\cK_{abba}$. 
An arrow flowing into a vertex indicates a $\p$-derivative with respect to the coordinate of the vertex, while an arrow flowing out of a vertex indicates a $\bar \p$-derivative with respect to the coordinate of the vertex. } \label{fig:6}
\end{center}
\end{figure}

The degeneration of the modular graph function $\cBZ_1$ involves substantially more complicated genus-one modular graph functions than its Kawazumi-Zhang or $\cBZ_2$ and $\cBZ_3$ counterparts. The complication arises from their higher loop order, including three and four loops, and the need for subtractions in some of the graphs, as will be explained below. The main source of the complication is the  integral \eqref{Kdef} appearing in the degeneration of $Z_1^{(a)}$ defined in 
the first line of \eqref{B20aa},
\bea
\label{KK}
\cK = {\tau_2^2  \over \pi^2} 
\int _\Sep   \kappa_1 ( z)  \int _\Sep \kappa_1 (w) \,  |\p_z f(z) \p_w f(w) |^2 \, g(z,w)^2
\eea
As is shown in Appendices \ref{sec_varia} and \ref{sec:Kfull}, the non-separating degeneration of $\cK$ consists of a polynomial of degree four in $t$, plus terms which are exponentially suppressed in $t$. To extract the polynomial in $t$,  we express $f$ in terms of the genus-one Green function $g$, and expand the integrand into 16 terms, which may be regrouped in terms of 5 distinct  building blocks,
\bea
\cK =  2 \cK_{abab}+ 2 \cK_{abba} + 2 \cK_{aabb}  - 8 \, \Re\left(  \cK_{aaab} \right )   + 2 \cK_{aaaa} 
\eea
The graphical representation of the functions $\cK_{abab}$ and $ \cK_{abba}$  is given in Figure \ref{fig:6}.   These functions are given by the following convergent integrals, 
\bea
\label{Kabab}
\cK _{abab} & = & {\tau_2^2 \over \pi^2}  
\int _{\Sigma_1} \! \kappa_1 ( z)  \int _{\Sigma_1} \! \kappa_1 (w)  \,
 \p_z g(z,p_a) \pbz g(z,p_b) \,  g(z,w)^2 \, \p_w g(w,p_a) \pbw g(w,p_b)  
\no \\
\cK _{abba} & = & {\tau_2^2 \over \pi^2}  
\int _{\Sigma_1} \! \kappa _1 ( z)  \int _{\Sigma_1} \! \kappa_1 (w) \,  
\p_z g(z,p_a) \pbz g(z,p_b)  \,  g(z,w)^2 \, \p_w g(w,p_b) \pbw g(w,p_a)  
\qquad 
\eea
and contribute a polynomial of degree zero in $t$, up to exponential corrections.

\begin{figure}[hbt]
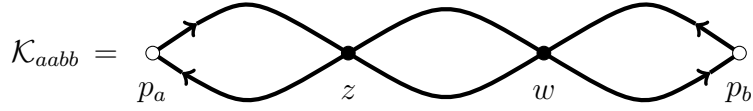

\begin{center}
\tikzpicture[scale=1.3]
\scope[xshift=0cm,yshift=0cm]
\draw (-0.9,0) node{$\cK_{aabb } \, =$};  
\draw[->,ultra thick] (0,0) -- (0.45,0.3);
\draw[ultra thick] (0.45,0.3) ..controls (1, 0.6) .. (2,0);
\draw[ultra thick] (0,0) -- (0.3,-0.2);
\draw[->,ultra thick]  (2.0,0) ..controls (1,-0.6) ..  (0.3,-0.2);
\draw[ultra thick]  (2.0,0) ..controls (3,0.6) ..  (4,0);
\draw[ultra thick]  (2.0,0) ..controls (3,-0.6) ..  (4,0);
\draw[->,ultra thick] (6,0) -- (5.55,0.3);
\draw[ultra thick] (5.55,0.3) ..controls (5, 0.6) .. (4,0);
\draw[ultra thick] (6,0) -- (5.7,-0.2);
\draw[->,ultra thick]  (4.0,0) ..controls (5,-0.6) ..  (5.7,-0.2);
\draw (0,0) [fill=white] circle(0.07cm) ;
\draw (6,0) [fill=white] circle(0.07cm) ;
\draw (2,0) [fill=black] circle(0.07cm) ;
\draw (4,0) [fill=black] circle(0.07cm) ;
\draw (0,-0.4) node{$p_a$};
\draw (6,-0.4) node{$p_b$};
\draw (2,-0.4) node{$z$};
\draw (4,-0.4) node{$w$};
\endscope
\endtikzpicture
\caption{Schematic representation of the modular graph function $\cK_{aabb}^0$. \label{fig:7}}
\end{center}
\end{figure}

\begin{figure}[h]
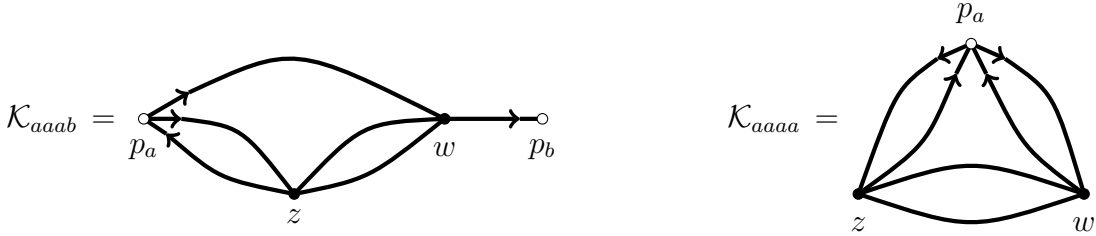

\begin{center}
\tikzpicture[scale=1]
\scope[xshift=0cm,yshift=0cm]
\draw (-1.1,1) node{$\cK_{aaab } \, =$};  
\draw[->,ultra thick] (0,1) -- (0.5,1);
\draw[ultra thick] (0.5,1) ..controls (1.3, 0.9) .. (2,0);
\draw[ultra thick] (0,1) -- (0.3,0.8);
\draw[->,ultra thick]  (2,0) ..controls (1,0.2) ..  (0.3,0.8);
\draw[ultra thick]  (2,0) ..controls (2.9,0.95) ..  (4,1);
\draw[ultra thick]  (2,0) ..controls (3,0.2) ..  (4,1);
\draw[->,ultra thick]  (0,1) -- (0.6,1.35);
\draw[ultra thick]  (0.6,1.35) ..controls (2,2) ..  (4,1);
\draw[ultra thick] (5.3,1) -- (5,1);
\draw[->,ultra thick]  (4.0,1) --  (5,1);
\draw (0,1) [fill=white] circle(0.07cm) ;
\draw (5.3,1) [fill=white] circle(0.07cm) ;
\draw (2,0) [fill=black] circle(0.07cm) ;
\draw (4,1) [fill=black] circle(0.07cm) ;
\draw (0,0.6) node{$p_a$};
\draw (5.3,0.6) node{$p_b$};
\draw (2,-0.3) node{$z$};
\draw (4,0.6) node{$w$};
\draw (8.5,1) node{$\cK_{aaa a} \, =$};  
\draw[ultra thick] (9.5,0) ..controls (10, 1.4) .. (10.55,1.8);
\draw[<-,ultra thick] (10.55,1.8) -- (11,2);
\draw[ultra thick] (12.5,0) ..controls (12.1, 1.35) .. (11.45,1.8);
\draw[<-,ultra thick] (11.45,1.8) -- (11,2);
\draw[ultra thick] (10.8,1.6) -- (11,2);
\draw[->,ultra thick] (9.5,0) ..controls (10.4, 0.65) .. (10.8,1.6);
\draw[ultra thick] (11.2,1.6) -- (11,2);
\draw[->,ultra thick] (12.5,0) ..controls (11.7, 0.6) .. (11.2,1.6);
\draw[ultra thick] (9.5,0) ..controls (11, 0.5) .. (12.5,0);
\draw[ultra thick] (9.5,0) ..controls (11, -0.5) .. (12.5,0);
\draw (11,2) [fill=white] circle(0.07cm) ;
\draw (9.5,0) [fill=black] circle(0.07cm) ;
\draw (12.5,0) [fill=black] circle(0.07cm) ;
\draw (11,2.4) node{$p_a$};
\draw (9.5,-0.4) node{$z$};
\draw (12.5,-0.4) node{$w$};
\endscope
\endtikzpicture
\caption{Schematic representation of $\cK_{aaab}^0$ and  $\cK_{aaaa}^0$.} \label{fig:8}
\end{center}
\end{figure}

The remaining three functions $\cK_{aabb}$, $\cK_{aaab}$ and $\cK_{aaaa}$ do have non-trivial polynomial $t$-dependence, and are represented schematically in Figures \ref{fig:7} and \ref{fig:8}. We isolate this dependence by splitting the integrals as follows,
\bea
\cK_{aabb} & = & \cK_{aabb}^0 + \cK_{aabb}^1
\no \\
\cK_{aaab} & = & \cK_{aaab}^0 + \cK_{aaab}^1
\no \\
\cK_{aaaa} & = & \cK_{aaaa}^0 + \cK_{aaaa}^1
\eea
where the contributions $\cK^0$ are constant in $t$, while the contributions $\cK^1$ are polynomials in $t$ with vanishing constant part, up to exponentially suppressed contributions.  The contributions  $\cK^0_{aabb}$ and $\cK^0_{aaab}$  are given by the following convergent integrals, while the $t$-dependent parts $\cK_{aabb}^1,  \cK_{aaab}^1$ and $\cK_{aaaa}^1$ will  be evaluated below.
\bea
\label{Kzero}
\cK _{aabb} ^0 & = & 
{\tau_2^2 \over \pi^2} \int _{\Sigma_1} \!\!\!\! \kappa_1 ( z)  \int _{\Sigma _1}  \!\!\!\! \kappa_1 (w) \,  |\p_z g(z,p_a) |^2  
| \p_w g(w,p_b) |^2 
 \\ && \hskip 1in \times
 \Big ( g(z,w)^2 - g(p_a,w)^2 -g(z,p_b)^2 + g(p_a, p_b)^2 \Big )    
\no \\
 \cK _{aaab} ^0 & = & 
{\tau_2^2 \over \pi^2}  \int _{\Sigma_1} \! \kappa_1 ( z)  \int _{\Sigma _1} \! \kappa_1 (w)   |\p_z g(z,p_a) |^2  
 \p_w g(w,p_a) \pbw g(w,p_b) \Big ( g(z,w)^2 - g(p_a,w)^2 \Big ) 
\no
\eea

Finally, the modular graph function $\cK_{aaaa}^0$ is given by the following convergent integral, 
\bea
\label{Kaaaa}
\cK^0_{aaaa} & = &  { \tau_2 \over \pi}  \int _{\Sigma _1} \kappa _1 (z) |\p_z g(z)|^2 \Big (W(z) - 4 \zeta (3) \Big )
\no \\
W(z) & = & { \tau_2 \over \pi}  \int _{\Sigma _1} \kappa _1 (w) |\p_w g(w)|^2 \Big (g(z,w) - g(z) \Big )
\Big (g(z,w) - g(w) \Big )
\eea
The integral over $z$ on the second line is convergent, but that the integral of each term in the parentheses separately is divergent due to the double pole of $ |\p_z g(z)|^2$ at $z=0$.

\sm

The functions $\cK_{aabb}^1$, $\cK_{aaab}^1$ and $\cK_{aaaa}^1$ are polynomials in $t$ whose coefficients are genus-one modular graph functions of the customary type. Their contribution to $\cK$ is given by, 
\be
\cK^t =   2 \cK_{abab}^1+ 2 \cK_{abba}^1 + 2 \cK_{aabb}^1  - 8 \, \Re \left (  \cK_{aaab}^1 \right )  
\ee
and computed in Appendix \ref{sec_varia}, using the variational method introduced in \cite[\S 3.6]{DHoker:2017pvk}. The functions  $\cK_{aabb}^0$, $\cK_{aaab}^0$ and $\cK_{aaaa}^0$  are more exotic genus-one modular graph functions, which are schematically represented in Figures \ref{fig:7} and \ref{fig:8} (these figures, however, do not indicate the subtractions in the integrand). The remainder  $\cK^c=\lim_{t\to\infty} (\cK-\cK^t)$  is given by, 
\bea
\label{defKc}
\cK^c & = & 
2 \cK_{abab} + 2 \cK_{abba} + 2 \cK_{aabb} ^0     - 4 \cK_{aaab}^0 - 4 (\cK_{aaab}^0)^* + 2 \cK_{aaaa}^0 
 \\ &&
+ 4 g(v) \left (  D_3   + 2 \zeta(3)  
  -    D_3 ^{(1)}(v) + {  \Delta _v F_4(v) \over 2 \pi}  \right )
- 3g(v)^4    -{ 7  \over 4} E_2^2  + { 5 \over 4} D_4 + { 3  \over 2}E_4 
\no
\eea
This concludes the explanation of the modular graph functions appearing in the 
minimal degenerations \eqref{bark3} of the string integrands $\cZ_i$ and $\cB_{(2,0)}$.

\newpage

\section{The separating degeneration} 
\setcounter{equation}{0}
\label{sec_sep}

In the separating degeneration, a genus-two Riemann surface $\Sigma$ tends to the union of two genus-one surfaces which  intersect at a common puncture. Denoting the corresponding  compact genus-one surfaces by $\Sigma _1$ and $\Sigma _1'$  with respective moduli $\tau$ and  $\sigma$,  the punctured surfaces are $\Sigma _1 \setminus \{ p\}$ and $\Sigma _1' \setminus \{ p'\}$ respectively, where the punctures $p$ and $p'$ are  identified with one another. We shall examine the degeneration of string invariants in a neighborhood of the separating degeneration, which we parametrize by the off-diagonal element $v$ of the period matrix.  We begin by presenting a review of the degeneration of the Abelian differentials, the canonical K\"ahler form, and the Arakelov Green function, and derive the degenerations of the Kawazumi-Zhang and $\cB_{(2,0)}$ example of higher invariants at genus two.

\subsection{Funnel construction of the separating degeneration}

A convenient parametrization of the neighborhood of the separating divisor is provided by the funnel construction given in \cite{fay73}. We shall carry out this construction here in the simplest case of a genus-two surface $\Sigma$ because this is the focus of the present paper, but the construction is easily generalized to arbitrary genus. 

\sm

For genus two, the starting point of the construction of $\Sigma$  in \cite{fay73} is provided by the compact genus-one surfaces $\Sigma _1$ and $\Sigma_1'$, to which we add punctures, respectively $p$ and~$p'$. Next, we introduce a system of local complex coordinates $(x,\bar x)$ and $(x', \bar x')$ on each surface, and denote the coordinates of the punctures simply by $p$ and~$p'$. We specify (simply connected) discs $\mD_i$  centered at $p$ with boundaries $\mC_i$ on $\Sigma _1$, and (simply connected) discs $\mD_i'$ centered at $p'$ with boundaries $\mC_i'$ on $\Sigma _1'$ for $i=1,2,3$, as shown in Figure \bp{10}.

\begin{figure}[h]
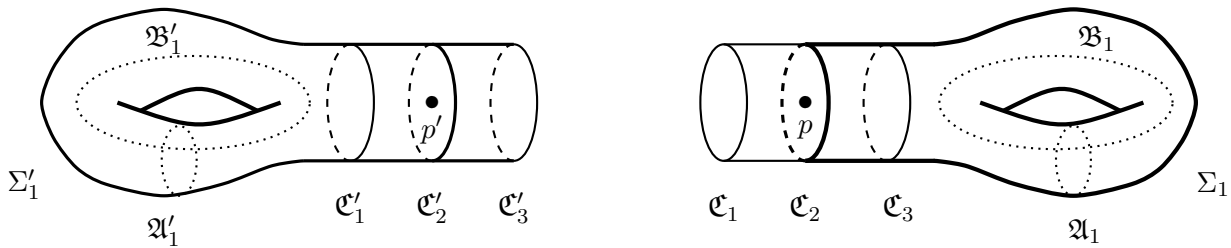

\begin{center}
\tikzpicture[scale=1.55]
\scope[xshift=-5cm,yshift=0cm]
\draw [thick] (0,-0.5) arc (0:360:0.2 and 0.5);
\draw [ultra thick] (0.5,-1) arc (-90:90:0.2 and 0.5);
\draw [very thick, dashed] (0.5,0) arc (90:270:0.2 and 0.5);
\draw [thick] (1.2,-1) arc (-90:90:0.2 and 0.5);
\draw [thick, dashed] (1.2,0) arc (90:270:0.2 and 0.5);
\draw [thick] (-0.2,0) -- (0.5,0);
\draw [thick] (-0.2,-1) -- (0.5,-1);
\draw [ultra thick] (0.5,0) -- (1.61,0);
\draw [ultra thick] (0.5,-1) -- (1.61,-1);
\draw[ultra thick] plot [smooth] coordinates {
(1.6,0) (1.7, 0.01) (1.9, 0.05) (2.3, 0.2) (2.8, 0.3) (3.3, 0.2) (3.55, 0.06) (3.75, -0.2)   
(3.85,-0.5) 
(3.75,-0.8) (3.55, -1.06) (3.3,-1.2) (2.8, -1.3) (2.3, -1.2) (1.9, -1.05) (1.7, -1.01)(1.6,-1)};
\draw[ultra thick] (2,-0.5) .. controls (2.7, -0.75) .. (3.4,-0.5);
\draw[ultra thick] (2.2,-0.57) .. controls (2.7, -0.32) .. (3.2,-0.57);
\draw [thick, dotted] (3.65,-0.5) arc (0:360:1 and 0.4);
\draw [thick, dotted] (2.95,-1) arc (0:360:0.15 and 0.3);
\draw [thick] (-2,-1) arc (-90:90:0.2 and 0.5);
\draw [thick, dashed] (-2,0) arc (90:270:0.2 and 0.5);
\draw [very thick] (-2.7,-1) arc (-90:90:0.2 and 0.5);
\draw [thick, dashed] (-2.7,0) arc (90:270:0.2 and 0.5);
\draw [thick] (-3.4,-1) arc (-90:90:0.2 and 0.5);
\draw [thick, dashed] (-3.4,0) arc (90:270:0.2 and 0.5);
\draw [very thick] (-2,0) -- (-3.8,0);
\draw [very thick] (-2,-1) -- (-3.8,-1);
\draw[very thick] plot [smooth] coordinates {
(-3.8,0) (-3.9, 0.01) (-4.1, 0.05)(-4.5, 0.2) (-5, 0.3) (-5.5, 0.2) (-5.75, 0.06) (-5.95, -0.2)   
(-6.05,-0.5) 
(-5.95,-0.8) (-5.75, -1.06) (-5.5,-1.2) (-5, -1.3) (-4.5, -1.2) (-4.1, -1.05) (-3.9, -1.01)(-3.8,-1)};
\draw[ultra thick] (-4,-0.5) .. controls (-4.7, -0.75) .. (-5.4,-0.5);
\draw[ultra thick] (-4.2,-0.57) .. controls (-4.7, -0.32) .. (-5.2,-0.57);
\draw [thick, dotted] (-3.75,-0.5) arc (0:360:1 and 0.4);
\draw [thick, dotted] (-4.72,-1) arc (0:360:0.15 and 0.3);
\draw (-0.2,-1.4) node{$\mC_1$};
\draw (0.5,-1.4) node{$\mC_2$};
\draw (1.3,-1.4) node{$\mC_3$};
\draw (0.5,-0.5) node{$\bullet$};
\draw (0.5,-0.73) node{{\small $p$}};
\draw (3,0.05) node{{\small $\mB_1$}};
\draw (2.9,-1.6) node{{\small $\mA_1$}};
\draw (4,-1.2) node{{\small $\Sigma_1$}};
\draw (-3.4,-1.4) node{$\mC_1'$};
\draw (-2.7,-1.4) node{$\mC_2'$};
\draw (-2,-1.4) node{$\mC_3'$};
\draw (-2.7,-0.5) node{$\bullet$};
\draw (-2.7,-0.73) node{{\small $p'$}};
\draw (-5,0.05) node{{\small $\mB_1'$}};
\draw (-5,-1.6) node{{\small $\mA_1'$}};
\draw (-6.2,-1.2) node{{\small $\Sigma_1'$}};
\endscope
\endtikzpicture
\caption{Funnel construction of a family of genus-two Riemann surfaces~$\Sigma$ near the separating divisor  in terms of two compact genus-one surfaces $\Sigma _1$ and $\Sigma _1'$. The circles $\mC_i$ and $\mC_i'$ for $i=1,2,3$ are centered respectively at the punctures $p$ and $p'$ and respectively bound the  discs $\mD_i$ and $\mD_i'$. The surface $\Sigma$ is constructed from the surfaces $\Sigma _1 \setminus \mD_1$  and $\Sigma _1' \setminus \mD_3'$ by identifying the annuli $[ \mC_1, \mC_3]$ and  $[ \mC_1', \mC_3']$.}
\end{center}
\end{figure}

The genus-two surface $\Sigma$ is obtained by identifying the annulus  $[\mC_1, \mC_3]$ with the annulus  $[\mC_1', \mC_3']$ with respective local complex coordinates  $x$ and $x'$ via the relation,
\bea
\label{fay1}
(x-p)(x'-p') = v_s
\eea
Here $v_s$ is a complex parameter governing the separating degeneration (which is referred to as $t$ in \cite{fay73}) and is such that the separating degeneration corresponds to the limit $v_s \to 0$. Customarily, the curves $\mC_i$ and $\mC_i'$ are defined to be circles in the local complex coordinates on the surfaces but here instead we shall use a more intrinsic definition, which will be given below.  Next, we shall construct the Abelian differentials and Green function on $\Sigma$ in terms of its genus-one data along with $v_s$.

\subsection{Global funnel construction}

In analogy with the construction of the neighborhood of the non-separating node, we may also here provide a convenient  intrinsic characterization of  $\mC_i$ and $\mC_i'$ as level-curves  of the scalar Green functions $g$ and $g'$ on the genus-one surfaces $\Sigma _1$ and $\Sigma _1'$, 
\bea
\mC_i & = & \{ x \in \Sigma _1 \hbox{ such that } g(x-p|\tau)=t_i \}
\no \\
\mC_i ' & = & \{ x' \in \Sigma _1' \hbox{ such that } g(x'-p'|\sigma)=t_i' \}
\eea
for sufficiently large values of $t_i, t_i'$ so that each level-set $\mC_i$ and $\mC_i'$ is connected.  The curves are related  by  the following relation between their $t_i$-values, valid for $i=1,2,3$,
\bea
t_i+t_i'= - \ln |2 \pi  v_s \, \eta (\tau)^2 \eta (\sigma )^2|^2 + \cO(v_s^2)
\eea
Here, we have used the short-distance expansion of the scalar Green function on the torus, given by 
$ g(z|\tau) = - \ln | 2 \pi z \, \eta (\tau)^2|^2 + \cO(z^2)$ to convert  (\ref{fay1}) into the  expression above. 

\sm

When performing integrals over the genus-two surface $\Sigma$, it will be convenient to 
decompose the integral into a sum of the contribution from $\Sigma_1 \setminus \mD_2$ plus the contribution from $\Sigma'_1 \setminus \mD_2'$ where the curves $\mC_2$ and $\mC_2'$ are defined so that $t_2=t_2'$. Under these conditions, the Abelian differentials $\om_1$ and $\om_2$ remain uniformly bounded throughout $\Sigma$ by a constant of order $\cO(v_s^0)$, with corrections which are suppressed by powers of $v_s$.

\subsection{Degeneration of Abelian differentials}

We choose  canonical homology bases for the genus-one surfaces by  $\mA_1, \mB_1 \subset \Sigma _1 \setminus \{p\}$ and $\mA_1',\mB_1' \subset \Sigma _1 ' \setminus \{p'\}$, and extend those to a canonical homology basis $\mA_I, \mB_I$ for $I=1,2$ for $\Sigma$ by setting $\mA_2=\mA_1', \mB_2=\mB_1' $. The genus-one holomorphic Abelian differentials  $\omega$ and $ \omega'$ respectively on  $\Sigma_1$ and $\Sigma'_1$  are normalized  as follows, 
\bea
\oint _{\mA_1} \om = \oint _{\mA_1'} \om' = 1
\hskip 1in 
\oint _{\mB_1} \om = \tau 
\hskip 1in 
\oint _{\mB_1'} \om' = \sigma 
\eea
To construct holomorphic 1-forms on the genus-two surface with period matrix,
\bea
\Omega = \left ( \begin{matrix} \tau & v \cr v & \sigma \cr \end{matrix} \right )
\eea
we extend $\om$ to a differential $\om_1$ on the genus-two surface $\Sigma$ and $\om'$ to a differential on the genus-two surface $\Sigma$ by using the identification (\ref{fay1}). Choosing  complex coordinates $x, x'$ on $\Sigma _1$ and $\Sigma _1'$  such that $\om= dx$ and $\om' = dx'$, we see that the differential $dx$ extends to $- v_s /(x'-p')^2dx' $ in $\Sigma _1'$ while the differential $dx'$ extends to $- v_s /(x-p)^2dx$ in $\Sigma _1$. Thus, the extensions are governed by meromorphic 1-forms with a double pole. The meromorphic 1-forms  $\varpi(x,y)$ and $\varpi'(x',y')$ respectively on $\Sigma _1$ and $\Sigma _1'$ are normalized to have vanishing $\mA$-periods and a double pole of unit strength at $x=y$ and $x'=y'$. Their $\mB$-periods are given by the Riemann bilinear relations,
\bea
\oint _{\mB_1} \varpi(x,y) = 2 \pi i \om(y)
\hskip 1in 
\oint _{\mB_1'} \varpi (x',y') = 2 \pi i \om' (y')
\eea
The holomorphic 1-forms $\om_1$ and $\om_2$ on the genus-two surface $\Sigma$, canonically normalized on $\mA_1$ and $\mA_2$-cycles, are then given as follows, 
\bea
\label{SDAb}
\omega_1 & = & \begin{cases}
\omega(x)& ~~ x\in \Sigma_1 \setminus \mD_1 \\
v  \,  \varpi'(x',p') / ( 2 \pi i \om' (p'))    & ~~ x'\in\Sigma'_1 \setminus \mD_3'
\end{cases}
\no \\
\omega_2 & = & \begin{cases}
 v \,  \varpi(x,p) /  (2 \pi i\omega(p) ) ~~ & ~~~  \, x\in\Sigma_1 \setminus \mD_1  \\
 \omega'(x') ~~~ & ~~~ \, x'\in \Sigma'_1 \setminus \mD_3'
\end{cases}
\eea
The parameter $v_s $ is related to the entry $v$ of the genus-two period matrix by, 
\bea
v = \oint _{\mB_1} \om _2 = \oint _{\mB_2} \om _1 = - 2 \pi i \, v_s \, \om(p) \, \om '(p')
\eea
The expressions  in (\ref{SDAb}) are valid up to corrections of order $\cO(v^2)$ which have been omitted.

\subsection{Degeneration of the Green function}

The degeneration of the string  Green function $G$ of (\ref{G})  on the genus-two Riemann surface~$\Sigma$ was obtained in \cite{D'Hoker:2013eea}, and is given by, 
\bea
G = \begin{cases}
g(x-y|\tau) + 2 \ln ( 2 \pi |\eta (\tau)|^2) & x,y \in \Sigma _1 \setminus \mD_1 \\
g(x'-y'|\sigma) + 2 \ln ( 2 \pi |\eta (\sigma)|^2) & x',y' \in \Sigma _1' \setminus \mD_3' \\
g(x-p|\tau) + g(y'-p'|\sigma) + \ln \big (  (2\pi)^3 |v \eta (\tau) \eta (\sigma)|^4 \big ) 
& x \in \Sigma_1\setminus \mD_1 , y' \in \Sigma _1' \setminus \mD_3'
\end{cases}
\eea
up to terms of order $\cO(v)$ which will be omitted in the sequel. The degeneration of the canonical K\"ahler form $\kappa$ of the genus-two Riemann surface $\Sigma$ is given as follows, 
\bea
\label{SDkap}
\kappa = \begin{cases}
\half \kappa _1 (x) = { i \over 4 \tau_2} \om(x)  \wedge \oom (x)& x \in \Sigma _1 \setminus \mD_1 \\
\half \kappa _1' (x')= { i \over 4 \sigma_2} \om'(x') \wedge \oom' (x') & x' \in \Sigma _1' \setminus \mD_3'
\end{cases}
\eea 
From these results we readily obtain the separating degeneration formulas for the  Arakelov Green function $\cG$ on the genus-two Riemann surface $\Sigma$, which are given as follows,
\bea
\label{SDArak}
\cG = \begin{cases}
- \half \ln |\hat v|  + g(x-y|\tau) - \half g(x-p|\tau) - \half g(y-p|\tau) & x, y \in \Sigma _1 \setminus \mD_1 \\
- \half \ln |\hat v|  + g(x'-y'|\sigma) - \half g(x'-p'|\sigma ) - \half g(y'-p'|\sigma ) & x', y' \in \Sigma _1' \setminus \mD_3' \\
\half \log|\hat v| +\half \gone(x-p|\tau)+\half \gone(y'-p'|\sigma)\ , & x\in\Sigma_1 \setminus \mD_1 , y' \in\Sigma'_1 \setminus \mD_3'
\end{cases}
\eea
where $\hat v$ is related to $v$ by the  Dedekind eta-function $\eta$,
\bea
\hat v = 2 \pi \, v \, \eta (\tau)^2 \eta (\sigma )^2
\eea
In the vicinity of the separating degeneration the genus-two  modular group $Sp(4,\ZZ)$  restricts to its $SL(2,\ZZ) \times SL(2,\ZZ)'$ subgroup which acts by  M\"obius transformations on $(\tau, \sigma)$, and $v$ by, 
\bea
\label{modSS}
\tau \to { a \tau + b \over c \tau + d} 
\hskip 0.8in 
\sigma \to { a' \sigma + b' \over c' \sigma  + d'}
\hskip 0.8in 
v \to { v \over (c \tau+d) (c' \sigma + d')}
\eea
with $a,b,c,d,a',b', c', d' \in \ZZ$ and $ad-bc=a'd'-b'c'=1$. Since $\eta (\tau)^2$ transforms as a one-form under $SL(2,\ZZ)$ acting on $\tau$, the combination $\hat v$ is invariant under $SL(2,\ZZ) \times SL(2,\ZZ)'$, \bp{as well as under exchange of $\tau$ and $\sigma$}.

\subsection{Degeneration of the genus-two Kawazumi-Zhang invariant}

As a warm-up, we consider the behavior of the genus-two KZ-invariant under separating degeneration. Instead of the expression (\ref{KZA}) for the KZ-invariant, it will be more convenient to use the expression given in \cite{D'Hoker:2013eea} and valid for any scalar Green function,
\bea
\varphi (\Omega) = -\frac18 \int_{\SM^2} P(x,y|\Omega )\, \cG(x,y|\Omega )
\eea
The bi-form $P(x,y|\Omega ) $ is symmetric in $x,y$ and defined by,
\be
P(x,y|\Omega ) =  \Big (2 (Y^{-1})^{IL} (Y^{-1}) ^{JK} - (Y^{-1}) ^{IJ} (Y^{-1})^{KL} \Big )
\omega_I(x)\, \oom_J(x)\, \omega_K(y)\, \oom_L(y)
\ee
Up to terms of order $\cO(v)$, the form $P(x,y|\Omega )$ degenerates as follows, 
\be
P(x,y|\Omega) \to \begin{cases}  
- 4 \, \kappa _1(x) \kappa _1 (y) &  x,y \in \Sigma_1 \setminus \mD_1 \\
- 4 \, \kappa _1' (x') \kappa _1'  (y') &  x',y' \in \Sigma'_1 \setminus \mD_3' \\
+ 4 \, \kappa _1(x) \kappa _1' (y') &  x\in\Sigma_1 \setminus \mD_1 , y' \in \Sigma'_1 \setminus \mD_3'
\end{cases} 
\ee
Combining the asymptotic behaviors to this order of the Arakelov Green function $\cG$ in (\ref{SDArak}) and of the differential $P$ in (\ref{SDAb}), we see that all the contributions of the genus-one Green functions in (\ref{SDArak}) integrate to zero against the degeneration of $P$, and only the contribution of the terms proportional to $\ln |\hat v|$ survive, giving,
\bea
\label{phikzsep}
\varphi
 =  -\ln |\hat v| + \cO(|\hat v|)
\eea
This result is consistent with part a) of the main Theorem in \cite{MR1105425} for $(h_1,h_2)=(1,1)$, and is identical to the more precise asymptotics derived in \cite{D'Hoker:2013eea}.

\subsection{Degeneration of the genus-two invariants $\cZ_i$ and $\cB_{(2,0)}$}

Our starting point is the expression for the genus-two modular graph functions $\cZ_i$ of (\ref{defcB1}), and its simplified form (\ref{defB2}) after some of the trivial integrals over points on the surface have been performed, along with the relation (\ref{totalcB}). To evaluate the degenerations of these invariants, neglecting terms of order {$\cO(|\hat v|)$}, we use the asymptotics of the Arakelov Green function in (\ref{SDArak}), of the canonical K\"ahler form $\kappa $ in (\ref{SDkap}), and of the combination involving the bi-holomorphic form $\Delta$,
\bea
\Delta = \begin{cases} 
0 & ~~ x, y \in \Sigma_1 \setminus \mD_1 \hbox{ or }  x',y' \in \Sigma'_1 \setminus \mD_3' \\
+ \om(x) \wedge \om'(y')  & ~~ x\in\Sigma_1 \setminus \mD_1 , ~ y'\in \Sigma'_1 \setminus \mD_3' \\
- \om'(x') \wedge \om(y)  & ~~ x'\in\Sigma_1' \setminus \mD_3' , ~ y\in \Sigma_1 \setminus \mD_1 
 \end{cases}
 \eea
The results are as follows,
\bea
\cZ _1 & = & 2\,  \ln^2 |\hat v|  + 4 E_2 (\tau) + 4 E_2 (\sigma) + \cO(|\hat v|)
\no \\
\cZ _2 & = & - 2\, \ln^2 |\hat v| - E_2 (\tau) - E_2 (\sigma) + \cO(|\hat v|)
\no \\
\cZ _3 & = & 2\,  \ln^2 |\hat v|  + \cO(|\hat v|)
\label{Zisep}
\eea
Summing the contributions gives,
\bea
\label{B20sep}
\cB_{(2,0)} = 4 \,\ln^2 |\hat v| + 3 E_2 (\tau) + 3 E_2 (\sigma) + \cO(|\hat v|)
\eea
Note in particular that $\cB_{(2,0)} -4 \,\f ^2$ is finite as $v\to 0$.

\subsection{Degeneration of general genus-two modular graph functions \label{sec_gensep}}

General classes of modular graph functions at genus two and beyond  were constructed in  subsection 2.8 of \cite{DHoker:2017pvk}, and are given as follows,
\bea
\cC [n_{ij}; c(\sigma)] 
=
c^{I_1 \cdots I_N} _{J_1 \cdots J_N} \int _{\Sigma ^N} \prod _{i=1}^N \om_{I_i} (z_i) \,\oom^{J_i} (z_i) 
\prod _{1 \leq i < j \leq N} \cG (z_i,z_j)^{n_{ij}}
\eea
Here, $n_{ij}\geq 0$ are integers, while  $c^{I_1 \cdots I_N} _{J_1 \cdots J_N}$ is an invariant modular tensor  built out of a linear combination of products of Kronecker $\delta$-symbols. Given these properties, it  may be expressed as follows, 
\bea
c^{I_1 \cdots I_N} _{J_1 \cdots J_N} 
= \sum _{\sigma \in \mS_N} c(\sigma) \prod _{i=1}^N \delta _{J_{\sigma(i)} } {}^{I_i}
\eea
where $c(\sigma)$ are constants which depend on the permutation $\sigma\in \mS_N$.
The weight $w$ of the modular graph function $\cC$ is given as follows, 
\bea
w = \sum _{1 \leq i < j \leq N} n_{ij}
\eea
We shall limit attention to the case of genus-two though the results extend to higher genus. The asymptotics of $\cC$ under separating degeneration  is given by the following theorem.

\vskip 0.2in

\noindent
{\bf Theorem 1} ~ {\sl The behavior of the modular graph function $\cC$ in a neighborhood of the separating degeneration node is given by a polynomial of degree $w$ in $ \ln |\hat v|$ plus terms that are suppressed by positive powers of {$|\hat v|$}, by the following expression,
\bea
\cC [n_{ij} , c (\sigma)] = \sum _{k=0}^w {\rho_k} (\tau, \sigma) \big ( - \ln |\hat v| \big )^k + \cO(|\hat v|^{1 - \ep} )
\eea
for any $\ep >0$. The expansion parameter $ \ln |\hat v| $ and coefficients $ {\rho_k}(\tau, \sigma)$ are invariant under the residual group $SL(2,\ZZ) \times SL(2,\ZZ)'$ acting  on $\tau, \sigma$ and $v$ as in (\ref{modSS}).}

\sm

The proof of the theorem proceeds as follows. Up to corrections of order {$\cO(|\hat v|)$}, the differential forms $\om_{I_i}(z_i) \,\oom ^{J_i}(z_i)$ on the genus-two surface reduces to a linear combination  of $\kappa_1(z_i)$ when $z_i \in \Sigma _1 \setminus \mD_1$ and $\kappa_1'(z_i)$ when $z_i \in \Sigma _1' \setminus \mD_3'$. Therefore, to order {$\cO(|\hat v|)$} the integral over $\Sigma^N$ reduces to a sum of integrals over these genus-one components of products of powers of $\ln |\hat v|$,  $g(z_i-p|\tau)$, $g'(z_i'-p'|\sigma)$, $g(z_i,z_j|\tau)$ and $g(z_i',z_j'|\sigma)$, all of which integrate to produce terms obeying the properties of the expansion announced in the Theorem. This part is straightforward.

\sm

The more delicate part of the proof consists in showing that the corrections of order {$\cO(|\hat v|)$}  to the leading contributions in the separating degeneration of the holomorphic Abelian differentials is indeed suppressed as indicated by the theorem. This part is not straightforward in view of the fact that the coefficient of $v$ is a meromorphic differential which has a double pole at the punctures, so that its contribution to various integrals near the punctures could overcome the $\cO(|\hat v|)$ suppression factor. 

\sm

To proceed, we shall represent the genus-two surface $\Sigma$ by the union of the genus-one surfaces with boundary, given by $\Sigma _1 \setminus \mD_2$ and $\Sigma _1' \setminus \mD_2'$, where we choose the curves $\mC_2$ and $\mC_2'$ to be defined by,
\bea
t_2 = t_2' = - \ln |2 \pi \, v \, \eta (\tau) ^2 \eta (\sigma )^2 | + \cO(|v|^2)
\eea
For sufficiently small $v$, the parameters $t_2$ and $t_2'$ are large and the curves $\mC_2$ and $\mC_2'$ are approximately circles centered at the punctures with radii squared of order $v$, 
\bea
\mC_2 & = & \{ z \in \Sigma _1 \hbox{ such that } |z-p| = |v|^\half  + \cO(|v|^{3 \over 2}) \}
\no \\
\mC_2' & = & \{ z' \in \Sigma _1' \hbox{ such that } |z'-p'| = |v|^\half  + \cO(|v|^{3 \over 2}) \}
\eea
The terms of fastest growth in the integrals required in Theorem 1 are as follows,
\bea
|v|^2 \int _{\Sigma _1 \setminus \mD_2} |\varpi (x,p) |^2 f(x)
\eea
The cases we need are when $f(x)$ is continuous throughout $\Sigma _1$ or behaves as a power of a logarithm near the puncture $p$. For $f$ continuous near $p$, the integral is of the form, 
\bea
|v|^2 \int _{|x-p|> |v|^\half} d^2x { f(x) \over |x-p|^4} = |v| \int _{|\tilde x |> 1} d^2\tilde x { f(p+|v|^\half \tilde x ) \over |\tilde x|^4}
\eea
where the equality was obtained by changing variables locally by setting $x=p + |v|^\half \tilde x$.
Thus, the contribution to the integrals from the Abelian differential with double pole is  suppressed by a power of $|v|$. The same scaling argument shows that upon multiplying $f$ by a factor of $g(x-p|\tau)^n$, the suppression factor is instead $|v| (\ln |v|)^n$. An analogous argument goes through for multiple integrations, say over variables $x,y$,  involving also powers of the Green function $g(x-y|\tau)^n$, as may be seen from the following double integral, 
\bea
|v|^4  \int _{(\Sigma _1 \setminus \mD_2)^2} \!\!\!
 |\varpi (x,p) |^2  |\varpi (y,p) |^2  g(x-y|\tau)^n \approx 
 |v|^2 \int _{|\tilde x |, |\tilde y| > 1}  { d^2\tilde x \, d^2\tilde y \over |\tilde x|^4 |\tilde y|^4}
 \Big ( - \ln |v||\tilde x - \tilde y|^2 \Big )^n
\eea
which is now suppressed by $|v|^2$ times powers of $\ln |v|$.   For any $n$, the integrals are therefore bounded by $|v|^{1 - \ep}$ for any $\ep >0$, which concludes the proof of the Theorem. An alternative proof may be given using the variational method on $\ln |v|$, which is closer in spirit to the proof given for the non-separating degeneration in terms of $t$.

\newpage

\section{The tropical degeneration}
\setcounter{equation}{0}
\label{sec_trop}

The complete  degeneration of a compact genus-two Riemann surface is obtained by letting the imaginary part of the period matrix $Y$ {scale} to $\infty$ while keeping the ratios of its entries fixed. We shall refer to this limit as the {\sl tropical degeneration} because maximal degenerations are generally described by tropical geometry \cite{Itenberg:2011}. The tropical degeneration provides a suitable framework for examining the relation between the integrands for amplitudes of superstring theory and those of the associated supergravity \cite{Tourkine:2013rda}. 
In this section, we shall review the geometry and symmetries of the tropical degeneration, and then obtain the corresponding asymptotic  behavior of the string invariant $\cB_{(2,0)}$. This study will prepare the ground for the comparison between string and supergravity amplitudes in  Section \ref{sec_twoloopsugra}.

\subsection{Geometry and symmetry of the tropical degeneration}

The geometry and symmetries of the tropical degeneration are most easily exposed by parametrizing the imaginary part of the period matrix $Y$, given in (\ref{OmY}) and (\ref{3a1})  in terms of a positive real variable $V$ and a parameter $S=S_1+iS_2$ in the Poincar\'e upper 
half-plane~\cite{Green:1999pu},
\be
Y = \frac{1}{V\ttau_2} \begin{pmatrix} 1 & \ttau_1 \\ \ttau_1 & |\ttau|^2 \end{pmatrix} 
\ee
The relation between the two systems of coordinates is given by, 
\bea
\label{YVtau}
V=\frac{1}{\sqrt{t\rrho_2}}= (\det Y)^{-\half}
\hskip 1in 
\ttau= u_2+ i  \sqrt{\frac{t}{\rrho_2}}
\eea
where we recall that  $v_2 = \tau_2 u_2$ and $t = \sigma _2 - \tau_2 u_2^2$. The tropical degeneration corresponds to  letting $V\to 0$ while keeping $S$ fixed. In terms of the original variables, it arises equivalently by taking  $t \tau_2 \to \infty$ while keeping $u_2$ and $t/\tau_2$ fixed and non-zero.

\sm

The subgroup of the genus-two modular group $Sp(4,\ZZ)$ which leaves the tropical degeneration invariant acts on $\Omega$ by $2 \times 2$ matrices $A,B$ with integer entries,
\bea
\label{ABOm}
\Omega \to \Omega ' = A (\Omega +B) A^t
\eea 
where $A \in GL(2,\ZZ)$ and $B$ is symmetric. Parametrizing the matrix $A$ by $a,b,c,d \in \ZZ$ as exhibited below, we find that $V$ is invariant, while $S$ transforms as follows,
\bea
\label{GL2act}
A = \left ( \begin{matrix} d & c \cr b & a \cr \end{matrix} \right )
 \hskip 0.6in 
S \to { aS + b \over cS +d} \bigg |_{ \det A = 1}
\hskip 0.6in
S \to { a \bar S + b \over c \bar S +d} \bigg |_{\det A = -1}
\eea
The modular subgroup of these transformations is $GL(2,\ZZ) \ltimes \ZZ^3 \subset Sp(4,\ZZ)$.\footnote{The parameterization  of two-loop supergravity in terms of the coordinates $S$ and $V$ was introduced in the analysis of properties of two-loop maximal supergravity in \cite{Green:1999pu, Green:2005ba, Green:2008bf},  where the complex coordinate $S$ was denoted by $\tau$.}

\sm

\begin{figure}[h]
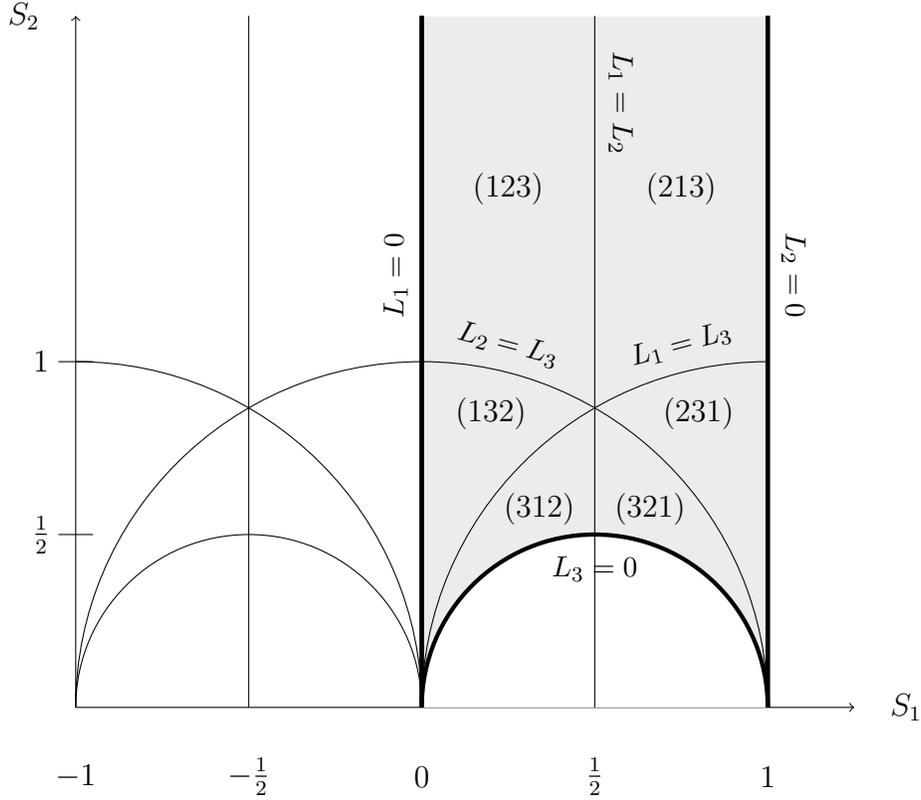

\begin{center}
\tikzpicture[scale=2.3]
\scope[xshift=-3cm,yshift=0.0]
\draw[white,fill=gray!15] (0,0) rectangle (2,4.0);
\draw (-2,0) -- (2,0) ;
\draw  [ultra thick] (0,0) -- (0,4.) ;
\draw   [ultra thick](2,0) -- (2,4) ;
\draw [->] (2,0) -- (2.5,0);
\draw (2.8, 0.0) node{$S_1$};
\draw [->] (-2,0) -- (-2,4.0) ;
\draw  (-2.3, 4.0) node{$S_2$};
\draw [fill=white,ultra thick] (2,0) arc(0:180:1.0) ;
\draw  (2,0) arc(0:90:2.0) ;
\draw (0,0) arc(180:90:2.0) ;
\draw (1,0) -- (1,4) ;
\draw (-2.1,1) -- (-1.9,1) ;
\draw (-2.1,2) -- (-1.9,2) ;
\draw  [ultra thick] (0,0) -- (0,4.) ;
\draw   [ultra thick](2,0) -- (2,4) ;
\draw (0,0) arc(0:180:1.0) ;
\draw  (0,0) arc(00:90:2.0) ;
\draw (-2,0) arc(180:90:2.0) ;
\draw (-1,0) -- (-1,4) ;
%
\draw (0, -0.4) node{$0$};
\draw (1, -0.4) node{$\frac{1}{2}$};
\draw (2, -0.4) node{$1$};
\draw (-1, -0.4) node{$-\frac{1}{2}$};
\draw (-2, -0.4) node{$-1$};

\draw (-2.2, 1.0) node{$\frac{1}{2}$};
\draw (-2.2, 2.0) node{$1$};

\draw (0.5, 3) node{{$(123)$}};
\draw (1.5, 3) node{{$(213)$}};
\draw (0.4, 1.7) node{{$(132)$}};
\draw (1.6, 1.7) node{{$(231)$}};
\draw (0.68, 1.15) node{{$(312)$}};
\draw (1.32, 1.15) node{{$(321)$}};
\draw (-0.15, 2.5) node[rotate=90]{\small $L_1=0$};
\draw (1.15, 3.5) node[rotate=-90]{\small $L_1=L_2$};
\draw (2.15, 2.5) node[rotate=-90]{\small $L_2=0$};
\draw (0.5, 2.1) node[rotate=-15]{\small $L_2=L_3$};
\draw (1.5, 2.1) node[rotate=15]{\small $L_1=L_3$};
\draw (1, 0.8) node{\small $L_3=0$};
\endscope

%
%
%
%
%
%
%
%
%
%

\endtikzpicture
\end{center} 
\caption{The extended fundamental domain $\hat\ccF$  (shaded in grey) is a six-fold cover of the  fundamental domain $\cF$ of $GL(2,\IZ)$ labeled by $(123)$. Here the labels $(ijk)$ denote the ordering $L_i<L_j<L_k$ of the Schwinger parameters.}
\label{fig:fundomain}
\end{figure}

We shall expand genus-two string invariants near the tropical degeneration within the approximation where all Laurent polynomial contributions in the components of $Y$ are retained but exponential contributions are neglected. This asymptotic expansion is the analogue of the one used for the non-separating degeneration where all exponential contributions in $t$ are neglected. Since periodicity forces the dependence on the moduli $\Re (\Omega)$ to be exponential, this dependence will vanish within the above approximation. Since the action of the transformation matrix $B$ given in (\ref{ABOm}) affects only $\Re(\Omega)$ and not $Y$, the $\ZZ^3$ components of the residual modular group acts trivially. Similarly, the center of $GL(2,\ZZ)$, which consists of the group of matrices $A=\pm I$ also acts trivially on $Y$. Therefore, the proper modular subgroup acting in the tropical degeneration will be $PGL(2,\ZZ) = GL(2,\ZZ) / \{ \pm I \}$. The corresponding  fundamental domain may be chosen as follows,
\bea
\ccF = \left \{ S \in \cH_1, \, 0 < S_1 < {\textstyle \half}, \,  |S| > 1 \right \}
\eea 
It will be convenient to consider the six-fold covering space $\hat \ccF$ of $\ccF$ defined as follows,
\bea
\hat \ccF = \left \{ S \in \cH_1, \, 0 < S_1 < 1, \,  |S-{\textstyle{\half }} | > {\textstyle {1 \over 4}} \right \}\ ,
\eea 
which happens to be a fundamental domain for the congruence subgroup $\Gamma_0(2)$ of matrices $A$ in \eqref{GL2act} with $\det A=1$ and $c=0$ modulo 2 (see Figure \ref{fig:fundomain}).
The corresponding deck transformations $\mS= \{ \Pi_0, \Pi_1, \Pi_2, \Pi_3, \Pi_4, \Pi_5  \}$ act on $\ttau$ by, 
\begin{align}
\label{mS}
\Pi_0(\ttau) & = \ttau & \Pi_2 (\ttau) & = 1-\ttau^{-1}  & \Pi _4 (\ttau) & = (1-\ttau)^{-1}
\no \\
\Pi_1(\ttau) & = 1- \bar \ttau & \Pi_3 (\ttau) & = \bar \ttau^{-1} & \Pi _5 (\ttau) & =  (1-\bar \ttau^{-1} )^{-1}
\end{align} 
and form a group $\mS$ which is isomorphic to the permutation group $\mS_3$, so that $\ccF= \hat \ccF/\mS$. The action of $PSL(2,\ZZ) \subset PGL(2,\ZZ)$ has isolated fixed points at $S=i \infty, i$ and $e^{2 \pi i/6}$ and their images under $SL(2,\ZZ)$. The set of fixed points $\cS$ of   transformations in $PGL(2,\ZZ)$ with $\det A=-1$ consists of the fixed line of the transformation $S \to - \bar S$ and its images under $SL(2,\ZZ)$.  The boundary components of $ \ccF$, located at $S_1=0$, $S_1 = {1 \over 2}$, and $|S |=1$ are fixed lines respectively  of  $S \to - \bar S$, $S \to 1 - \bar S$, and $S \to 1/\bar S$.  The fundamental domain $\ccF$ has been defined as an open subset of $\cH_1$ which excludes the points in~$\cS$. 

\sm

A more physical interpretation of the fixed lines may be obtained by changing variables from $V, S$ to real variables $L_1, L_2, L_3 >0$ related to one another as follows \cite{Green:1999pu},
\bea
Y= 2 \pi \left (  \begin{matrix} L_1+L_2 & L_1 \cr L_1 & L_1+L_3 \cr \end{matrix} \right )
\hskip 1in 
V = { 2 \pi \over ( L_1L_2+L_2L_3+L_3L_1)^\half}
\eea
The variable $S$ takes the following form, 
\bea
\label{defS12}
S_1= { L_1 \over L_1+L_2} 
\hskip 0.5in 
S_2 = {( L_1L_2+L_2L_3+L_3L_1)^\half \over L_1+L_2}
\eea
As will be further explained in the next section, the $L_i$'s arise as Schwinger parameters in supergravity Feynman diagrams.  The domain where all $L_i$'s are positive coincides with the domain $\hat \ccF$ and the  group $\mS$ of (\ref{mS}) acts by permuting the $L_i$'s (namely, $\Pi_1, \Pi_3, \Pi_5$ exchange
$(L_1,L_2),$  $(L_3, L_1)$, $(L_2,L_3)$, respectively while $\Pi_2$, $\Pi_4$ act by circular permutations). The domain $\cF$ corresponds to the particular choice of ordering $L_1<L_2<L_3$.
The boundary components of~$\hat \ccF$, namely $S_1=0$, $S_1=1$ and $|S-\half |={1 \over 4}$ respectively correspond to the vanishing of $L_1, L_2$, and $L_3$.  The intersection of the tropical degeneration with the non-separating degeneration ($t \to \infty$ for fixed $\tau$ and $v$),  corresponds to $L_3/L_1, L_3/L_2  \to \infty$ while keeping $L_1/L_2$ fixed, or equivalently $V \to 0$  and $S_2 \to \infty$ keeping $VS_2$ and $S_1$ fixed. The intersection of the tropical degeneration with the separating degeneration ($v \to 0$ for fixed $\tau$ and $\sigma$),  corresponds to $V, L_1  \to 0$ while keeping $L_3/L_2$ fixed, or equivalently $V, S_1 \to 0$  keeping $S_2$  fixed.

\subsection{Tropical limit of string invariants}
\label{sec_tropinv}

The asymptotic behavior of genus-two modular graph functions near the tropical degeneration consists of a Laurent expansion in the variable $V$ with coefficients which are functions only of $S$ (and $\bar S$), plus exponentially suppressed contributions, which we neglect. Since the original modular graph function is invariant under $Sp(4,\ZZ)$ and the expansion parameter $V$ is invariant under its $PGL(2,\ZZ)$ subgroup, it follows that the expansion coefficients of the Laurent polynomial in $V$ must also be invariant under $PGL(2,\ZZ)$. If the genus-two string invariant is real-analytic away from the separating degeneration locus $v=0$ (and $Sp(4,\IZ)$ images thereof), then each expansion coefficient will be a real-analytic function of $\ttau$ away from the locus $\cS$ given by the union of images of the line $\ttau_1=0$ under  the action of $GL(2,\IZ)$.The Laurent coefficients are real-analytic modular invariant functions on $\cH_1 \setminus \cS$, a class of functions  known as {\sl modular local polynomials}, first encountered in the study of two-loop supergravity amplitudes \cite{Green:2008bf} and further developed in the mathematics literature \cite{zbMATH06251204,bringmann2014modular}. 
We postpone a general discussion of these functions to subsection~\ref{sec_local}, and concentrate
here on the specific examples of the first two non-trivial genus-two string invariants, $\cB_{(0,1)}$ and  $\cB_{(2,0)}$.

\subsection{Tropical limit of non-separating degenerations}

In terms of  the variables $t,\tau_2,u_2$ introduced in \eqref{3a1}, the fundamental domain $\ccF$ covers the region $t>\tau _2(1-u_2^2)$, $0<u_2<\frac12$ which includes the non-separating degeneration $t\to\infty$ for $\tau_2$ fixed. We can therefore access the tropical limit $V\to 0$ for $\ttau_2$ near the cusp of $\ccF$ by starting
from the asymptotic series \eqref{LaurentBpq} in the non-separating degeneration limit $t\to\infty$.   In taking this limit, we shall retain only terms which are power-behaved in $\tau_2$, since exponentially suppressed terms will not contribute to the Laurent expansion around $V=0$. 
Due to the modular graph nature of the coefficients in the large $t$ expansion \eqref{LaurentBpq}, it will turn out that in the limit $\rrho_2\to\infty$  keeping $u_2=v_2/\tau_2\in]0,1/2[$ fixed, each of these coefficients reduces to a Laurent polynomial  in $\rrho_2$, with coefficients given by Bernoulli polynomials in $u_2$. After transcribing these results in terms of $V$ and $\ttau$, we will be able to express the Laurent coefficients around $V=0$ in terms of a family of local modular functions $A_{i,j}(\ttau)$ defined in the next subsection (see \eqref{defAij}).  This process is rather involved and the derivations are relegated to Appendix~\eqref{sec:D}, where the results are first expressed in terms of the variables $t, \tau_2$ and $u_2$. 

\sm

The tropical limit of the Kawazumi-Zhang invariant $\f$, first obtained 
at leading order in  \cite{DHoker:2014oxd} and then extended to all orders 
in \cite{Pioline:2015qha}, is derived by letting $\rrho_2\to 0$ at fixed $u_2$  in the expansion \eqref{nonkz}, and retaining only power-like terms in $\rrho_2$. The result reads
\bea
\label{tropkz2}
\varphi^{(t)} = \frac{5\pi}{6V} A_{1,0} +  \frac{5V^2 }{4 \pi^2} \zeta(3) \, A_{0,0}
\eea
where $A_{0,0}=1$ and 
\be
\label{defA10}
A_{1,0} (S) &=& \frac{S_2}{5}+  \frac{6}{5S_2}\, B_2(S_1)   + {1 \over S_2^3}  \left(\frac{1}{30}+ B_4(S_1)\right)\ .
\ee
where $B_{2k}(S_1)$ are Bernoulli polynomials of even index. These expressions are valid in the extended fundamental domain $\hat\ccF$ only, and can be extended to {continuous (but non differentiable) functions on} the full upper half plane by $GL(2,\IZ)$ invariance. As we shall see in the next section, the leading Laurent coefficient in $\f$ of order $1/V$ matches the two-loop supergravity integrand, while the sub-leading term, proportional to $\zeta(3) V$, can be traced to a two-loop diagram with a higher derivative $\cR^4$ interaction on one vertex. 

\sm

Similarly, the tropical limit of the genus-two modular graph functions $\cBZ_2, \cBZ_3$ is given by, \\
\bea
\cBZ^{(t)}_2 &=& {32\pi^2 \over V^2} \left[
\frac{1}{504} A_{0,0} - \frac{1}{1008} A_{0,2} -\frac{5}{792} A_{1,1} - \frac{17}{960} A_{2,0} \right] 
\nn\\ &&
- \frac{5 V \zeta(3)}{2\pi}\,  A_{1,0} -\frac{7 V^3 \zeta(5) }{4\pi^3} A_{0,1}
\label{Z2trop}
 \\
\cBZ^{(t)}_3 &=& {32\pi^2 \over V^2} \left[
-\frac{11}{7560} A_{0,0} + \frac{1}{1512} A_{0,2}+\frac{1}{792} A_{11} + \frac{17}{576} A_{2,0} \right] 
\nn\\ &&
+ \frac{5V\zeta(3)}{6\pi}  A_{1,0} +  \frac{11V^4  \zeta(3)^2}{8\pi^4}  A_{0,0}
\label{Z3trop}
\eea
where the  functions $A_{i,j}=A_{i,j}(S)$ are given in the extended fundamental domain $\hat\ccF$ by the following expressions, 
 \bea
  \label{arewrite}
A_{0,1} (S) &=&  S_2 + {1 \over S_2 }\left( \frac{5}{6} + B_2\right)
\no \\
A_{0,2} (S) &=& S_2^2+\left(\frac{2}{3} + 2B_2 \right) + {1 \over S_2^2} \left(\frac{7}{10} + 2 B_2 + B_4 \right)
 \nn\\
A_{1,1} (S) &=&\frac{ S_2^2}{7}+ \left (\frac{1}{70}  + {9 \over 7}  B_2 \right) + {1 \over S_2^2} \left(\frac{9}{7}  B_2   +  \frac{15}{7} B_4 \right)  
+\frac{1}{S_2^4}  \left(\frac{11}{420} + \frac{3}{2}  B_4  +  B_6 \right) 
\nn\\
A_{2,0} (S) &=&\frac {S_2^2}{33}+ \left( \frac{20}{693} + \frac{20}{33}   B_2 \right)+ {1 \over S_2^2} \left(\frac{20}{33} B_2+\frac{70}{33}  B_4\right) 
 \nn\\
    && + \frac{1}{S_2^4}\, \left(\frac{20}{11}  B_4 + \frac{28}{11} B_6 \right) 
    + \frac{1}{S_2^6} \left(\frac{1}{630} + \frac{4}{3}  B_6+ B_8 \right) 
    \quad
\eea
where for brevity we denote $B_{2n}=B_{2n}(S_1)$. After extending them to the full 
upper-half plane by modular invariance, 
the $A_{i,j}$'s become eigenfunctions of the Laplace-Beltrami  operator $\Delta=S_2^2 \left ( \p_{S_1}^2 + \p_{S_2}^2 \right )$ on $\cH_1\backslash \cS$,
\be
\Delta  A_{i,j} - n(n+1)\, A_{i,j}=0 \ ,\quad n=3i+j
\ee
up to a delta function source supported on the singular locus $\cS$. 
They provide a basis for the class of functions encountered in the
low energy expansion of the two-loop supergravity amplitude computed to high order in \cite{Green:2008bf}. In particular, they are invariant under the group of permutations $\mS_3$ on $L_1,L_2,L_3$. This particular basis was constructed by  Zagier \cite{ZagierPC} and
will be reviewed in the next subsection. 

\sm

For the remaining string invariant $\cBZ_1$, in principle we would need 
to compute the tropical limit of the
integral $\cK_{aaaa}^0$ defined in \eqref{Kaaaa}, which appears to be a project in its own right.
By construction however, $\cK_{aaaa}^0$ is only a function of $\tau$, independent of the variable $v$,
and therefore the tropical limit of $\cK_{aaaa}^0/t^2$ cannot be written as a linear combination of 
$V^\alpha A_{i,j}$ without spoiling the coefficients of the higher powers of $t$. Fortunately, there are other offending terms coming from the tropical limit of $\cK^c$ which are also independent of $u_2$. Collecting these terms together,
we obtain\footnote{
We thank A. Basu for pointing 
out a factor of 2 mistake in the coefficient of $\cK_{aaaa}^0$ in \eqref{Z1trop}, \eqref{Kctrop}, \eqref{cB1trop} and 
\eqref{B20trop}, which lead to erroneous values for the coefficient of $\zeta(3)^2$ in earlier
versions of this article. }
\bea
\cBZ^{(t)}_1 &=& { 32\pi^2 \over V^2} \left[
-\frac{1}{315} A_{0,0} + \frac{1}{252} A_{0,2}-\frac{1}{792} A_{1,1} + \frac{23}{960} A_{2,0} \right] 
\nn\\ &&+ {V \zeta(3) \over \pi} \, \left[ \frac{18}{5}\, A_{0,1}-\frac{1}{2} A_{1,0} \right]
-\frac{V^3 \zeta(5)}{2\pi^3} \,  A_{0,1} - \frac{3 \zeta (3)^2 V^4}{16\pi^4} A_{0,0}
\nn\\ && + 
\frac{1}{8\pi^2 t^2}\left[ 2\cK_{aaaa}^0 -  \frac{2 y^4}{945}+\frac{8 y \zeta
   (3)}{5}+\frac{145 \zeta (5)}{6 y}+\frac{3 \zeta (7)}{4 y^3}\right]
   \label{Z1trop}
\eea
For consistency with the symmetries of the tropical limit, the bracket on the last line must be proportional to $V^4 A_{0,0}$ with no dependence on $S$. We conclude that  the tropical limit of $\cK_{aaaa}^0$ must be given by 
\be
\label{Kaaaa0predict}
\cK_{aaaa}^0 = \frac12\left[ \frac{2 y^4}{945}-\frac{8 y \zeta
   (3)}{5}-\frac{145 \zeta (5)}{6 y}-\frac{3 \zeta (7)}{4 y^3}
   + \beta \frac{\zeta (3)^2}{y^2} \right]+\cO(e^{-2y})
   \ee
where the coefficient $\beta$  is unknown at this stage. Note that the naive evaluation of the integral \eqref{Kaaaa} by replacing $g(z)$ by its polynomial approximation $\mg_1(z)$ and ignoring the term proportional to $\zeta(3)$ correctly produces the
leading term in \eqref{Kaaaa0predict}.

\sm

We conclude that the tropical degeneration of the string invariant $\cB_{(2,0)}$ is given by,
\bea 
\cB_{(2,0)}^{(t)}&=& \half \cBZ_1^{(t)} - \cBZ_2^{(t)}+ \half \cBZ_3^{(t)}\nn\\
&=& {32\pi^2 \over V^2}\left[
{-\frac{13}{3024} A_{0,0} + \frac{5}{1512} A_{0,2}+\frac{5}{792} A_{1,1} + \frac{2}{45} A_{2,0}}
 \right] 
\label{B20trop}\\
&&  +{V \over \pi}  \zeta(3)\ \left[ {\frac{9}{5}  A_{0,1} + \frac{8}{3} A_{1,0}} \right]
+\frac{3V^3 }{{2\pi^3}}\zeta(5) A_{0,1} +{  \left(\beta+\frac{19}{2}\right)
 \frac{ \zeta (3)^2 V^4 }{16\pi^4}} A_{0,0} \nn
\eea
where the coefficient  $\beta$ could in principle be determined by a full analysis of 
the tropical limit of the integral $\cK_{aaaa}^0$ defined in \eqref{Kaaaa}, which we leave
for future work.

\subsection{Modular local  polynomials}
 \label{sec_local}
 
In this subsection we review the construction of the space of modular local polynomials,
following \cite{ZagierPC} and expanding thereon. Our goal is to construct functions $A(S)$
on the Poincar\'e upper half-plane, which are invariant under the action \eqref{GL2act}, real-analytic
away from the locus $\cS$, and given in each connected domain of $\cH_1\backslash \cS$ by 
a Laurent polynomial in $S_2$, with coefficients which are polynomial in  $S_1$. 
We further require that these functions satisfy the Laplace eigenvalue equation, 
\be
\label{DeltaA}
\left[\Delta   - n(n+1) \right] \, A=0
\ee
away from $\cS$, with $n\geq 0$ integer. Since the differential operator $D_k = \p_S + \frac{k}{S-\bar S}$
satisfies $\Delta_{k+2} \cdot D_k - D_{k} \Delta_k = -k D_k$, where 
{$\Delta_k=4 D_{k-2} \circ (S_2^2 \p_{\bar S})$} is the Laplacian acting on weight $k$ modular forms, it is clear that the function given locally by $A=D_{-2n}^{(n)} P$ where $D_{-2n}^{(n)}$ is the iterated derivative operator 
\be
\label{Dn}
D_{-2n}^{(n)} = {(-2i)^n n! \over (2n)!} D_{-2} \circ D_{-4} \circ \cdots \circ D_{-2n+2} \circ D_{-2n}
\ee
will satisfy \eqref{DeltaA} whenever $P$ is annihilated by the Laplacian $\Delta_{-2n}$,
in particular when $P$ is a holomorphic function of $S$. Since the operator \eqref{Dn} can be written
as 
\be
D_{-2n}^{(n)} = {(-2i)^n n! \over (2n)!} 
\sum_{m=0}^n \begin{pmatrix} n \cr m \end{pmatrix}\frac{(-n-m)_m}{(\ttau -\bar \ttau )^m}\, \frac{\partial^{n-m}}{\partial \ttau ^{n-m}}
\ee
where {$(k)_m=k(k+1)\dots (k+m-1)$ is the ascending} Pochhammer symbol, it is also clear that whenever
$P$ is a polynomial in $S$, the resulting
function $A$ will be a Laurent polynomial in $S_2$ with coefficients which are polynomial in $S_1$.
In order for $A(S)$ to be invariant under the action \eqref{GL2act}, the polynomial $P$ should transform according to, 
\bea 
\label{PGL}
P\vert_{\gamma} (S)   = 
\begin{cases} 
(cS+d)^{-2n} P \left ( { a S + b \over cS+d} \right )  & \mbox{if}\quad {\det \gamma =+1} \\
\\
(c S+d)^{-2n} \overline{P \left ( { a \bar S + b \over c \bar S+d} \right ) }
&  \mbox{if}\quad {\det \gamma =-1}
\end{cases}
\qquad \mbox{where}\quad 
\gamma = \left ( \begin{matrix} a & b \cr c & d \cr \end{matrix} \right )
\eea 
It is easy to check that this action preserves the space $V_{2n}$ of polynomials of degree at most $2n$, while polynomials of higher degree are mapped into rational functions. Thus, the functions
of interest are of the form $A=D_{-2n}^{(n)} P$ where $P(S)$ is a polynomial in $S$ of degree at most $2n$. 

\sm

Since the extended fundamental domain $\hat\ccF$ covers a single connected component
of $\cH_1\backslash \cS$, we must restrict to functions are invariant under the deck transformations \eqref{mS}. We claim that this  amounts to requiring that $P(S)$ is a sum of monomials $\sum_{i} C_{i} u^i v^{n-3i}$,  where  $C_i$'s are real constants and we set, 
\bea
u = S^2 (1-S)^2\ ,\qquad  v =  S^2 -S+1
\eea
This result may be established by considering the generating function $(S +t)^{2n}$ for the space of polynomials in $V_{2n}$ parametrized by $t \in \RR$. The operator \eqref{Dn} acts on this function by, 
\bea
D_{-2n}^{(n)}\, (S+t)^{2n} = |S + t |^{2n} S _2^{-n}
\eea
The deck transformations $\Pi_i \in \mS$ acting on the functions {$|S + t |^{2n} S_2^{-n}$} lift
to purely holomorphic transformations $\tilde \Pi_i$ acting on the functions {$(S+t)^{2n}$},
\bea
 \Pi _i \left ( |S + t |^{2n} S _2^{-n} \right ) = D_{-2n}^{(n)}\, \left \{ \tilde \Pi _i \left ( (S+t)^{2n} \right ) \right \} 
\eea
where
\begin{align}
\tilde \Pi _0 (S+t)^{2n} & = (S+t)^{2n} & \tilde \Pi_1 (S +t)^{2n} & =  (S-t-1)^{2n}
\no \\
\tilde \Pi _2 (S+t)^{2n} & = (S+t S -1)^{2n} & \tilde \Pi_3 (S+t)^{2n} & =  (t S  + 1 )^{2n}
\no \\
\tilde \Pi _4 (S+t)^{2n} & = (t S -t-1)^{2n} & \tilde \Pi_5 (S+t)^{2n} & =  (S+t S - t )^{2n}
\end{align}
The projection of the generating function $(S+t)^{2n}$ onto the space $\cP^\mS_n$ of 
weight $2n$ $\mS$-invariant polynomials is given by summing over all images,
\bea
P_t(S) = {1 \over 6} \sum _{i=0}^5 \tilde \Pi _i  (S+t)^{2n} 
\eea
Invariance under $S \to 1-S$  trivially implies that $P_t (S)$ is a polynomial in $S (1-S)$,
and therefore a polynomial in $u,v$. It remains to show that the only allowed monomials 
are those of the form $u^i v^j$ with $n=3i+j$. 

\sm

For this purpose, we  linearize  the action of $\mG$ by introducing a set of three complex variables, $z_1, z_2, z_3$ in terms of which $u$ and $v$ are given by symmetric polynomials,
\bea
z_1 + z_2 + z_3 & = & 0
\no \\
z_1^2 + z_2^2 + z_3^2 & = & 2 v
\no \\
z_1^3 + z_2^3 + z_3^3 & = & 3 \sqrt{u}
\eea
The projected generating function $P_t(S)$ can then be obtained as 
\bea
 z_2^{2n} P_t (-z_1/z_2) = F_t(z_1,z_2,z_3)
\eea 
where  $F_t$ is the following polynomial, which is  homogeneous of degree $2n$ in the variables $z_i$, and invariant under permutations of the $z_i$'s, 
\bea
F_t (z_1,z_2,z_3) & = & 
(t z_1-z_2)^{2n} + (t z_1-z_3)^{2n} +
(t z_2-z_1)^{2n} 
\no \\ && 
+ (t z_2-z_3)^{2n} +(t z_3-z_1)^{2n} + (t z_3-z_2)^{2n} 
\eea
Since $F_t$ is a symmetric polynomial in $z_i$, it may be expressed as a polynomial in $v$
and $\sqrt{u}$. Under the parity transformation $z_i \to -z _i$, the polynomial $F_t$ is invariant, while $v$ is even but $\sqrt{u}$ is odd.  Since the polynomials $u$, $v$, and $F_t$ have respective 
homogeneity degree weights 2, 6, and $2n$, we have the decomposition,
\bea
F_t (z_1,z_2,z_3)
= \sum _{i,j\geq 0; \, 3i+j=n} C_i(t) \,u^i v^j
\eea
for some polynomials $C_i(t)$ in $t$ with real coefficients, thus proving the announced result.  

\sm

We are now ready to define the family of functions $A_{i,j}$ whose first few
members appeared in the previous subsection: they are simply the descendents of the monomials
\bea
\label{defAij}
A_{i,j}(S) = D_{-2n}^{(n)} ( u^i v^j) \hskip 1in n= 3 i +j \quad \hbox{ with } \quad i,j \geq 0
\eea
 In the  fundamental domain $\hat\ccF$, the modular function $A_{i,j}(S)$ takes the following form,
\be
\label{Aijexp}
A_{i,j}(\ttau) =\sum_{k=0}^{2i+j} A_{i,j}^{(k)}(\ttau_1)\, \ttau_2^{i+j-2k}
\ee
where $A_{i,j}^{(k)}(S_1) $ is a polynomial of degree  $k$ in $S_1(1-S_1)$, and thus of degree $2k$ in $S_1$. Since it is invariant under $S_1\mapsto 1-S_1$, it may be expressed as a linear combination of Bernoulli polynomials $B_{2k}(S_1)$ of even index. After expressing $S_1, S_2$ in terms 
of $L_1,L_2,L_3$ using \eqref{defS12}, the function $A_{i,j}$ is  then by construction a homogenous function of the $L_i$'s, invariant under permutations. Multiplying $A_{i,j}$ by a power $V^\alpha$ 
and expressing it in terms of the variables $t,{\tau_2},u_2$, we see that $V^\alpha A_{i,j}$ has a Laurent expansion near $t=\infty$ with powers ranging from $\frac12(i+j-\alpha)$ to $-\frac12(3i+j+\alpha)$.
This Laurent expansion is compatible with that of a genus-two modular graph function with weight $w$ only when 
\be
\label{Ruleij}
i+j\leq 2w+\alpha\ ,\quad 3i+j\leq 2w-\alpha \ ,\quad 
|\alpha|\leq 2w\ ,\quad i+j-\alpha \ \mbox{even} 
\ee
For $w=2$ and $\alpha=-2$, 
this constraint singles out the functions $A_{0,0}, A_{0,2}, A_{1,1},A_{2,0}$ appearing in the leading term in \eqref{B20trop}.  The subleading terms in the same equation also satisfy the requirement \eqref{Ruleij}, but it is worth mentioning that terms with $\alpha=-4,-3,0,2$ are in principle allowed, although they do not occur in practice. In particular, agreement with supergravity requires $\alpha\geq -w$. 
Finally, since $A_{i,j}$ satisfies \eqref{DeltaA} with $n=3i+j$, it easily follows 
that $V^{\alpha} A_{i,j}$ is an eigenmode of the Laplacian $\Delta_{\cH_2}$ on the Siegel upper half plane with eigenvalue $\frac12[n(n+1)+\alpha(\alpha+3)]$, away from the separating degeneration locus.

\newpage

\section{Low energy expansion in two-loop supergravity}
\setcounter{equation}{0}
\label{sec:4}

The amplitudes of closed superstring theory are related at energy scales   $\ll (\alpha')^{-1/2}$ to amplitudes in maximal supergravity.  At tree level this connection is easy to demonstrate, 
but at loop level the connection to higher genus string amplitudes is more subtle due to 
ultraviolet divergences occurring in supergravity loop amplitudes. Still, the maximal degeneration of the {\it integrand} of the genus-$h$ superstring amplitude is expected to be related to the sum of integrands of the corresponding supergravity amplitude, and in fact provides an efficient
reorganisation of the sum over Feynman diagrams 
\cite{Bern:1991aq,Schmidt:1994zj,Roland:1996np}.  In this section we will compare the low energy expansion of the integrands in maximal supergravity with the genus-two string theory results of the preceding sections. Our discussion will highlight the fact that  the integrands of the Feynman diagrams do not capture the full content of the tropical limit that was analyzed in the last section -- the terms 
{proportional to odd zeta values}  do not arise from the field theory expression.

 \sm
 
\begin{figure}[h]
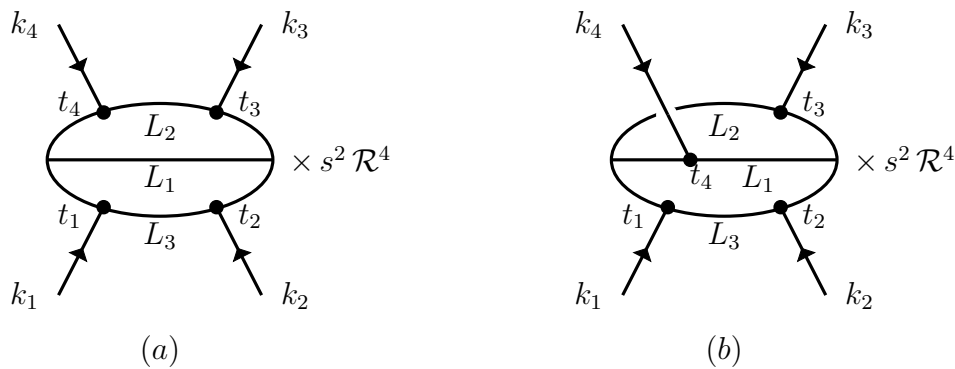

\begin{center}
\tikzpicture[scale=1.5]
\scope [very thick, every node/.style={sloped,allow upside down},xshift=-2.50cm,yshift=0cm]
\draw [draw=black, ]        (0,0) circle (1 and .5);
\draw (-1,0) -- (1,0) ;
\draw  (-0.9, 1.20)--  node {\midarrow}  (-0.5,0.42);
\draw  (0.9, 1.20) --  node {\midarrow}  (0.5,0.42);
\draw (0.9, -1.20) --  node {\midarrow}   (0.5,-0.42);
\draw    (-0.9, -1.20)--  node {\midarrow}   (-0.5,-0.42);
\draw ((-0.8, -0.5)node{$t_1$};
\draw ((0.8, -0.5)node{$t_2$};
\draw ((0.8,0.5)node{$t_3$};
\draw ((-0.8,0.5)node{$t_4$};
\draw (-1.2, -1.20) node{$k_1$};
\draw (1.2, -1.20) node{$k_2$};
\draw (1.2, 1.20) node{$k_3$};
\draw (-1.2, 1.20) node{$k_4$};
\draw (1.6, 0.0) node{$\times\, s^2 \, \cR^4$};
\draw (0,-.17) node{$L_1$};
\draw (0, .29) node{$L_2$};
\draw (0, -.67) node{$L_3$};
\draw [fill=black] (-0.5,0.42) circle [radius=.05];
\draw [fill=black] (0.5,0.42) circle [radius=.05];
\draw [fill=black] (-0.5,-0.42) circle [radius=.05];
\draw [fill=black] (0.5,-0.42) circle [radius=.05];
\draw (0, -1.7) node{$(a)$};
\endscope
\scope [very thick, every node/.style={sloped,allow upside down},xshift=2.50cm,yshift=0cm]
\draw [draw=black, ]        (0,0) circle (1 and .5);
\draw (-1,0) -- (1,0) ;
\draw (-0.51,0.40) [white, fill=white] circle(0.08cm) ;
\draw  (-0.9, 1.20) --   node {\midarrow}(-0.5,0.4);
 \draw (-0.5,0.4)--  (-0.3, 0.0);
\draw   (0.9, 1.20) --  node {\midarrow}  (0.5,0.42);
\draw  (0.9, -1.20)  --  node {\midarrow}  (0.5,-0.42);
\draw  (-0.9, -1.20) --  node {\midarrow}  (-0.5,-0.42);
\draw ((-0.8, -0.5)node{$t_1$};
\draw ((0.8, -0.5)node{$t_2$};
\draw ((0.8,0.5)node{$t_3$};
\draw ((-0.2,-0.15)node{$t_4$};
\draw (-1.2, -1.20) node{$k_1$};
\draw (1.2, -1.20) node{$k_2$};
\draw (1.2, 1.20) node{$k_3$};
\draw (-1.2, 1.20) node{$k_4$};
\draw (1.6, 0.0) node{$\times\, s^2 \, \cR^4$};
\draw (.3, -.17) node{$L_1$};
\draw (0, .29) node{$L_2$};
\draw (0, -.67) node{$L_3$};
\draw [fill=black] (0.5,0.42) circle [radius=.05];
\draw [fill=black] (-0.3,0.0) circle [radius=.05];
\draw [fill=black] (-0.5,-0.42) circle [radius=.05];
\draw [fill=black] (0.5,-0.42) circle [radius=.05];
\draw (0, -1.7) node{$(b)$};
\endscope
\endtikzpicture
\end{center} 
\caption{(a) A ``planar'' Feynman diagram with a pair of external states connected to
two different lines of a two-loop vacuum diagram.
(b)  A  ``non-planar''  Feynman diagram in which one pair of external states is attached to a single line
and the other states are each attached to separate lines. 
}
\label{fig:skeleton}
\end{figure}

  The Feynman diagrams contributing to the two-loop four-graviton amplitude in maximal supergravity were expressed in an efficient manner in \cite{Bern:1998ug}, where it was demonstrated that they could be reduced to a sum of diagrams of the form shown in Figure~\ref{fig:skeleton}. As indicated in that figure, each diagram has the structure of a graph in $\phi^3$ quantum field theory multiplied by a kinematic factor of $s^2\, \cR^4$.  The full amplitude, which is  symmetric in the external states is obtained by summing over the diagrams with the three inequivalent permutations of the external particles, which involve kinematic factors of $t^2\, \cR^4$ and $u^2\, \cR^4$ in addition to the $s^2\, \cR^4$ term shown in Figure~\ref{fig:skeleton}. Note that graphs in which more than two vertices are attached to a single line are absent.
   
\sm

After integrating over the loop momenta the expression for each Feynman integral involves integration over seven ``Schwinger'' parameters.  These may be interpreted as the positions of the four vertices, $t_i$ ($i=1,2,3,4$), and the parameters, $L_1$, $L_2$ and $L_3$, which label the lengths of the lines, and which take values in the range $0\le L_i\le \infty$. These real parameters of the Feynman integrand can be understood as the moduli of tropical Riemann surfaces \cite{Tourkine:2013rda} and are  analogous to the seven complex parameters that enter into the integrand of the genus-two superstring amplitude, \eqref{gen2amp} and \eqref{B2g}, which label the complex positions of the  four vertex operators and the three complex moduli of the compact genus-two surface.

\sm

Expanding the sum of Feynman diagrams in powers of $s$, $t$ and $u$ and integrating each term over $t_i$ leads to a  power series  of the form \eqref{Bpq} with the string coefficient functions  $\cB_{(p,q)}(\Omega)$ replaced by their supergravity counterparts,  $\cB^{(sg)}_{(p,q)} (L_1,L_2,L_3)$\footnote{The superscript ${}^{\sugra}$ will be used to label the supergravity versions of the various quantities in the following equations.}.  These functions can be obtained reasonably straightforwardly up to any given order in the low energy expansion by expanding the Feynman diagrams.  In \cite{Green:2008bf} these coefficients were explicitly evaluated up to terms with $2p+3q=6$ (terms of order $s^6$), and it is straightforward to generate them to much higher order.  The expressions for   $\cB^{(sg)}_{(p,q)} (L_1,L_2,L_3)$ are sums of the local modular functions $A_{i,j}(S)$ (that were defined in the last section) with rational coefficients, multiplied by a factor of  $V^{-w}$ (where $w$ was defined in \eqref{wdef}).

\sm

We are here interested  in studying the detailed correspondence between the tropical limit of the genus-two string amplitude and the supergravity expression.  For this purpose we would like to express the Feynman diagrams  in terms of sums over word-lines in  a manner that mimics the expression of string theory amplitudes as sums  over world-sheets.   Such a world-line procedure was described in the context of scalar field theory in \cite{Schmidt:1994zj,Roland:1996np,Dai:2006vj} and in the context of the two-loop four-graviton amplitude in maximal supergravity in \cite{Green:2008bf}.  In the latter reference  it was shown that  the low energy expansion of the sum of supergravity Feynman diagrams is reproduced by the sum of word-line diagrams.  In other words, the coefficient   $\cB_{(2,0)}^{(sg)}$ of the term at order $\sigma_2^2$ in the low energy expansion is given by a world-line expression analogous to the world-sheet expression in \eqref{defcB}.

\sm 

Such a world-line formulation will allow us to evaluate quantities $\cBZ^{(sg)}_1, \cBZ^{(sg)}_2, \cBZ^{(sg)}_3$ that are the supergravity analogues of the integrals of bilinears in the world-sheet Green function that were defined by \eqref{totalcB} and  \eqref{defcB1}.  We will then compare them with  the tropical limit of the string invariants, $\cBZ^{(t)}_1, \cBZ^{(t)}_2, \cBZ^{(t)}_3$ that were  computed in  Section~\ref{sec_trop}.   The form of the supergravity expressions,  when expressed in the world-line formalism, is given by,
  \bea
\label{defB1sugra}
\cBsugra_1 &=&\int _{\Gamma^4} \frac{\Deltasugra(1,3)\Deltasugra(2,4)}{8\, (\det \Ysugra)^2}
   \GAsugra(1,2) ^2 \\ 
\label{defB2sugra}
\cBsugra_2 &=&\int _{\Gamma^4} \frac{\Deltasugra(1,3)\Deltasugra(2,4)}{8\, (\det \Ysugra)^2}
\GAsugra(1,2) \, \GAsugra(1,4) \\
\label{defB3sugra}
\cBsugra_3 &=&\int _{\Gamma^4} \frac{\Deltasugra(1,3)\Deltasugra(2,4)}{8\, (\det \Ysugra)^2}
\GAsugra(1,2) \, \GAsugra(3,4)  
\eea
where $\GAsugra(i,j)=\GAsugra(t_i,t_j)$ now denotes the Arakelov Green function on the two-loop graph $\Gamma$ with 
three edges of length $L_1,L_2,L_3$  and $t_i$, $t_j$ ($i=1,2,3,4$) label the positions of the vertex operators that are to be integrated over the network of world-lines in the graph (as shown in Figure \ref{fig:skeleton}).  We normalize the (real) period matrix of  the graph, $\Ysugra$, so that it agrees  with the imaginary part of the period matrix of the Riemann surface $\Sigma$  in the maximal degeneration limit  (where we have made a particular choice for the arbitrary overall normalisation)
\be
\label{YL123}
\Ysugra=2\pi \begin{pmatrix} L_1 + L_2 & L_1 \\ L_1 & L_1 + L_3 \end{pmatrix}
\ee
The measure in \eqref{defB1sugra}--\eqref{defB3sugra}  involves factors of $\Deltasugra(i,j)$, each of which is a two-form on $\Gamma\times \Gamma$, and is the limit of the corresponding factor $|\Delta(z_i,z_j)|^2$ in the string measure defined in \eqref{defcB1}.  It reduces to $\pm 4 \, d t_i\, d t_j$ if the points $i,j$ are on different edges, and zero otherwise.

 \sm

As in the string computation, the expression for the total coefficient 
\be
\cB_{(p,q)}^{(sg)}=\frac12\cBsugra_1-\cBsugra_2+\frac12\cBsugra_3= 
\frac12\Bsugra_1-\Bsugra_2+\frac12\Bsugra_3
\ee
is independent of whether one uses the world-line Green function, 
$\Gsugra(t_i,t_j)$ (as in \cite{Dai:2006vj,Green:2008bf}) or the Arakelov Green function, $\GAsugra(t_i,t_j)$, but   the individual contributions $\cBZ_i$ and $\BZ_i$ differ in the two cases. In order to compare these with the tropical limit of the string calculation, it is crucial that we use the Arakelov Green function in the following.  As stressed earlier, the  use of the Arakelov Green function guarantees the conformal invariance of each individual component $\cBZ_i$.

\sm

However, just as in the string case, it is far more convenient to first compute the diagrams with the world-line Green function $\Gsugra(t_i,t_j)$, giving rise to contributions $\Bsugra_i$, which may then be transcribed into $\cBsugra_i$ by using the relation between the Green functions.  We will see that the supergravity results $\cBsugra_i(L_1,L_2,L_3)$ reproduces the leading term in the tropical limit $\cBZ^{(t)}_i(\Omega)$ of the string invariant, upon identifying the graph period matrix \eqref{YL123} (up to an overall scale factor) with the imaginary part of the period matrix $\Omega$. However,
the tropical
limit of the string amplitude also contains subleading terms which do not arise in the supergravity calculations, but can nevertheless be understood as two-loop amplitudes with higher-derivative vertices.

\subsection{Green functions on graphs \label{sec_worldline}}

We first recall the general definition of the world-line propagator from \cite{Dai:2006vj}, and its relation to the Arakelov-Green function. We consider a graph $\Gamma$ with $h$ loops, and denote the edges by $e_i$. We  choose a basis $a_I$, $I=1\dots h$ of homology cycles in $H^1(\Gamma,\ZZ)$. The dual one-forms $\omega_I\in H_1(\Gamma,\ZZ)$ are given by $\pm \de t$ on the edge $e_j$ if the edge $e_j$ lies on the cycle $a_I$, with a sign depending on the orientation of $e_j$ along $a_I$, or zero of $e_j$ does not belong to $a_I$. The period matrix of the graph is defined by $\Ysugra_{IJ}= 2\pi \int_{a_I} \omega_J$ and is a symmetric positive real matrix.  The world-line Green function on $\Gamma$ is given by\footnote{We use a somewhat unusual
normalization of the world-line Green function such that it agrees with \bp{the} string Green function
\eqref{G} in the tropical limit.}
\be
\Gsugra(t,t') = - s(p(t,t')) + 2\pi {\Ysugra}^{IJ} \int_{p(t,t')} \omega_I\, \int_{p(t,t')} \omega_J
\ee
where $s(t,t')$ is the length of the path $p(t,t')$ and ${\Ysugra}^{IJ}$ is the inverse of $\Ysugra_{IJ}$. Note that $\Gsugra(t,t')$ \bp{depends on the choice of homology basis and and a choice of path, which we fix by cutting the graph along $h$ edges such that it becomes simply connected.} Moreover it satisfies, 
\be
\label{delGsugra}
\partial^2_t \Gsugra(t,t') = -2 \delta(t-t') + 2h\, \ksugra(t)
\ ,\quad 
\ee
where $\ksugra(t)$ is the Arakelov one-form on $\Gamma$, given on the edge $e_i$ by 
\be
\ksugra_i = \frac{2\pi}{h} s_{IJ} {\Ysugra}^{IJ} \de t
\ee
where $s_{IJ}=\pm 1$ if the edge $e_i$ belongs to both $a_I$ and $a_J$, or $s_{IJ}=0$ if
it does not. Alternatively, $\ksugra_i=\de t/ (L_i+ r_i)$, where $L_i$ is
the length of the edge $e_i$ and $r_i$ is the effective resistance between the two endpoints
of $e_i$ once the edge is removed from $\Gamma$. By Foster's theorem from electric
network theory, $\sum_i \frac{L_i}{L_i+r_i}=h$ so  $\int_\Gamma \, \ksugra =1$. 
Note that unless $h=1$, the
r.h.s. of \eqref{delGsugra} does not integrate to zero. The Arakelov Green function $\GAsugra(t,t')$ is  obtained from $\Gsugra(t,t')$ using the relation,
\be
\label{GtoGA}
\GAsugra(t,t') = \Gsugra(t,t') - \gsugra(t) - \gsugra(t') +\ggsugra 
\ee
where
\be
\gsugra(t) = \frac12 \int_{\Gamma} \Gsugra(t,t') \, \ksugra(t')\ ,\quad \ggsugra=\int_\Gamma\, \gsugra(t)\, \ksugra(t) 
\ee
The Arakelov Green function satisfies,
\be
\label{delGAsugra}
\partial^2_t \GAsugra(t,t') = -2 \delta(t-t') +  \ksugra(t)\ ,\quad \int_\Gamma\, \GAsugra(t,t') \, \ksugra(t') = 0
\ee
so, unlike the world-line Green function, its integral using the Arakelov measure, vanishes.

\subsection{World-line evaluation of two-loop supergravity invariants \label{sec_twoloopsugra}}

We now apply the previous formulae to the diagrams of Figure \ref{fig:skeleton} with three edges of $e_i$ length $L_i$, $i=1,2,3$. We choose a basis of $H_1(\Gamma)$ such that the loop $B_1=e_1-e_2$, and $B_2=e_2-e_3$. Then, on the edge $e_1$, the Abelian differentials $(\omega_1,\omega_2)$ reduce to $(\de t,0)$; on $e_2$ to $(\de t,-\de t)$; and on $e_3$
to $(0,\de t)$. The period matrix of the graph is given in \eqref{YL123}. The canonical volume form is then 
\be
\ksugra(t)= \frac{(L_j+L_k)\,\de t}{2\Delta_L} 
\quad (t\in e_i)
\ee
where $\Delta_L=L_1 L_2+L_2 L_3+L_3 L_1=V^{-2}$ {and $\{i,j,k\}=\{1,2,3\}$}. It is straightforward to check 
that $\int_\Gamma \ksugra=1$. 
The world-line Green function is given by \cite{Green:2008bf}
\be
\label{Gsugra}
\Gsugra(t,t')=
\begin{cases}
 -\frac12|t-t'| + \frac{L_j+L_k}{2\Delta_L}(t-t')^2  & t,t' \in e_i\\
 -\frac12(t+t') + \frac{(L_j+L_k)t'^2}{2\Delta_L}
 +\frac{(L_i+L_k)t^2}{2\Delta_L}+\frac{2L_k tt'}{2\Delta_L} & t\in e_i, t'\in e_j
 \end{cases}
\ee 
 From this it follows that 
 \bea
\gsugra(t) &=& \frac{L_j+L_k}{4\Delta_L} t(t-L_i) - \frac{4L_1L_2L_3+L_i^2(L_j+L_k)+L_i(L_j^2+L_k^2)+
L_j L_k (L_j+L_k)}{24\Delta_L}  \nn\\
\ggsugra &=&  -\frac{(L_1+L_2)(L_2+L_3)(L_3+L_1)}{16\Delta_L} 
\eea
{where as before $t\in e_i$ and $\{i,j,k\}=\{1,2,3\}$,}
from which it is easy to obtain  $\GAsugra(t,t')$.

\sm

Using \eqref{GtoGA} it is easy to show
that the supergravity invariants \eqref{defB1sugra}--\eqref{defB3sugra} are related to their
counterparts $\Bsugra_i$ defined using the world-line Green function \eqref{Gsugra} via
the analogue of \eqref{BtocB}, 
\bea
\label{BtocBsugra}
\cBsugra_1&=& \Bsugra_1 + 32 (\ggsugra)^2 -64 \gggsugra  \nn\\
\cBsugra_2 &=&\Bsugra_2  -32 \ggggsugra
-32  \gggsugra+ 32 (\ggsugra)^2  \\
\cBsugra_3  &=& \Bsugra_3 -64  \ggggsugra + 32 ( \ggsugra)^2
\nn
\eea
where
\bea
\gggsugra &=&  \int_\Gamma \gsugra(t)^2 \ksugra(t) 
=\frac{1}{2880\Delta_L^2}
\left[ 13 \sum_{i\neq j} L_i^4 L_j^2 + 20 \sum_{i<j} L_i^3 L_j^3 \right. \nn\\
&&\left.
\qquad +  L_1 L_2 L_3 \left( 26 \sum_i L_i^3 + 67  \sum_{i\neq j} L_i^2 L_j\right)
 +106 L_1^2 L_2^2 L_3^2 \right]
\\
\ggggsugra&=& \int_{\Gamma^2}  \gsugra(t)  \gsugra(t') \, \frac{\Deltasugra(t,t')}
{(\det \Ysugra)^2} 
\nn\\
&=&
\frac{1}{576\Delta_L}
\left[ 2 \sum_{i\neq j} L_i^3 L_j +5 \sum_{i<j} L_i^2 L_j^2+
7  L_1 L_2 L_3 \sum_i L_i^3  \right]
\eea

\sm

The integrals $\Bsugra_i$ and $\cBsugra_i$ may now be evaluated using the above procedure.
  In the case of \eqref{defB1sugra}, if both $t_1,t_2$ are on the same edge $e_1$, then the integral over $t_3,t_4$ along the edges $e_2,e_3$ produces $2(L_2+L_3)^2/\Delta_L^2=8\ksugra(t_1) \ksugra(t_2)$. If $t_1,t_2$ are instead on distinct edges $e_1,e_2$, then the integral over $t_3,t_4$ along the edges $e_2,e_3$ produces $2(L_1+L_3)(L_2+L_3)/\Delta_L^2=8\ksugra(t_1) \ksugra(t_2)$. 

\sm

In the interest of brevity we will only display the results for $\cBZ_1,\cBZ_2,\cBZ_3$ which are based on the Arakelov Green function, suppressing the intermediate results for the $\BZ_i$'s, which are based on the \bp{world-line} Green function. We shall express the result both in terms of the Schwinger parameters $L_i$, and in terms of the local modular functions $A_{i,j}$ introduced in Section \ref{sec_local}, in order to facilitate comparison with the tropical limit of the string integrand. 

\medskip

\noindent $\bullet$\ For the supergravity integrand associated with the graph 1 in Figure \ref{fig:1}, we find,
\bea
\label{B201Asugra}
\cBsugra_1  &=&  \frac{2\pi^2}{9}
\left[ -\frac{4}{5}\Delta_L +
\frac{13}{20}(L_1+L_2+L_3)^2
-\frac{17}{10}(L_1+L_2+L_3)\frac{L_1L_2L_3}{\Delta_L}
+\frac{69}{20}\left( \frac{L_1L_2L_3}{\Delta_L}\right)^2 \right] \nn
\\
&=&
 { 32\pi^2 \over V^2} \left[
-\frac{1}{315} A_{0,0} + \frac{1}{252} A_{0,2}-\frac{1}{792} A_{1,1} + \frac{23}{960} A_{2,0} \right]
 \eea
This result precisely reproduces the leading term in the tropical limit  $\cBZ^{(t)}_1$ in \eqref{cB1trop} of the string invariant, up to subleading terms proportional to  odd zeta values,
namely 
\be
\cBZ^{(t)}_1= \cBsugra_1 
+  \zeta(3)\,V\, \left[ \frac{18}{5\pi}\, A_{01}-\frac{1}{2\pi} A_{01} \right]
-\zeta(5)\, \frac{V^3}{2\pi^3}  A_{01} + \beta  \frac{\zeta(3)^2 V^4}{\pi^4}
\ee

\noindent $\bullet$\ For the supergravity integrand associated with the graph 2 in Figure \ref{fig:1}, we get instead,
\bea
\cBsugra_2 &=& \frac{2\pi^2}{9}
\left[ \frac{1}{5}\Delta_L -
\frac{7}{20}(L_1+L_2+L_3)^2
+\frac{23}{10}(L_1+L_2+L_3)\frac{L_1L_2L_3}{\Delta_L}
-\frac{51}{20}\left( \frac{L_1L_2L_3}{\Delta_L}\right)^2 \right] \nn
\\
&=&{32\pi^2 \over V^2} \left[
\frac{1}{504} A_{0,0} - \frac{1}{1008} A_{0,2} -\frac{5}{792} A_{1,1} - \frac{17}{960} A_{2,0} \right]  
\eea
Again, this result precisely matches the tropical limit  $\cBZ^{(t)}_2$ in \eqref{cB2trop} of the string invariant, up to subleading terms proportional to odd zeta values, 
\be
\cBZ_2^{(t)}= \cBsugra_2
- \frac{5}{2\pi}\zeta(3)\, V\, A_{10} -\frac{7}{4\pi^3}\zeta(5) V^3 A_{01}
\ee

\sm

\noindent $\bullet$\ Finally, for  the supergravity integrand associated with the graph 3 in Figure \ref{fig:1} we find,
\bea
\cBsugra_3 &=& \frac{2\pi^2}{9}
\left[ 
\frac{1}{4}(L_1+L_2+L_3)^2
-\frac{5}{2}(L_1+L_2+L_3)\frac{L_1L_2L_3}{\Delta_L}
+\frac{17}{4}\left( \frac{L_1L_2L_3}{\Delta_L}\right)^2 \right] \nn
\\
&=&
 {32\pi^2 \over V^2} \left[
-\frac{11}{7560} A_{0,0} + \frac{1}{1512} A_{0,2}+\frac{1}{792} A_{11} + \frac{17}{576} A_{2,0} \right] 
\eea
Comparing with the tropical limit of $\cBZ_3$ in \eqref{cB3trop},  we have
\be
\cBZ^{(t)}_3 = \cBsugra_3 
+ \frac{5}{6\pi}\zeta(3) V A_{10} +  \frac{11\zeta(3)^2}{8\pi^2} V^4  
\ee
Combining these results, we find that the total supergravity invariant 
 which determines the $D^8\cR^4$ coupling is given by 
\bea
\cB_{(2,0)}^{(sg)}
&=&   \frac12(\cBsugra_1-2\cBsugra_2+\cBsugra_3)
\nn \\
&=& \frac{2\pi^2}{9}
\left[ -\frac{3}{5}\Delta_L +
\frac45(L_1+L_2+L_3)^2-\frac{22}{5}(L_1+L_2+L_3)\frac{L_1L_2L_3}{\Delta_L}
+\frac{32}{5}\left( \frac{L_1L_2L_3}{\Delta_L}\right)^2 \right] \nn
\\
&=& {32\pi^2 \over V^2}\left[
-\frac{13}{1512} A_{0,0} + \frac{5}{756} A_{0,2}+\frac{5}{396} A_{1,1} + \frac{4}{45} A_{2,0} \right] 
\eea
which agrees (up to an overall normalization convention) with the result in \cite{Green:2008bf}. Comparing with the
tropical limit of the string invariant given in \eqref{B20trop}, we find
\bea
\label{tropB}
 \cB_{(2,0)}^{(t)}=\cB_{(2,0)}^{(sg)}
  +V\, \zeta(3)\ \left[ \frac{18}{5\pi}  A_{01} + \frac{16}{3\pi} A_{10} \right]
+\frac{3}{\pi^3}\zeta(5) V^3 A_{01}
+\bp{(\beta+11)} \frac{ \zeta (3)^2 V^4 }{16\pi^4}
\eea
The leading term proportional to $1/V^2$ in the string integrand is therefore exactly reproduced 
by the supergravity computation. The subleading terms proportional to $V, V^3$ and $V^4$
are stringy corrections which can be interpreted as two-loop Feynman diagrams with one gravity vertex (or two vertices)  replaced by higher derivative couplings.

\newpage
\appendix

\section{Genus-one basics and integration formulas}
\setcounter{equation}{0}
\label{sec:A}

In this appendix, we summarize various definitions and results for functions and forms on a compact genus-one surface $\Sigma _1$, including the  volume form $\kappa _1$, the Green function $g$ and its successive convolutes $g_n$, the non-holomorphic Eisenstein series $E_a$ and its associated modular forms $D_{a,b}$. We shall also evaluate  various integrals on the genus-one surface with two boundary components $\Sigma_{ab}$ and reduce them to integrals on $\Sigma _1$ which have a smooth limit as $ t \to \infty$ and $\Sigma _{ab}$ tends to $\Sigma _1$ with punctures.

\subsection{Genus-one differentials and scalar Green function}
\label{sec:A1}

We parametrize a genus-one surface $\Sigma_1= \CC/(\ZZ+\ZZ \tau)$  with modulus  $\tau \in \cH_1$  by a complex coordinate $z=\alpha+ \beta \tau$ where $\alpha, \beta \in \RR/\ZZ$. We choose canonical $\mA_1$ and $\mB_1$ homology cycles along the identifications $z\approx z+1$ and $z\approx z+ \tau$ respectively, and denote the holomorphic Abelian differential dual to the $\mA_1$-cycle by $\om_1(z) = d z$. The volume form $\kappa _1$ of unit area, and the corresponding ``coordinate" Dirac $\delta$-function are as follows,
\bea
\kappa_1 (z)  = { i \over 2 \tau_2} dz \wedge d \bar z = d\alpha \wedge d\beta
\hskip 1in 
\delta ^{(2)} (z) = {1 \over \tau_2} \delta (\alpha) \delta (\beta)
\eea
The derivatives with respect to $z$ are related to those with respect to $\alpha , \beta$ by,
\bea
\p_z = { 1 \over 2 i \tau_2} (\p_\beta - \bar \tau \p_\alpha)
\hskip 1in
\pbz = -{ 1 \over 2 i \tau_2} (\p_\beta - \tau \p_\alpha)
\eea
while the (negative) Laplace operator in $z$ is given by,
\bea
\Delta _z =  4 \tau_2 \pbz \p_z = { 1 \over \tau_2} (\p_\beta - \tau \p_\alpha)(\p_\beta - \bar \tau \p_\alpha)
\eea
The scalar Green function $g$ is defined by,
\bea
\Delta _z \gone(z|\tau) = -4 \pi \tau_2 \delta ^{(2)} (z) +4 \pi 
\hskip 1in \int _{\Sigma_1} \kappa_1 (z) \gone(z|\tau)=0
\eea
It may be expressed as a double Fourier series {(which converges provided $z\notin \IZ+\IZ\tau$)},
\bea
\label{g1}
\gone(z|\tau) = \sum _{(m,n) \not= (0,0)} { \tau_2\over \pi |m + n \tau|^2} \, e^{2 \pi i (m\beta -n\alpha)}
\eea
or in terms of  the Jacobi $\tet$-function, and  the Dedekind $\eta$-function,
\bea
\label{5c1}
\gone(z|\tau) = - \ln \left | { \tet _1 (z |\tau) \over \eta (\tau) } \right |^2 + { 2 \pi \over \tau_2} \left ( \Im z \right )^2
\eea
The Green function $\gone(z|\tau)$ is doubly periodic in $z$ with periods $\ZZ+ \ZZ\tau$ and   is invariant under $SL(2,\ZZ)$ modular transformations, as given in (\ref{modgk}).

\subsection{Kronecker-Eisenstein series and elliptic polylogarithms}
\label{sec:A2}

Iterated integrals of the scalar Green function and its derivatives give non-holomorphic Eisenstein series and the elliptic polylogarithm functions $D_{a,b}(z|\tau)$  introduced in \cite{zbMATH04144378}.  In this subsection, we shall provide the precise normalizations of these integrals, and often replace  $D_{a,a}$ by a  more transparent modular function $g_a$. These functions are defined as follows using the notation $z=\alpha +\tau \beta$, for $\alpha, \beta \in \RR$, and for $b - a \in \ZZ$, 
\bea
\label{defEgD}
g_a(z|\tau) & = & \sum _{(m,n) \not= (0,0)} { \tau_2^a \,  e^{2 \pi i (m\beta -n\alpha)} \over \pi^a |m + n \tau|^{2a} }
\no \\
D_{a,b} (z|\tau ) & = & 
{ (2 i \tau_2)^{a+b-1} \over 2 \pi i} \sum _{(m,n) \not= (0,0)} { e^{2 \pi i (m\beta -n\alpha )} \over (m + n \tau)^a
(m + n \bar \tau)^b } 
\eea
Clearly, we have $g_1(z|\tau) = g(z|\tau)$ and  $g_a(0|\tau)  =  E_a (\tau)$, and 
\bea
\label{5e2}
D_{a,a} (z|\tau) = (- 4 \pi \tau_2)^{a-1} g_a(z|\tau)
\eea
The functions $g_a(z |\tau)$  satisfy the  modular transformation law of $g$ in (\ref{modgk}), while $D_{a,b}$ for $a\not=b$ transforms as a modular form. They satisfy the following integral relations,
\bea
\label{gconvol}
g_{a_1+a_2} (z |\tau) & = & \int _{\Sigma_1} \kappa_1 (w) \, g_{a_1}(z-w|\tau) \, g_{a_2} (w |\tau)
\no \\
D_{a_1+a_2, b_1+b_2} (z |\tau) & = & - 4 \pi \tau_2 
\int _{\Sigma_1} \kappa_1(w) \, D_{a_1, b_1} (z-w |\tau) D_{a_2, b_2} (w |\tau)
\eea
The functions $g_a(z|\tau)$ and $D_{a,b}(z|\tau)$ satisfy the following differential equations, 
\bea
\label{diffD}
\p_z^n g_a(z|\tau) & = & (2 \pi i)^n (- 4 \pi \tau_2)^{1-a} D_{a,a-n}(z|\tau)
\no \\
\Delta _z g_a (z |\tau) & = & -4 \pi g_{a-1} (z |\tau)  
\no \\
\p_z D_{a,b} (z|\tau) & = & + 2 \pi i D_{a,b-1}(z |\tau)
\no \\
\pbz D_{a,b} (z|\tau) & = & - 2 \pi i D_{a-1,b}(z |\tau)
\no \\
\Delta _z D_{a,a} (z |\tau) & = & 16\pi^2 \tau_2  D_{a-1,a-1} (z |\tau) 
\eea
The differential relations given thus far were with respect to the parameter $z$. Actually, several differential relations with respect to the modulus $\tau$ will also be useful in the sequel and will be derived here. The basic differential equation for $D_{a,b}$  in the modulus, from which all others may be deduced,  is given by,
\bea
\label{A28}
2 i \tau _2 \p_\tau D_{a,b} (v |\tau) = a D_{a+1, b-1} (v |\tau) +(b-1) D_{a,b} (v |\tau)
\eea
We record the standard normalization the  Laplace operator on the upper half plane, 
\bea
\Delta_\tau = 4 \tau_2^2 \p_\tau \p_{\bar \tau}
\eea
For given $\alpha, \beta$ the function $g_a(\alpha + \tau \beta|\tau)$  is an eigenfunction of $\Delta _\tau$  with eigenvalue $a(a-1)$.

\subsection{Reducing integrals on $\Sigma _{ab}$ to integrals on $\Sigma _1$}
\label{sec:A3}

Extracting the power behavior in $t$ of integrals of various products of Green functions is achieved by recasting integrals over the genus-one surface $\Sep$ with boundary  to a sum over integrals over the compact genus-one surface $\Sigma _1$ without boundary. The key result, obtained in Section 3.5 of \cite{DHoker:2017pvk}, states the following relation between integrals for $I,J\in \{1,t\}$,
\bea
\int _\Sep \om _I \wedge \oom_J \, \psi = \int _{\Sigma _1} \om _I \wedge \oom_J \, \psi + \cO(e^{-2 \pi t})
\eea
provided $\psi$ is smooth near the punctures $p_a,p_b$ and $(I,J) \not= (t,t)$. The relation also holds when $I=J=t$ provided $\psi$ vanishes at both punctures. Several of the integrals below were derived in Section 4.4 of \cite{DHoker:2017pvk}. In the remainder of this appendix, we shall no longer indicate the exponentially suppressed terms, which will always be understood.

\subsection{Integrals involving two punctures}

We refer to integrals involving two punctures as those whose integration measure is singular at both punctures.
The following integrals \cite{DHoker:2017pvk} are valid for any integer $n \geq 0$, 
\bea
\label{oddf}
\int  _{\Sep }\, \om _t \wedge \oom _1 \, f^n  = \int  _{\Sep } \om _t \wedge \overline{\om _t } \, f^{2n+1}  = 0
\eea
and 
\bea
\label{regs}
\int _{\Sep } \om _t \wedge \overline{\om _t } \, \, f^{2n}  =  - { 2 i \over 2n+1} (2 \pi)^{2n} t^{2 n+1}
\eea
Throughout, it will be convenient to use the following notation,
\bea
\label{fnz}
f_n (z) = g_n(z-p_b) - g_n(z-p_a)
\eea
where for $n=1$ we recover $f_1(z)=f(z)$. We have the following integrals,
\bea
\label{gD}
\int _{\Sep }\om_t(z) \wedge \oom_1(z) \, g_n(z-w)  
 =  {\tau_2 \over  \pi } \, \p_w  f_{n+1} (w) 
\eea  
For any function  $\psi(z)$ which is smooth on $\Sep $, and whose Laplacian $\p_z\pbz \psi(z)$ is smooth on $\Sep$,  but which does not need to extend to a smooth function at the punctures $z=p_a,p_b$, we have the following integral formula,
\bea
\label{fint}
\int _\Sep \om _t  \wedge \overline{\om_t } \, f^n \psi
& = & 
- \,  { i \, (2 \pi t )^{n+1}  \over 4 \pi^2 (n+1)} \int _0 ^{ 2 \pi} d \theta 
\Big ( \psi \left (p_b^\theta   \right ) 
+ (-)^n \psi \left (p_a^\theta  \right ) \Big ) 
\no \\ &&
- {i \tau_2 \over 2 \pi^2(n+1)(n+2) }  \int _{\Sigma _1} \kappa_1 (z) \,  f(z)^{n+2} \, \p_z \pbz \psi (z)
\eea
where $p_a^\theta, p_b ^\theta$ are defined in terms of the variable $R$ as follows, 
\bea
\label{defT}
- 2 \ln R & = & 2 \pi t + \lambda(\tau)  + g(v|\tau)
\no \\
p_a^\theta & = & p_a + R \, e^{ i \theta}
\no \\
p_b^\theta & = & p_b + R \, e^{ i \theta}
\eea
and $\lambda$ is  given by, 
\bea
\label{deflambda}
g(z|\tau ) & = & - \ln |z|^2 - \lambda(\tau) + \cO(z) 
\no \\ 
\lambda (\tau) & = &  \ln \left | 2 \pi  \eta(\tau)^2 \right |^2
\eea
In particular, the following special cases will be used in the sequel,
\bea
\label{ompat}
\int _{\Sep } \om _t(z)  \wedge \oom_t (z) \, f(z) \, \gone(z,w)
& = & - i \pi t^2 f(w) + { i \over 12 \pi} f(w)^3
\no \\
\int _{\Sep } \om _t(z)  \wedge \oom_t (z) \, f(z)^{2n} \, \gone(z,w)
& = &
- { i \, (2 \pi t)^{2n+1} \over 4 \pi^2 (2n+1) }\int _0 ^{2 \pi} d \theta  \Big ( g(w,p_a^\theta) + g(w, p_b^\theta) \Big )
\no \\ &&
-{ i \, (2n)! \over 2\pi} F_{2n+2} + { i \, f(w)^{2n+2} \over 4 \pi (2n+1) (n+1)}
\eea
When $\psi$ extends to a smooth function at the punctures, the function $\psi$ inside the $\theta$-integral in (\ref{fint}) is constant up to exponentially suppressed corrections, and simplifies as follows,
\bea
\label{fgint}
\int _\Sep \om _t  \wedge \oom_t  \, f^n \psi
& = & -i \,  {  (2 \pi)^n t^{n+1}  \over n+1} 
\Big ( \psi (p_b)  + (-)^n \psi (p_a) \Big )
\no \\ &&
- {i \tau_2 \over 2 \pi^2(n+1)(n+2) }  \int _{\Sigma _1} \kappa_1(z) \,  f(z)^{n+2} \, \p_z \pbz \psi (z)
\eea

\subsection{Integrals involving at most one puncture}

We refer to integrals involving at most one puncture as those whose integrand is singular at most at only one puncture. The following integrals are valid for any integer $n \geq 0$,
\bea
\label{G1}
{\tau_2 \over \pi} \int _\Sep \kappa_1 (z) |\p_z g(z,p_a)|^2 \, g(z,p_a)^{n-1}
 =   { 1 \over n } \Big ( T ^n - D_n \Big )
\eea
where  we have used the parameter $T$, defined by,
\bea
\label{mt}
T= - 2 \ln R - \lambda = 2 \pi t + g(v)
\eea
We also use the following  integrals,
\bea
\label{G25}
{\tau_2 \over \pi} \int _{\Sigma_1}  \kappa_1(z) |\p_z g(z,p_a)|^2 \Big ( g(z,p_b)^2 - g(p_a,p_b)^2 \Big ) 
& = &
-  D_3 ^{(1)} (v) - 2  D_4 ^{(a)}(v)
\no \\
 \label{G4}
{\tau_2\over \pi}  \int _\Sep \kappa_1(z) \pbz g(z,p_a)  \p_z g(z,p_b) g(z,p_a)^{n-1}
& = & {1 \over  n} \Big ( g(v)^n -  D_n \Big )
\eea
where the function $D_n^{(1)} (v) $ was defined in (\ref{Dlk}) and $D_4 ^{(a)}(v)$ is defined by,
\bea
\label{D4a}
D_4 ^{(a)} (z |\tau) = { \tau_2 \over \pi} \int _{\Sigma _1} \kappa _1 (x) g(z+x) \p_x g(z+x) g(x) \pbx g(x)
\eea
It may be expressed as the Laplacian in the variable $z$ of the function $D_4^{(2)}$ as follows,
\bea
\label{D4adiff}
\Delta _z D_4^{(2)} (z|\tau) = - 16 \pi D_4 ^{(a)} (z|\tau)
\eea
Alternatively, it may also be expressed in terms of the Laplacian in $z$ of $F_4$, using the above identity, (\ref{F4}), and (\ref{delDk}), and we find, 
\bea
D_4^{(a)} (v) = {1 \over 3} g_1^3 - {1 \over 3} D_3 -{1 \over 4 \pi} \Delta _v F_4(v)
\eea
It is in this form that we shall present the final results involving $D_4 ^{(a)}$.

\sm

Furthermore, we have the following integrals for $n \geq 1$ which involve the Green function $g(z,w)$ at a generic point $w$ on $\Sep$, 
\bea
\label{G2}
&&
{\tau_2 \over \pi} \int _\Sep \kappa_1 (z) |\p_z g(z,p_a)|^2 g(z,p_a)^{n-1} g(z,w) \
\\ && \quad =
{ 1 \over n(n+1)} \Big ( D_{n+1}  - g(w,p_a)^{n+1} \Big )
- {1 \over n} D_{n+1}^{(1)} (w-p_a)
+ {T^n \over 2\pi n}   \int _0 ^{2 \pi} d \theta  \, g(w,p_a^\theta ) 
\no
\eea
In the special case $n=1$, we have $D_2=E_2$ and $D_2^{(1)}(w-p_a) = g_2(w-p_a)$.

\newpage

\section{Non-separating degeneration of $\BZ_i$ and $\cBZ_i$}
\setcounter{equation}{0}
\label{sec:B}

In this appendix we will present some of the core calculations of this paper and compute the three contributions to $\cB_{(2,0)}$ defined  in (\ref{totalcB}),  (\ref{defcB1}) in terms of the functions $\cBZ_i$ and in \eqref{defB1} and (\ref{defBx}) in terms of the functions $\BZ_i$. In the process, we shall evaluate the intermediate functions of (\ref{BtocB}) as well. The first ingredient in this evaluation is the relation (\ref{limG}) between the Green function $G$ on the genus-two Riemann surface $\Sigma$, in terms of which the integrals in $\BZ_i$ are expressed, and its representation in terms of the genus-one surface $\Sigma _{ab}$, which we repeat here for convenience,
\bea
\label{Gg}
G(x,y) = \gone(x,y) + \frac{1}{8\pi t} \Big (f(x) - f(y) \Big )^2 + \cO(e^{-2\pi t}) 
\eea
The second ingredient is the analogous expression for the integration measure, which may be decomposed in terms of the following factors,
\bea
\label{del}
|\Delta (z_i, z_j)|^2 = | \om_1 (z_i) \wedge \om_t (z_j) - \om_t(z_i) \wedge \om_1(z_j)  |^2 = 
\nu _{ij} ^- - \nu_{ij}^+
\eea
where the forms $\nu_{ij}^\pm$ have been separated according to their parity  in the form $\om_t$ and its complex conjugate at each point. These forms are given explicitly by, 
\footnote{For the sake of brevity, we shall often abbreviate the points $z_i$  by $i$ in the arguments of functions and forms, and we shall omit the wedge in the product of forms.}
\bea
\label{nupm}
\nu_{ij} ^+ & = & 
\omega_1(i) \wedge \oom_1(i) \wedge \om_t(j) \wedge \oom_t(j) 
+ \omega_t(i) \wedge \oom_t(i) \wedge \om_1(j) \wedge \oom_1(j) 
\no \\
\nu_{ij} ^- & = & 
\om_1(i) \wedge \oom_t(i) \wedge  \om_t(j) \wedge \oom_1(j) 
+ \omega_t(i) \wedge \oom_1(i) \wedge \om_1(j)\wedge \oom_t(j) 
\eea
The third ingredient is the representation of the genus-two Arakelov K\"ahler form $\kappa$ in terms of data on $\Sigma _{ab}$, given by (\ref{limkappa}) which we repeat here for convenience,
\bea
\label{limk}
\kappa = \half  \kappa_1  +  \frac{i}{4 t} \omega_t \wedge \bar\omega_t + \cO(e^{-2\pi t})
\hskip 1in \kappa _1 = \frac{i}{2 \tau_2} \om_1 \wedge \oom_1
\eea
The determinant is given by $\det Y = t \tau_2$. Finally, we shall extract the $t$-power dependence of the integrals over $\Sep$ and cast the result in terms of a Laurent polynomial in $t$ with coefficients given by convergent integrals over $\Sigma _1$. A very useful tool will be Stokes theorem on the surface $\Sigma _{ab}$ for a $(1,0)$ form $\om = \om_z(z) dz$, formulated as follows,
\bea
\label{stokes}
\int _\Sep \kappa_1 (z) \, \pbz \om_z (z) = -{ i \over 2 \tau_2} \oint _{\p \Sep} dz \, \om _z(z)
= { i \over 2 \tau_2} \left ( \oint _{\mC_a} + \oint _{\mC_b} \right )  \om _z(z) dz
\eea
We shall carry out these procedures in increasing order of difficulty and complexity. We begin with $\BZ_3$, then $\BZ_2$ and finally compute $\BZ_1$.

\subsection{Degeneration of $\BZ_3$}
\label{B3sec}

We start with the simplest modular graph function $\BZ_3$ defined in \eqref{defB1}, which corresponds to the disconnected diagram on the right of Figure~\ref{fig:1}.  Using \eqref{Gg}, it will be convenient to expand   $\BZ_3 $, in terms of the number of $f$ functions,  into a sum of 3 terms,
\bea
\BZ_3^{(a)} &=& \frac{1}{8t^2 \tau_2^2}\int _{\Sigma _{ab} ^4}|\Delta(1,3)\Delta(2,4)|^2\, 
\gone(1,2) \, \gone(3,4) 
\no \\
\BZ_3^{(b)} &=& \frac{1}{32\pi t^3 \tau_2^2}\int _{\Sigma_{ab} ^4}|\Delta(1,3)\Delta(2,4)|^2\, 
\gone(1,2) \, \big ( f(3)-f(4) \big )^2 
\no \\
\BZ_3^{(c)} &=& \frac{1}{512 \pi^2 t^4 \tau_2^2}\int _{\Sigma_{ab} ^4}|\Delta(1,3)\Delta(2,4)|^2\, 
\big ( f(1)-f(2) \big ) ^2 \, \big ( f(3)-f(4)\big )^2 
\eea
where it is understood that $\Delta$ is expressed in terms of $\om_1$ and $\om_t$ using (\ref{del}).
We will start with the last of these integrals since it is the simplest.

\subsubsection{Evaluating $\BZ_3^{(c)}$}

Thanks to the property \eqref{oddf}, the contributions of $\nu_{13}^-$ and $\nu_{24}^-$ in (\ref{del}) integrate to zero. Further using symmetry and the property that terms linear in $f$ integrate to zero against either $|\omega_1|^2$ or $|\omega_t|^2$ (see  \eqref{oddf} and \eqref{regs}),  we can replace $(f(1)-f(2))^2(f(3)-f(4))^2$
appearing in $\BZ_3^{(c)}$ by $2(f(1)^2+f(2)^2) f(3)^2$ to obtain,
\bea
\BZ_3^{(c)}=
\frac{1}{256\pi^2 t^4 \tau_2^2} \int_{\Sigma _{ab} ^4} \, \nu_{13}^+ \, \nu _{24}^+
\Big ( f(1)^2+f(2)^2 \Big ) f(3)^2
\eea
The integral over the point 4 can be computed using  \eqref{oddf} and \eqref{regs}, 
\bea
\label{nuint}
\int _{\Sigma _{ab} ^{(4)}} \nu_{24}^+ = - 2i t |\om_1(2)|^2 - 2 i \tau_2 |\om_t(2)|^2 = - 8 t \tau_2 \kappa (2)
\eea
Using the function $F_\ell$ of (\ref{Fn}) and (\ref{F2}) as well as equation \eqref{regs} to compute the remaining integrals successively, we arrive at,
\bea
\BZ_3^{(c)}= \frac{\pi ^2  t^2}{9}+F_2+\frac{F_2^2}{4  \pi ^2 t^2} +\cO(e^{-2\pi t})
\eea
where $F_2(v)= E_2 - g_2(v)$ as is familiar by now.

\subsubsection{Evaluating $\BZ_3^{(b)}$}

In $\BZ_3^{(b)}$, the contributions of $\nu_{13}^-$ and $\nu_{24}^-$ similarly  integrate to zero.  
Using symmetry again and the fact that terms linear in $f(3)$ and $f(4)$ integrate to zero to replace $(f(3)-f(4))^2$ by $2f(3)^2$. In this way we find,
\bea
\BZ_3^{(b)} =
\frac{1}{16\pi t^3 \tau_2^2} \int_{\Sigma _{ab} ^4}  \nu_{13}^+ \, \nu_{24}^+ \, 
f(3)^2\, \gone(1,2)
\eea
Integrating over point $z_4$ using (\ref{nuint}), we have, 
\bea
\BZ_3^{(b)} =
- \frac{1}{2\pi t^2 \tau_2} \int_{\Sigma _{ab} ^3}  \nu_{13}^+ \, \kappa (2)  \,  f(3)^2\, \gone(1,2)
\eea
The part proportional to $\kappa_1(1)$ in $\nu_{13}^+$ and  the part proportional to $\kappa_1(2)$ in $\kappa(2)$  integrate to zero. Taking these simplifications into account, the measure in point 3 is proportional to $\kappa_1(3)$ and the integral over this point may be performed, simplifying the result to give,
\bea
\BZ_3^{(b)} =
- \frac{F_2}{2\pi t^3} \int_{\Sigma _{ab} ^2}  \om_t (z) \, \oom_t(z) \, \om_t (w) \, \oom_t(w) \,  \gone(z,w)
\eea
To evaluate the integrals, we use \eqref{regs} and \eqref{ompat} to arrive at,
\bea
\BZ_3^{(b)}= \frac{2}{3} F_2 +\frac{2 \gone F_2}{\pi   t}+\frac{F_2^2}{\pi ^2 t^2} +\cO(e^{-2\pi t})
\eea
where $g$ is shorthand for $ g= g(v)$.

\subsubsection{Evaluating $\BZ_3^{(a)}$}

The computation of $\BZ_3^{(a)}$ is slightly more complicated. In contrast with the previous two cases, it is now the contributions from $\nu_{13}^+$ and $\nu_{24}^+$ that  integrate to zero against the  Green functions $\gone(1,2) \gone(3,4)$. The remaining contribution is given as follows, 
 \bea
\BZ_3^{(a)} = 
{ 1 \over 8 t^2 \tau_2^2}  \int _{\Sigma_{ab} ^4} \nu_{13}^- \, \nu_{24}^- \, \gone(1,2) \, \gone(3,4)
\eea
This integral is manifestly convergent when extended to the punctures, so that $\Sigma _{ab}$ may be replaced by $\Sigma _1$, up to exponentially suppressed corrections which we neglect. We carry out the integrals over the points 2 and 4 using the following relation (\ref{gD}) and its complex conjugate, for the special case $n=1$.  The contributions arising from the two terms in $\nu_{24}^-$ are pairwise equal, and we may simplify the result as follows,
 \bea
\BZ_3^{(a)} =  
-  { 1 \over 4 \pi^2 t^2 }  \int _{\Sigma _{ab} ^2} \nu_{zw}^- \,  \pbz f_2(z) \, \p_w f_2(w)
\eea
where $z$ and $w$ respectively stand for the point $z_1$ and $z_3$,  and  $f_n$ was defined in (\ref{fnz}). To evaluate the remaining integrations over the points $1$ and $3$, we use (\ref{gD}), and we find,  
\bea
\BZ_3^{(a)} = 
{ F_2^2 \over  \pi^2 t^2} 
+  { \tau_2^2 \over  \pi^4 t^2} \big  | \p_w^2 f_3(w)  \big |^2 \Big | _{w=p_a} 
\eea
Expressing the result in terms of  $D_{3,1}$ using (\ref{diffD}), we have, 
\bea
\BZ_3^{(a)} = 
{ F_2^2 \over  \pi^2 t^2} 
+  { 1 \over 16 \pi^4 t^2 \tau_2^2} \Big | D_{3,1}(v|\tau) - D_{3,1}(0|\tau) \Big |^2
\eea
Using the differential relation (\ref{A28}) for $a=b=2$, and a suitable rearrangement formula,
\bea
\label{A30}
D_{3,1} (v |\tau) - D_{3,1} (0 |\tau) & = &  4 \pi i \tau_2^2 \p_\tau F_2 (v |\tau)
\no \\
\Delta _\tau F_2^2 - 4 F_2^2 & = & 8 \tau_2^2 |\p_\tau F_2|^2
\eea
we simplify the final expression for $\BZ_3^{(a)}$ as follows, 
\bea
\BZ_3^{(a)} =    { 4 F_2^2 + \Delta _\tau F_2^2  \over 8 \pi^2 t^2 }
\eea
Collecting the three contributions $\BZ_3^{(a,b,c)}$, we arrive at our final formula,
\bea
\label{B3minsep}
\BZ_3 = 
\frac{\pi^2 t^2}{9}+\frac{5 F_2}{3}+\frac{2\gone F_2}{ \pi  t}
 +\frac{\Delta _\tau F_2^2 + 14 F_2^2 }{8\pi^2t^2} 
+\cO(e^{-2\pi t})
\eea

\subsection{Degeneration of $\BZ_2$}
\label{B2sec}

The modular graph function $\BZ_2$ is defined in \eqref{defB1} and corresponds to the L-shape diagram  of Figure~\ref{fig:1}.  Using \eqref{limG}, one  decomposes $\BZ_2 $ into a sum of 3 terms,
\bea
\BZ_2^{(a)} &=& \frac{1}{8t^2 \tau_2^2}\int _{\Sigma_{ab}  ^4}|\Delta(1,3)\Delta(2,4)|^2\, 
\gone(1,2) \, \gone(1,4) 
\no \\
\BZ_2^{(b)} &=& \frac{1}{32\pi t^3 \tau_2^2}\int _{\Sigma_{ab}  ^4}|\Delta(1,3)\Delta(2,4)|^2\, 
\gone(1,2) \, \big ( f(1)-f(4) \big ) ^2 
\no \\
\BZ_2^{(c)} &=& \frac{1}{512 \pi^2 t^4 \tau_2^2}\int _{\Sigma_{ab}  ^4}|\Delta(1,3)\Delta(2,4)|^2\, 
\big ( f(1)-f(2) \big ) ^2 \, \big ( f(1)-f(4) \big )^2 
\eea
We proceed to evaluating these integrals again in order of increasing difficulty.

\subsubsection{Evaluating $\BZ_2^{(c)}$}

The contributions from $\nu_{13}^-$ and $\nu_{24}^-$ vanish upon integrating with respect to points 3 and 4. As a result, and using the symmetries of the integrand, the integral reduces to, 
\be
\BZ_2^{(c)} = 
\frac{1}{8^3\pi^2 t^4 \tau_2^2} \int_{\Sigma_{ab} ^4} \, \nu_{13}^+ \, \nu_{24}^+ \, 
\Big ( f(1)^4+ 2f(1)^2 f(4)^2 + f(2)^2 f(4)^2 \Big )
\ee
The integral over point 3 may be carried out with the help of (\ref{nuint}) while the integral over point 4 can be computed using \eqref{regs}. Performing also the integrals over the remaining points 1 and 2, we find, 
\bea
\BZ_2^{(c)} =\frac{14 \pi ^2 t^2}{45}+\frac{2}{3}F_2+\frac{6F_4+ F_2^2}{4 \pi ^2   t^2} +\cO(e^{-2\pi t})
\eea
where $F_4$ was defined in (\ref{Fn}) and given explicitly in (\ref{F4}).

\subsubsection{Evaluating $\BZ_2^{(b)}$}

The contributions from $\nu_{13}^-$ and $\nu_{24}^-$ similarly integrate to zero in $\BZ_2^{(b)}$. Using \eqref{regs} we find,
\bea
\BZ_2^{(b)} = 
\frac{1}{32\pi t^3 \tau_2^2} \int_{\Sigma_{ab} ^4} \, \nu_{13}^+ \, \nu_{24}^+\, 
\Big (  f(1)^2+ f(4)^2 \Big ) \, \gone(1,2)
\eea
The integral over point 3 may be performed using (\ref{nuint}), while the one over point 4 may be performed using (\ref{regs}), and we find, 
 \bea
\BZ_2^{(b)} = 
{ 2 \over \pi t} \int _{\Sigma _{ab}^2} \kappa (1) \kappa(2) f(1)^2 g(1,2) 
+ { i F_2 \over \pi t^2} \int _{\Sigma _{ab}^2} \kappa (1) \om_t (2) \oom_t(2) \,  g(1,2) 
\eea
The contribution of $\kappa_1$ in $\kappa$ cancels out for point 2 in the first integral and point 1 in the second integral. The remaining integrals may be evaluated using (\ref{ompat}) for both the integrals in points 1 and 2, and we find, 
\bea
\BZ_2^{(b)}  
=
\frac{3 \pi ^2 t^2}{10}
+\frac{2}{3} \pi t \gone 
+\frac{F_2}{2} 
+\frac{D_3-D_3^{(1)}+2 \gone F_2}{2 \pi  t}
 +\frac{3 F_2^2 -5 F_4}{4 \pi ^2  t^2}
 +\cO(e^{-2\pi t})
\eea
The function $D_3^{(1)}$ was defined in (\ref{Dlk}), while $F_2$ and $F_4$ were given  in (\ref{Fn}).

\subsubsection{Evaluating $\BZ_2^{(a)}$}

The contribution from this integral is simplified by integrating over the point 3, and we have, 
 \bea
 \BZ_2 ^{(a)}  =   { i \over 4 t^2 \tau_2^2} 
\int _{\Sigma _{ab}^3} \Big ( t \, |\om_1 (1)|^2 + \tau_2 \, |\om_t (1) |^2 \Big )  
\Big ( \nu_{24}^- - \nu_{24}^+ \Big )  \gone(1,2) \gone(1,4) 
\eea
We begin by carrying out the integrals over points 2 and 4. The terms proportional to $|\om_1 (2)|^2$ and $|\om_1(4)|^2$ in $\nu_{24}^+$ integrate to zero in view of the normalization of $g$, leaving only the contribution from $\nu_{24}^-$. To evaluate the integrals over the points 2 and 4, we use the relation derived earlier in (\ref{gD}) for $n=1$, and its complex conjugate, and we find,
  \bea
 \BZ_2 ^{(a)}   =   - { i \over  2\pi^2  t } 
\int _{\Sep  } \om_1 (w) \oom_1 (w)  | \p_w f_2(w) |^2 
-{ i \tau_2 \over  2\pi^2  t^2 } 
\int _{\Sep  }  \om_t (w) \overline{\om_t(w)}  | \p_w f_2 (w)  |^2
\eea
where $w=z_1$ represents the remaining integration point 1. The first integral is evaluated by integrating by parts in $w$, and using the following Laplacian relation,
\bea
\label{dg2}
\tau_2 \pbw \p_w f_2 (w)   = -  \pi  f(w)
\eea
The remaining integral is carried out using the definitions of $E_3$ and $g_3$, and we find,
  \bea
 \BZ_2 ^{(a)}  =   -{ 2 \over   \pi  t  } 
 \Big ( E_3(\tau)  - g_3 (v|\tau) \Big )
 -{ i  \tau_2 \over  2 \pi^2  t^2 }  \int _{\Sep }  \om_t (w) \, \oom_t(w) \, | \p_w f_2 (w)  |^2
\eea
To evaluate the remaining integral we use (\ref{fint}) for $n=0$ and $\psi (w) = | \p_w f_2 (w)  |^2$. Since $\psi$  is  regular at the punctures, we may use the simplified formula (\ref{fgint}).  We evaluate $ | \p_w f_2 (w)  |^2$ at the punctures in terms of the function $D_{2,1}$, by using the first line of (\ref{diffD}).
Taking into account the fact that $D_{2,1}(0|\tau)=0$, we find, 
\bea
\label{B30}
 \int _{\Sep }  \om_t (w) \, \oom_t(w) \, | \p_w f_2 (w)  |^2
 = - 2 it  |\p_v g_2 (v)|^2 -{ i \tau_2 \over 4 \pi^2} \int_\Sep  \kappa _1 (w) f(w)^2 \p_w \pbw |\p_w f_2|^2
\eea
The Laplacian of $\psi$ may be simplified with the help of (\ref{dg2}) and is given by,
\bea
\p_w \pbw \psi(w)  =   
{ \pi^2 \over \tau_2^2} f(w)^2
+  |  \p_w^2 f_2(w)  |^2
- {\pi \over \tau_2}  \p_w  f_2(w) \, \pbw f(w)
- {\pi \over \tau_2}  \pbw  f_2(w)\,  \p_w f(w)
\eea
Integrating by parts in the last two terms above so as to regroup in terms of $\p_w \pbw f_2(w)$ and then using (\ref{dg2}), the integral in (\ref{B30}) takes the following form, 
\bea
\int_\Sep  \kappa _1 (w) f(w)^2 \p_w \pbw |\p_w f_2|^2
 = { 8 \pi^2 \over \tau_2^2} F_4 + \int_\Sep  \kappa _1 (w) f(w)^2  |\p_w^2 f_2|^2
 \eea
Combining the derivative relations in $\p_w^2 g_2$ and $\p_\tau g$ in (\ref{diffD}), we obtain,
\bea
\p_w^2 f_2(w) = 2 \pi i \p_\tau f(w)
\eea
Using furthermore the relation $\p_{\bar \tau } \p_\tau f(w)=0$, we may rearrange the integral as follows, 
\bea
\int_\Sep  \kappa _1 (w) f(w)^2 \p_w \pbw |\p_w f_2|^2
 = { 8 \pi^2 \over \tau_2^2} F_4 +{ 2 \pi^2 \over \tau_2^2} \Delta _\tau F_4
 \eea
Using also the relation,
\bea
\label{pvg2}
8\tau_2 |\p_v g_2(v)|^2 = \Delta _v F_2(v)^2 -  8\pi  g(v) F_2(v)
\eea
we obtain the following  expression, 
\bea
\BZ_2^{(a)}=\frac{1}{\pi t}(2g_3-2E_3+ g F_2 )-{ \Delta _v F_2^2  \over 8 \pi^2 t } 
- \frac{(\Delta _\tau+4) F_4 }{4\pi^2 t^2} +\cO(e^{-2\pi t})
\eea
Collecting the three contributions $\BZ_2^{(a,b,c)}$, we arrive at our final result,
\bea
\label{B2minsep}
\BZ_2 &=&
\frac{11 \pi^2 t^2}{18}
+\frac{2 \gone \pi t}{3}
+\frac{7 F_2 }{6}
+\frac{1}{2 \pi t} \left (D_3-D_3^{(1)}+4 g F_2+4 g_3 -4   E_3 - { \Delta _v F_2^2 \over 4 \pi} \right )
\nn \\
   &&+\frac{4 F_2^2 -(\Delta _\tau+3) F_4 }{4 \pi ^2 t^2}+\cO(e^{-2\pi t})
\eea

\subsection{Degeneration of $\BZ_1$}
\label{B1sec}

The modular graph function $\BZ_1$ was defined in \eqref{defB1} and corresponds to the one-loop graph on the left of Figure~\ref{fig:1}.  Using \eqref{limG}, one finds that $\BZ_1 $ decomposes into a sum of 3 terms,
\bea
\label{B20aa}
\BZ_1^{(a)} &=&8 \int _{\Sigma _{ab} ^2} \, \kappa(1) \, \kappa(2) \, \gone(1,2)^2  
\no \\
\BZ_1^{(b)} &=& \frac{2}{\pi t} \int _{\Sigma_{ab} ^2} \kappa(1)\, \kappa(2)  \, \gone(1,2)  \,  \big( f(1) - f(2) \big )^2 
\no \\
\BZ_1^{(c)} &=& \frac{1}{8\pi^2 t^2} \int _{\Sigma_{ab} ^2} \kappa(1)\, \kappa(2)  \, \big ( f(1) - f(2) \big )^4 
\eea
The contributions $\BZ_1^{(c)}$ and $\BZ_1^{(b)}$ are routine, but $\BZ_1^{(a)}$ will involve some rather serious analysis.

\subsubsection{Evaluating $\BZ_1^{(c)}$}

Substituting $\kappa$ by its expression on $\Sigma _{ab}$, given in \eqref{limkappa}, the integrals in \eqref{B20aa} can be evaluated  successively using \eqref{regs}, and we find,
\bea
\BZ_1^{(c)} = \frac{11\pi^2 t^2}{15}+F_2+\frac{12 F_4 +3F_2^2}{4\pi^2t^2} +\cO(e^{-2\pi t})
\eea

\subsubsection{Evaluating $\BZ_1^{(b)}$}

Using symmetry under the exchange of the points 1 and 2,  $\BZ_1^{(b)}$ may be decomposed into 
a sum of two terms, 
\bea
\label{B20ab1}
\BZ_1^{(b,1)} &=& \frac{4}{\pi t} \int _{\Sigma_{ab} ^2} \kappa(1) \, \kappa(2) \gone(1,2) \, 
f(1)^2  \\
\BZ_1^{(b,2)} &=& - \frac{4}{\pi t} \int _{\Sigma_{ab} ^2} \kappa(1)\, \kappa(2)  \,  \gone(1,2) \, 
f(1)f(2)
\eea
The integral over point 2 in $\BZ_1^{(b,1)}$ may be computed using \eqref{ompat}, 
\bea
\label{B20ab2}
\BZ_1^{(b,1)}  =
\frac1{2\pi t^2} \int_{\Sep } \kappa (z) f(z)^2 \left (  F_2(v) - \frac12 f(z)^2 + 
 {t \over \pi} \int _0 ^{2\pi} d \theta \Big ( \gone(z,p_a^\theta )+  \gone(z,p_b^\theta) \Big )
\right) 
\eea
where the quantities $p_a^\theta$ and $p_b^\theta$ were introduced in (\ref{defT}). The remaining integrals in $z$ may be computed using (\ref{ompat}) again, and we find, 
\bea
\label{b1b1}
\BZ_1^{(b,1)}  =
\frac{3\pi^2 t^2}{5} +\frac{4\pi t}{3} \gone + \frac{F_2}{3} +\frac{D_3-D_3^{(1)}}{\pi t}
+\frac{F_2^2 - 5F_4}{2\pi^2 t^2}
\eea
The second integral $\BZ_1^{(b,2)}$ is a sum of three terms obtained by decomposing each  $\kappa$ into its $\kappa_1$ and its $\om_t \wedge \oom_t$ parts. The resulting integrals may be carried out using (\ref{ompat}), 
\bea
\BZ_1^{(b,2)} = -   {8  \over 15}  \pi^2 t^2  -2 F_2 + { 2 \over  \pi t }  (g_3- E_3) + { 2 F_4 \over  \pi^2 t^2 } 
\eea
Combining these results with the contribution \eqref{b1b1} from $\BZ_{1}^{(b,2)}$, we find,
\be
\BZ_1^{(b)} = \frac{\pi^2 t^2}{15}
+\frac{4\pi t}{3} \gone -\frac{5}{3}F_2
+ \frac{D_3-D_3^{(1)}-2E_3+2g_3}{\pi t} 
+ \frac{F_2^2- F_4}{2\pi^2 t^2} 
\ee

\subsubsection{Evaluating $\BZ_1^{(a)}$}

We finally come to the most difficult part of the computation, namely the evaluation of $\BZ_1^{(a)}$.
Substituting the volume form $\kappa$ by \eqref{limk}, we decompose $Z_1^{(a)}$ as follows, 
\bea
\label{defZ1a}
\BZ_1 ^{(a)}  =  
2 \int _{\Sigma_{ab}  ^2} \kappa_1(1) \, \gone(1,2)^2 \left ( \kappa_1(2) \, +{ i \over t}  \om_t (2) \, \overline{\om_t(2)} \right )
 + { \cK \over 8 \pi^2 t^2} 
\eea
where $\cK$ is given by, 
\bea
\label{Kdef}
 \cK = 
-  4 \pi^2 \int _{\Sigma_{ab}  ^2} \om_t (1) \overline{\om_t(1)} \om_t (2) \overline{\om_t(2)} \, \gone(1,2)^2
 \eea
The factor of $8 \pi^2 t^2$ has been extracted for later convenience. Carrying out the integral over point 1 in the  integral in the first term of (\ref{defZ1a}) gives a result that is independent of point 2, namely $E_2$. Performing the remaining integral, we find,  
 \bea
\BZ_1 ^{(a)}= 6 E_2  + { \cK \over 8 \pi^2 t^2} 
 \eea
The degeneration of $\cK$ is  complicated, but the calculation of its variation in $t$ up to exponentially suppressed corrections may be obtained  using the variational method developed in Section 3.6 of \cite{DHoker:2017pvk} and is relatively simple. Therefore, we shall split the function $\cK$ into a sum of two contributions, 
\bea
\cK = \cK^c + \cK^t + \cO(e^{-2\pi t}) 
\eea
where $\cK^c$ is independent of $t$, and $\cK^t$ is a polynomial in $t$ of degree four, with vanishing constant term. The variational method will allow us to compute $\cK^t$ completely in the next subsection but does not give us access to $\cK^c$, which will have to be computed by other methods in the subsequent subsection.
The result for $\cK^t$ will be found to be, 
\bea
\label{KK0}
{\cK^t  \over 8 \pi^2 t^2} = 
\frac{2\pi^2 t^2}{3} + \frac{4\pi t}{3} \gone +2 \gone^2
+{ 1 \over \pi t} \left ( D_3 - D_3^{(1)} + 2 \zeta (3) + { \Delta _v F_4 \over 2 \pi} \right )
\eea
Collecting the three contributions $\BZ_1^{(a)}, Z_1^{(b)}, Z_1 ^{(c)}$, we arrive at the final formula, 
\bea
\label{B1minsep}
\BZ_1&=& 
\frac{22 \pi^2 t^2}{15} 
+ \frac{8 \pi t}{3} \gone 
+  6E_2-{2 \over 3} F_2+2\gone^2
+ { 2 \over \pi t} \left ( g_3 - D_3 ^{(1)} + 2 \zeta (3) + {\Delta _v F_4  \over 2 \pi}\right)
 \no \\
&& + {\cK^c + 20 F_4 + 10 F_2^2 \over 8 \pi^2 t^2}  +\cO(e^{-2\pi t})
\eea
where $\cK^c$ is independent of $t$. Next, we proceed to the calculation of $\cK^t$ using the variational method the next subsection and then of the constant contribution $\cK^c$  in  subsection \ref{sec:Kfull}.

\subsection{Variational calculation of $\cK^t$ \label{sec_varia}}

In this subsection, we shall calculate the variation $\delta  \cK^t= \delta \cK$ under an infinitesimal variation $\delta t$  holding all other moduli fixed.\footnote{Throughout, we shall neglect all contributions which are exponentially suppressed  in~$t$.}   We begin by recasting the defining formula (\ref{Kdef}) for $\cK$ in the following form,
 \bea
 \cK = {\tau_2^2 \over \pi^2} \int_\Sep  \kappa_1(w) \int _\Sep \kappa_1(z) \,  |\p_z f(z)|^2 \,  |\p_w f(w)|^2 \, \gone(w,z)^2
 \eea
The integrand is independent of $t$, so that all $t$-dependence arises from the dependence on $t$ of the  integration domain, $\Sep = \{ z \in \Sigma _1, \, |f(z)| \leq 2 \pi t \}$. As  a result,  $\delta \cK$  is given entirely by  the effects of varying  the integration regions for  both $z$ and $w$ (which contribute equally) with $t$, and we have,
\bea
\delta \cK = {2 \tau_2^2 \over \pi^2}  \int _\Sep \kappa_1 ( z) |\p_z f|^2 \left [ \int _{\delta \mD_a \cup \, \delta \mD_b} \kappa_1(w) |\p_w f|^2 g(z,w)^2 \right ]
\eea
The infinitesimal  integration domains $\delta \mD_a, \delta \mD_b$ are defined as follows,
\bea
\delta \mD_a & = & \{ w \in \Sigma _1,   \, - 2 \pi (t+\delta t) \leq f(w) \leq - 2 \pi t \}
\no \\
\delta \mD_b & = & \{ w \in \Sigma _1,  \,  + 2 \pi t \leq f(w) \leq 2 \pi (t+\delta t) \}
\eea
The $w$-integrals may be simplified as follows. We begin with the contribution from $\delta \mD_b$, the one from $\delta \mD_a$ being analogous. We parametrize $w$ in $\delta \mD_b$ as follows, 
\bea
2 \pi t \leq g(w,p_b) - g(w,p_a) \leq 2 \pi (t + \delta t)
\eea
Up to exponential corrections, which we neglect, $g(w,p_a)$ equals $g(p_b,p_a) = g(v)$ for $ w \in \delta \mD_b$. Furthermore, the Green function in the funnel is given by (\ref{deflambda}), 
\bea
g(w,p_b) = - \ln |w-p_b|^2 - \lambda + \cO(w-p_b)
\eea
where $\lambda$ was defined as well. In terms of the variable $R$ introduced in (\ref{defT}), the domain 
$\delta \mD_b$ consists of the points $w$ restricted by,
\bea
R \, e^{ - \pi \delta t} \leq |w-p_b| \leq R
\eea
and may be parametrized  by two real coordinates $x,y$,
\bea
w= p_b + R \, e^{-x-i \theta }  \hskip 1in 
0 \leq x \leq \pi \delta t 
\hskip 0.5in  0 \leq \theta \leq 2 \pi
\eea
With this parametrization, the  integral in $x$ may be evaluated at $x=0$, so that we find the simplified formulas, 
\bea
\tau_2 \int _{\delta \mD_{a,b}} \kappa_1 (w) |\p_w f|^2 g(z,w)^2
 = 
 \pi \delta t  \int _0 ^{2 \pi} d \theta \, g  (z, p_{a,b}^\theta  )^2
\eea
To evaluate the $z$-integrals, we split up the calculation of $\delta \cK$ into three parts,
\bea
\delta \cK = \delta \cK_{(m)} + \delta \cK_{(a)} + \delta \cK_{(b)}
\eea
where
\bea
\delta \cK_{(m)} & = & 4   \tau_2 \,  \delta t \int _\Sep \kappa_1( z) |\p_z f|^2 \Big ( g(z,p_a)^2 + g(z,p_b)^2 \Big )
\no \\
\delta \cK_{(a,b)} & = & {2 \tau_2\over \pi} \, \delta t \int _\Sep \kappa_1( z) |\p_z f|^2 \int _0 ^{2 \pi}  d \theta 
\Big ( g (z, p_{a,b}^\theta   )^2 - g(z,p_{a,b})^2 \Big )
\eea
The purpose of this rearrangement is to simplify the integrand for the most complicated part of the calculation, namely in $\delta \cK_{(m)}$, and be left with $\delta \cK_{(a)}, \delta \cK_{(b)}$ which receive contributions only from the funnel parts. We shall now evaluate each part in turn.

\begin{figure}
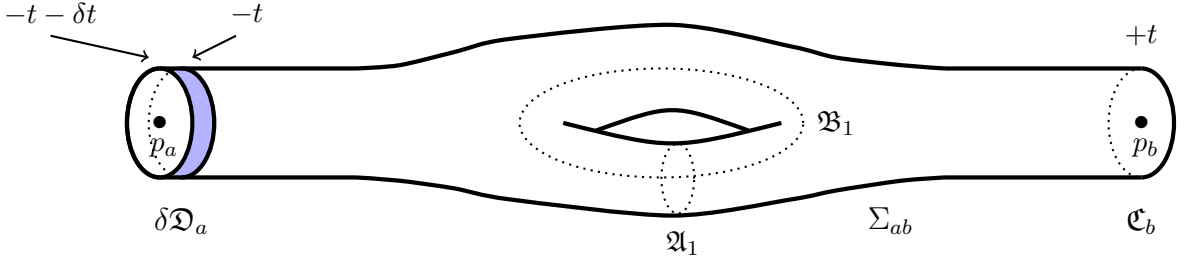

\begin{center}
\tikzpicture[scale=1.45]
\scope[xshift=-5cm,yshift=0cm]
\draw [ultra thick, fill, color=blue!30] (2,-1) arc (-90:270:0.3 and 0.5);
\draw [ultra thick, fill, color=white] (1.8,-1) arc (-90:270:0.3 and 0.5);
\draw [ultra thick] (2,-1) arc (-90:90:0.3 and 0.5);
\draw [ultra thick] (1.8,-1) arc (-90:90:0.3 and 0.5);
\draw [ultra thick] (1.8,0) arc (90:270:0.3 and 0.5);
\draw [thick, dotted] (2,0) arc (90:270:0.3 and 0.5);
\draw [ultra thick] (1.8,0) arc (90:270:0.3 and 0.5);

\draw [ultra thick] (1.8,0) -- (3.6,0);
\draw [ultra thick] (1.8,-1) -- (3.6,-1);
\draw[ultra thick] (5.5,-0.5) .. controls (6.5, -0.75) .. (7.5,-0.5);
\draw[ultra thick] (5.8,-0.57) .. controls (6.5, -0.32) .. (7.2,-0.57);
\draw [ultra thick] (10.8,-1) arc (-90:90:0.3 and 0.5);
\draw [thick, dotted] (10.8,0) arc (90:270:0.3 and 0.5);
\draw [ultra thick] (9,0) -- (10.8,0);
\draw [ultra thick] (9,-1) -- (10.8,-1);
\draw[ultra thick] plot [smooth] coordinates {(3.6,0) (4,0.03) (4.5, 0.15) (5, 0.27) (6.5,0.4)  (7.6, 0.2) (8, 0.10) (8.6,0.03) (9,0)};
\draw[ultra thick] plot [smooth] coordinates {(3.6,-1) (4,-1.03) (4.5, -1.10)  (5, -1.2) (6.5,-1.35) (7.6, -1.2) (8, -1.12) (8.6,-1.03) (9,-1)};
\draw [thick, dotted] (7.7,-0.5) arc (0:360:1.3 and 0.5);
\draw [thick, dotted] (6.7,-1.01) arc (0:360:0.15 and 0.32);
\draw (2,-1.4) node{$\delta \mD_a$};
\draw (0.8,0.5) node{\small $-t-\delta t$};
\draw[->, thick] (0.8,0.3) -- (1.7, 0.1);
\draw (2.6,0.5) node{\small $-t$};
\draw[->, thick] (2.5,0.3) -- (2.1, 0.1);
\draw (10.8,-1.4) node{$\mC_b$};
\draw (10.8,0.3) node{\small $+t$};
\draw (1.8,-0.5) node{$\bullet$};
\draw (10.8,-0.5) node{$\bullet$};
\draw (1.84,-0.73) node{{\small $p_a$}};
\draw (10.84,-0.73) node{\small $p_b$};
\draw (8,-0.5) node{{\small $\mB_1$}};
\draw (6.6,-1.6) node{{\small $\mA_1$}};
\draw (8.5,-1.4) node{{$\Sep$}};
\endscope
\endtikzpicture
\caption{The variational method evaluates the contribution from varying the boundary cycles of  $\Sep$ through a variation of $t$, here represented for the variation of the cycle $\mC_a$.}
\end{center}
\label{fig:var}
\end{figure}

\subsubsection{Calculating $\delta \cK_{(m)}$}

We begin by expanding the factor $|\p_z f|^2$, 
\bea
\label{deltaK0}
\delta \cK_{(m)} & = & 
4  \tau_2 \, \delta t \int _\Sep \kappa_1 ( z) |\p_z g(z,p_a)|^2 \Big ( g(z,p_a)^2 + g(z,p_b)^2 \Big )
\\ && 
+ 4  \tau_2 \, \delta t \int _\Sep \kappa _1 ( z) |\p_z g(z,p_b)|^2 \Big ( g(z,p_a)^2 + g(z,p_b)^2 \Big )
\no \\ &&
- 4  \tau_2 \, \delta t \int _\Sep \kappa_1 ( z) \, \pbz g(z,p_a) \p_z g(z,p_b) \,  
\Big ( g(z,p_a)^2 + g(z,p_b)^2 \Big ) + \hbox{c.c}
\no 
\eea
where addition of the complex conjugate applies only to the last line. To evaluate the first two lines, we use the integral (\ref{G1}) in terms of the parameter $T$ introduced in (\ref{mt}). Using also the integrals of (\ref{G25}), and putting all together,  we have,  
\bea
\delta \cK_{(m)} =
4  \delta T \left ( { 1 \over 3} T^3 +  T \, g(v)^2 + { 1\over 3} D_3
- { 2 \over 3} g(v)^3   -  D_3 ^{(1)}  -  2  D_4 ^{(a)}  \right )
\eea
Integrating the above equation, we obtain,
\bea
\cK_{(m)} =
 { T^4  \over 3} + 2 T^2 g(v)^2 + { 4  \over 3} T \, D_3 - {8  \over 3} T g(v)^3
  - 4  T \, D_3 ^{(1)} - 8  T D_4 ^{(a)}  + \cO(t^0)
\eea

\subsubsection{Calculation of $\delta \cK_{(a)}$ and $\delta \cK_{(b)}$}

The bulk contributions to $\cK_{(a)}$ and $\cK_{(b)}$ are exponentially suppressed, as is manifest by Taylor expanding the Green function. To evaluate the contributions from the funnel,  we may approximate all functions by their form strictly in the funnel, and extend their functional form arbitrarily beyond the funnel. 
 
 \sm
 
 To evaluate $\cK_{(b)}$, we extend the range of $z$ near $p_b$ by  requiring only the condition $f(z) \leq 2 \pi t$ and dropping the lower minimum condition on $f(z)$. Furthermore, we use the approximations suitable for the funnel for the Green functions $g(z,w)$ and $g(w,p_b)$,
\bea
g(z,w) & = & - \ln |z-w|^2 - \lambda 
\no \\
g(w,p_b) & = & - \ln |w-p_b|^2 - \lambda 
\eea
Within this approximation, the integration domain for $z$ then becomes $R \leq |z-p_b| $, 
and may be parametrized by two real coordinates $\alpha, \beta$,
\bea
z= p_b + R \, e^{ \alpha + i \theta}
 \hskip 1in  
 0 \leq \alpha 
\hskip 0.5in
0 \leq \theta \leq 2 \pi
\eea
The integral over $y$ combines with the integral over $\theta$, and we find after some simplifications, 
\bea
\delta \cK_{(b)} = 4  \delta t \int _0 ^\infty d \alpha \int _0 ^{2 \pi } d \theta 
\left ( \Big ( 2 \alpha - T +  \ln | 1 - e^{-\alpha - i \theta} |^2 \Big )^2 -  ( 2 \alpha - T )^2 \right )
\eea
Using the vanishing  for $\alpha >0$ of the  integral over $\theta$ of a single power of $\ln | 1 - e^{-\alpha - i \theta} |^2$,  we are left with performing the integral of the square of the logarithm, which may be done by Taylor expanding each logarithm, performing the integrals over $\theta$ and then performing the integrals over $\alpha$. The result is given by  $\delta \cK_{(b)}  = 8 \pi \zeta (3) \, \delta t $. Upon integration in $T$, we find,
\bea
\cK_{(a)}+ \cK_{(b)} = 8 \zeta (3) \, T + \cO(t^0)
\eea
Putting all of this together, we find, 
\bea
\label{Kvar}
\cK =
 { T^4 \over 3} + 2  T^2 g(v)^2 + { 4 \over 3} T D_3 - {8 \over 3} T g(v)^3  + 8 \zeta(3) T
  - 4  T D_3 ^{(1)}- 8  T D_4 ^{(a)}  + \cO(t^0)
\eea
Using the definition of $T= 2 \pi t + g(v)$, and retaining only the $t$-dependent terms in the above formula for $\cK$ we readily obtain (\ref{KK0}).

\subsection{Calculation of $\cK^c$}
\label{sec:Kfull}

Having calculated the non-constant  $t$-dependence $\cK^t$ of $\cK$ in the preceding subsection, we define $\cK^c$ by the following limit,
\bea
\cK^c  = \lim _{t \to \infty} \left ( \cK  -  \cK^t  \right )
\quad
\eea
We shall compute all contributions to $\cK$, cancel the ones with non-constant dependence on~$t$, and then extract the remainder in the limit. Although this procedure duplicates the variational method to some extent, the confirmation of the validity of the $t$-dependent terms will be of value in this rather tricky calculation. 

\subsubsection{Partitioning the integral}

We recall the starting formula for $\cK$, with the function $f$ expanded into its two contributions, 
\bea
\label{K1}
\cK = {\tau_2^2  \over \pi^2} 
\int _\Sep \!\!\!\!  \kappa_1 ( z)  \int _\Sep \!\!\!\! \kappa_1 (w)   \Big |\p_z g(z,p_a) - \p_z g(z,p_b) \Big |^2 
 \Big |\p_w g(w,p_a) - \p_w g(w,p_b) \Big |^2 g(z,w)^2
 \quad
\eea
We decompose $\cK$ into a sum over sixteen contributions,
\bea
\cK = \sum _{\alpha, \bar \alpha  , \beta , \bar \beta \in\{ a,b \} } (-)^{\#(a)} \cK _{\alpha \bar \alpha \beta \bar \beta}
\eea
obtained by expanding both absolute-value-squared factors in (\ref{K1}) into the following basis,
\bea
\cK _{\alpha \bar \alpha \beta \bar \beta}= { \tau_2^2 \over \pi^2}  
\int _\Sep \!\!\!\!  \kappa _1 ( z)  \int _\Sep \!\!\!\! \kappa_1 (w) \, 
 \p_z g(z,p_\alpha) \, \pbz g(z,p_{\bar \alpha}) \,  \p_w g(w,p_\beta) \pbw g(w,p_{\bar \beta}) \,  g(z,w)^2
\eea
Here, $\#(a)$ is the number of $a$-labels amongst $\alpha, \bar \alpha, \beta, \bar \beta$ (which equals the number of $b$-labels modulo two). Swapping $z$ with $w$ and complex conjugating produce the following relations,
\bea
\cK _{\alpha \bar \alpha \beta \bar \beta} = \cK _{\beta \bar \beta \alpha \bar \alpha }
= \left (  \cK _{\bar \alpha  \alpha  \bar \beta \beta } \right )^* 
\eea
Taking also into account the symmetry under swapping $p_a$ and $p_b$, the sum over 16 terms reduces to a sum over 5 irreducible terms,
\bea
\cK = 2 \cK_{abab}+ 2 \cK_{abba} + 2 \cK_{aabb} - 4 \cK_{aaab}- 4 \cK_{aaab}^*   + 2 \cK_{aaaa}  
\eea
Two of these integrals are finite in the limit $t \to \infty$, and may thus be extended to finite integrals over the compact torus $\Sigma_1$, up to exponential corrections which we neglect,
\bea
\cK _{abab} & = & {\tau_2^2 \over \pi^2}  
\int _{\Sigma_1} \! \kappa_1 ( z)  \int _{\Sigma_1} \! \kappa_1 (w)  \,
 \p_z g(z,p_a) \pbz g(z,p_b) \,  g(z,w)^2 \, \p_w g(w,p_a) \pbw g(w,p_b)  
\no \\
\cK _{abba} & = & {\tau_2^2 \over \pi^2}  
\int _{\Sigma_1} \! \kappa _1 ( z)  \int _{\Sigma_1} \! \kappa_1 (w) \,  
\p_z g(z,p_a) \pbz g(z,p_b)  \,  g(z,w)^2 \, \p_w g(w,p_b) \pbw g(w,p_a)  
\quad 
\eea
They are both three-loop Feynman diagrams represented in Figure~\ref{fig:6} on page \pageref{fig:6}.
The remaining integrals do have non-trivial polynomial $t$-dependence. 

\sm

To express the remaining contributions, $\cK_{aabb}$, $\cK_{aaab}$ and $\cK_{aaaa}$, in terms of a polynomial in $t$ whose coefficients are convergent integrals over the compact surface $\Sigma _1$, we proceed as follows. We split the integrals into a part $\cK^0$ which is given by a convergent integral as $t \to \infty$, and a part $\cK^1$ which has non-trivial polynomial $t$ dependence and which is easier to evaluate than the original integral,
\bea
\cK_{aabb} & = & \cK_{aabb}^0 + \cK_{aabb}^1
\no \\
\cK_{aaab} & = & \cK_{aaab}^0 + \cK_{aaab}^1
\no \\
\cK_{aaaa} & = & \cK_{aaaa}^0 + \cK_{aaaa}^1
\eea
We shall begin by discussing the first two functions above, and then proceed to the most intricate case of the last function.

\subsubsection{Decomposing $\cK_{aabb}$ and $\cK_{aaab}$}

The functions $\cK_{aabb}$ and $\cK_{aaab}$ are schematically represented by three-loop Feynman diagrams in Figures \ref{fig:7} and \ref{fig:8} respectively.  The functions $\cK^0_{aabb}$ and $\cK^0_{aaab}$ are defined by, 
\bea
\label{Kzero}
\cK _{aaab} ^0 & = & 
{\tau_2^2 \over \pi^2}  \int _\Sep \! \kappa_1 ( z)  \int _\Sep \! \kappa_1 (w)   |\p_z g(z,p_a) |^2  
 \p_w g(w,p_a) \pbw g(w,p_b) \Big ( g(z,w)^2 - g(p_a,w)^2 \Big ) 
\no \\
\cK _{aabb} ^0 & = & 
{\tau_2^2 \over \pi^2} \int _\Sep \!\!\!\! \kappa_1 ( z)  \int _\Sep \!\!\!\! \kappa_1 (w) \,  |\p_z g(z,p_a) |^2  
| \p_w g(w,p_b) |^2 \, \Big ( g(z,w)^2 - g(p_a,w)^2
\no \\ && \hskip 3.2in 
 -g(z,p_b)^2 + g(p_a, p_b)^2 \Big )    
\eea
They do have finite limits as $t \to \infty$, so that the integration domains may be smoothly extended from $\Sep$ to $\Sigma _1$, and the resulting integrals evaluate to generalized modular graph functions.  The finiteness of the integrals in $\cK_{aaab}^0$ and $\cK_{aabb}^0$ as $ t \to \infty$ may be proven conveniently by using the variational method, but we shall not present these proofs here. 

\sm

The functions  $\cK^1_{aabb}$ and $\cK^1_{aaab}$ are defined by,
\bea
\label{Kone}
 \cK _{aabb} ^1 & = & 
{\tau_2^2 \over \pi^2}   \int _\Sep \!\!\!\! \kappa_1 ( z)  \int _\Sep \!\!\!\! \kappa_1 (w)   |\p_z g(z,p_a) |^2  
| \p_w g(w,p_b) |^2  
 \Big (  g(p_a,w)^2 + g(z,p_b)^2 - g(p_a, p_b)^2 \Big ) 
\no \\
\cK _{aaab} ^1 & = & 
{\tau_2^2 \over \pi^2}  \int _\Sep \!\!\!\! \kappa _1( z)  \int _\Sep \!\!\!\! \kappa_1 (w) \,   |\p_z g(z,p_a) |^2  
 \p_w g(w,p_a) \pbw g(w,p_b) \, g(p_a,w)^2 
\eea
 The integrals  $\cK_{aabb}^1$ and $\cK_{aaab}^1$ do have  non-trivial dependence on $t$ which may, however, be easily evaluated.  To compute $\cK_{aaab}^1$ we note that its integral over $z$ may be performed using (\ref{G1}), while to compute $\cK_{aabb}^1$ we use the fact that the $z$-integral may be similarly computed for the first and third term in the parentheses of the integrand, while for the second term it is the $w$-integral that may be readily computed. The results are as follows,
\bea
\label{Kone1}
 \cK _{aabb} ^1 & = & 
{\tau_2 T \over \pi}   \int _\Sep \!\!\!\! \kappa_1 (w)   
| \p_w g(w,p_b) |^2    \Big ( 2 g(w,p_a)^2   -  g(p_a,p_b)^2 \Big ) 
\no \\
\cK _{aaab} ^1 & = & 
{\tau_2 T \over \pi}  \int _\Sep \!\!\!\! \kappa_1 (w) \,  
 \p_w g(w,p_a) \pbw g(w,p_b) \, g(w,p_a)^2 
\eea
The remaining $w$-integrals are readily evaluated, and we obtain, 
\bea
\cK^1_{aabb} & = &   T^2 g(v)^2 - 2 T D_3^{(1)} - 4 T D_4^{(a)} + \cO(e^{- 2 \pi t})
\no \\
\cK _{aaab} ^1  & = &  { T \over 3} (g(v)^3 - D_3) + \cO(e^{- 2 \pi t})
\eea

\subsubsection{Decomposing $\cK_{aaaa}$}

The remaining integral $\cK_{aaaa}$ is given by the four-loop Feynman diagram in Figure \ref{fig:8},
\bea
\cK_{aaaa}  =  
{\tau_2^2 \over \pi^2}  \int _\Sep \! \kappa _1 ( z)  \int _\Sep \! \kappa_1 (w) \,   |\p_z g(z,p_a) |^2  
| \p_w g(w,p_a) |^2 \,  g(z,w)^2 
\eea
We rearrange $\cK_{aaaa}$ as the sum of three integrals,
\bea
\cK_{aaaa} = K_1+K_2+K_3
\eea
where each part is defined as follows,
\be
\begin{split}
K_1  = & 
{\tau_2^2 \over \pi^2} \!\!\! \int _\Sep \!\!\!\! \kappa_1( z)  \!\!\! \int _\Sep \!\!\!\! \kappa_1(w) 
|\p_z g(z,p_a) |^2   | \p_w g(w,p_a) |^2  
\Big ( g(z,w) - g(z,p_a) \Big )\Big ( g(z,w) - g(p_a,w) \Big )
 \\
K_2  = & 
{2 \tau_2^2  \over \pi^2}  \int _\Sep \!\!\!\! \kappa_1( z)  \int _\Sep \!\!\!\! \kappa_1(w) 
 |\p_z g(z,p_a) |^2   | \p_w g(w,p_a) |^2   g(z,w)  g(z,p_a) 
 \\
K_3  = & 
- {\tau_2^2 \over \pi^2}   \int _\Sep \!\!\!\!  \kappa_1 ( z)  \int _\Sep \!\!\!\! \kappa_1 (w)  
|\p_z g(z,p_a) |^2   | \p_w g(w,p_a) |^2    g(z,p_a)  g(w,p_a) 
\end{split}
\ee
The purpose of the rearrangement is to expose the last two factors in the integrand of $K_1$ which vanish at $z=p_a$ and $w=p_a$, and decrease the orders of the poles at these points. For fixed $z$ away from $p_a$, the $w$-integral is absolutely convergent. However, the multiple integration over both $z,w$ will not, in fact, be convergent yet. In quantum field theory language, $K_1$ has no sub-divergences, but it has a primitive divergence, with which we shall deal shortly. The remaining integrals $K_2$ and $K_3$ are simpler and can be evaluated exactly. It is straightforward to evaluate $K_3$ using (\ref{G1}) and we obtain, 
\bea
K_3 = - {1 \over 4} (T^2 - E_2)^2 + \cO(e^{- 2 \pi t})
\eea
To evaluate $K_2$, we use a formula analogous to (\ref{ompat}) to carry out the integral over $w$, 
\bea
{2\tau_2 \over \pi} \int _\Sep \kappa _1 (w) |\p_w g(w,p_a)|^2 g(z,w)
= 
E_2 - 2g_2(z,p_a) - g(z,p_a)^2 + { T \over  \pi} \int d \theta g(z,p^\theta _a)
\eea
where $p_a^\theta$ was defined in (\ref{defT}).  We find, 
\bea
K_2 & = & 
{\tau_2 \over \pi}   \int _\Sep \! \kappa_1 ( z)  |\p_z g(z,p_a) |^2 g(z,p_a) \left ( -E_2   - g(z,p_a)^2 + 2 T g(z,p_a) \right )
\nn \\ && 
+ {2 \tau_2 \over \pi}   \int _\Sep \! \kappa_1 ( z)  |\p_z g(z,p_a) |^2 g(z,p_a) \left ( E_2 - g_2(z,p_a)   \right )
\\ && 
+ { \tau_2 T \over \pi^2}   \int _\Sep \! \kappa_1 ( z)  |\p_z g(z,p_a) |^2 g(z,p_a) \int ^{2 \pi } _0 d \theta  \Big ( g(z,p_a^\theta) - g(z,p_a) \Big )
\no
\eea
The first line is evaluated with the help of (\ref{regs}). To evaluate the second line, we integrate by parts twice and use the Laplace equations for $g$ and $g_2$, taking into account that any $\delta (z,p_a)$ vanishes since the surface $\Sigma _{ab}$ does not contain the point $p_a$.  The integral on the last line receives contributions only from the region where $w$ is in the funnel. Evaluating these contributions by using for $g$ the Green function on the plane, we find that the contribution vanishes. Adding all up, we find, 
\bea
K_2+K_3 = 
{  T^4 \over 6}   -{ 2 T \over 3} D_3  
-{ 7  \over 8} E_2^2 + { 5  \over 8} D_4 + { 3 \over 4}E_4 + \cO(e^{- 2 \pi t})
\eea
It remains to analyze $K_1$. We do this by recasting it in the following way, 
\bea
\label{Wz}
K_1 & = &  {\tau_2 \over \pi}   \int _\Sep \! \kappa_1( z)   |\p_z g(z,p_a) |^2  W_{ab}(z)
\no \\
W_{ab}(z) & = & {\tau_2 \over \pi}  \int _\Sep \! \kappa_1 (w)    | \p_w g(w,p_a) |^2  
\Big ( g(z,w) - g(z,p_a) \Big )\Big ( g(z,w) - g(p_a,w) \Big )
\eea
and first studying the function $W_{ab}(z)$. When $z$ is in the funnel,  the entire contribution of the integral in $w$ arises for $w$ in the funnel as well, in view of the second factor in parentheses in the integrand. Thus, we use the following parametrization,
\bea
z = p_a+R \, e^{ x + i y} 
\hskip 0.6in
w = p_a+ R \, e^{ \alpha + i \beta} 
\hskip 1in  0 \leq y,\beta \leq 2 \pi
\eea
where $x,y$ are data determined by $z$, and $\alpha, \beta$ are to be integrated over. 
The condition $w \in \Sep$ restricts $\alpha $ to be positive and bounded above by a quantity of order $- \ln R$. The integral inside the funnel will rapidly converge as $\alpha$ becomes large and may be extended to $\infty$.  For $z \in \Sep$ we require $x >0$. We shall also be interested in continuing the point $z$  to the interior of the disc $\mD_a$ in $\Sigma _1$ outside of $\Sep$ where  $x<0$.  Under these assumptions, and using for $g$ the expression suitable for the funnel, the integral reduces to,
\bea
W_{ab}(z)  =  { 1 \over \pi} \int _0 ^\infty d \alpha \int _0 ^{2 \pi} d \beta 
\left ( - \ln \left | 1 - e^{\alpha-x+i\beta -iy} \right |^2 \right )
\left ( - \ln \left | 1 - e^{x-\alpha-i\beta +iy} \right |^2 \right )
\quad
\eea
The integral is independent of $y$ by translation invariance in $\beta$. For $x >0$, we split the integration region for $\alpha$ into two regions, $0\leq \alpha \leq x$ and $x \leq \alpha$. Expanding the logarithms into absolutely convergent Taylor series, the integrals over $\beta$ and $\alpha$ may be carried out, 
\bea
W_{ab}(z) = 4  \zeta(3) \theta (x)  - 2 \ep (x)  \sum _{m=1}^\infty {e^{-2m|x|} \over m^3} 
\eea
where $\theta (x)$ is the Heaviside step function and $\ep(x)=\theta(x) - \theta (-x)$ is the sign function. The function $W_{ab}(z)$ is continuous and differentiable once at $x=0$, and vanishes as $z \to p_a$ as expected from its defining integral in (\ref{Wz}). 

\sm

Using this result to evaluate the contribution to $K_1$ from the region where $z$ is in the funnel, we see that the $\zeta(3)$ term produces $ 4 T \zeta(3)$ upon integrating over $z \in \Sep$. The $z$-integral of the second term above is localized  in the funnel thanks to the exponential suppression. The sum of its contributions from the funnel and insider the disc $\mD_a$ cancel.  Therefore, we may extend the domains of integration for $W(z)$ and $K_1$ from $\Sep$ to $\Sigma _1$,  
\bea
\cK_{aaaa}^0 & = &  K_1 - 4 T \zeta (3) 
 =  {\tau_2 \over \pi}   \int _{\Sigma_1} \! \kappa_1 ( z)   |\p_z g(z) |^2  \Big (  W(z) - 4 \zeta (3) \Big )
 \no \\
 W(z) & = & {\tau_2 \over \pi}  \int _{\Sigma _1}  \! \kappa_1 (w)    | \p_w g(w) |^2  
\Big ( g(z,w) - g(z) \Big )\Big ( g(z,w) - g(w) \Big )
\eea
Here, we have set $p_a=0$ by translation invariance on $\Sigma _1$. While the integrand of $W_{ab}(z)$ vanishes at $z=0$ for  $w$ away from $w=0$, a careful analysis shows that the integral over $w \in \Sigma _1$ evaluates as follows $\lim _{z\to 0} W_{ab} (z) = 4 \zeta(3)$.  This limit arises from the original  integral over $\Sep$ by taking  the limit as $t \to 0$ of $W(z) =  W_{ab}(z) + \cO(e^{-2\pi t})$ for $ z\not =0$.
In terms of $\cK_{aaaa}^0$ we have,
\bea
\cK_{aaaa} = 
{ T^4 \over 6}   -{ 2 T \over 3} D_3  + 4 T \zeta (3)  
-{ 7  \over 8} E_2^2 + { 5 \over 8} D_4 + { 3  \over 4}E_4 + \cK^0_{aaaa} + \cO(e^{- 2 \pi t})
\eea
This concludes the decomposition of $\cK_{aaaa}$.

\subsubsection{Summary of results for $ \cK$}

Putting all together, we find, 
\be
\begin{split}
\cK = & 
 { T^4 \over 3} + 2 T^2 g(v)^2 + { 4 T \over 3} D_3 - {8 T \over 3} g(v)^3  + 8 T \zeta(3) 
  - 4 T  D_3 ^{(1)} - 8 T D_4 ^{(a)}   -{ 7  \over 4} E_2^2 
  \\ &
 + { 5 \over 4} D_4 + { 3  \over 2}E_4
+ 2 \cK_{abab} + 2 \cK_{abba} + 2 \cK_{aabb} ^0     - 4 \cK_{aaab}^0 - 4 (\cK_{aaab}^0)^* + 2 \cK_{aaaa}^0  
+ \cO(e^{- 2 \pi t})
\end{split}
 \ee
Splitting the contributions into $\cK^t$ and $\cK^c$ we recover precisely $\cK^t$ of (\ref{KK0}),
a fact which provides a double check on the calculations since (\ref{KK0}) was obtained by the variational method, and allows us to compute the constant part,
\bea
\cK^c & = & 
2 \cK_{abab} + 2 \cK_{abba} + 2 \cK_{aabb} ^0     - 4 \cK_{aaab}^0 - 4 (\cK_{aaab}^0)^* + 2 \cK_{aaaa}^0 
 \\ &&
  + 4 g(v) \left ( D_3   + 2  \zeta(3)   -    D_3 ^{(1)} + { \Delta _v F_4 \over 2 \pi}  \right ) - 3g(v)^4
  -{ 7  \over 4} E_2^2  + { 5 \over 4} D_4 + { 3  \over 2}E_4 
\no
\eea
This result completes the calculation of the functions $\BZ_i(\Omega)$.

\subsection{Calculation of the functions $\gamma_i$ and $\cBZ_i$}

The conversion of the functions $\BZ_i(\Omega)$ to the genuine modular graph functions $\cBZ_i(\Omega)$ is achieved with the help of the formulas (\ref{BtocB}). The functions $\gamma (x)$ and $\gamma _1$ govern the conversion of the standard string Green function $G$ into the Arakelov Green function $\cG$ and were given in (\ref{gamx1}). 
The function $\gamma (x)$ is readily obtained from its definition and, In the non-separating degeneration limit, is obtained by substituting \eqref{limkappa} and \eqref{limG} in \eqref{gamma},
\bea
\gamma (x) = \int _\Sep \left ( \half \kappa_1(y) +{ i \over 4 t} \om_t \wedge \oom_t(y) \right )
\left (g(x,y) + { (f(x)-f(y))^2 \over 8 \pi t} \right ) +\cO(e^{-2\pi t})
\eea
The integral of $\kappa _1$ against $g$ vanishes while against the term in $f^2$ it may be evaluated in terms of $F_2$. The remaining integrals are evaluated using (\ref{regs}) and the second equation in (\ref{ompat}) for $n=0$ and generic $w=y$, so that we recover the result for $\gamma (x)$,
\bea
\gamma (x) =   \frac{\pi t}{12}  + \frac14 \gone(x,p_a)+ \frac14 \gone(x,p_b)  + \frac{f(x)^2}{16\pi t}    + \frac{F_2(v)}{4\pi t} +\cO(e^{-2\pi t})
\eea 
announced in (\ref{gamx1}), and repeated here for convenience. To obtain also $\gamma _1$ simply requires a further integration using the formulas of (\ref{regs}) and (\ref{ompat}) for $n=0$.
Using these results, we shall now also evaluate the remaining functions $\gamma _2$ and $\gamma _3$,
\bea
\gamma _2 & = & \int _\Sep \left ( \half \kappa _1 (x) + { i \over 4 t} \om_t \wedge \oom_t (x) \right ) \gamma (x)^2
\no \\
\gamma _3 & = & \int _{\Sigma _{ab}^2} \left ( \nu_{xy} ^- - \nu_{xy}^+ \right ) \gamma (x) \gamma (y)
\eea
where $\gamma (x) $ is given in (\ref{gamx1}), and $\nu_{xy}^\pm$ was defined in (\ref{nupm}).
The calculation of $\gamma _2$ may be performed using the following integrals, in addition to those of (\ref{regs}) and (\ref{ompat}), 
\bea
{\tau_2\over \pi} \int _\Sep \kappa_1 (z) |\p_z f(z)|^2 g(z,p_a)
& = &
 {T^2 \over 2}  +  T g(v)  -{ 3 \over 2} g(v)^2 +  F_2(v)
 \\
{\tau_2\over \pi} \int _\Sep \kappa_1 (z) |\p_z f(z)|^2 g(z,p_a) g(z,p_b)
& = &
T^2 g(v) + {1 \over 3} D_3 - {1 \over 3} g(v)^3 - D_3 ^{(1)} (v) - 2 D_4 ^{(a)}(v)
\no
\eea
as well as the rearrangement $(g(x,p_a)+g(x,p_b))^2 = f(x)^2 + 4 g(x,p_a,) g(x,p_b)$. The calculation of $\gamma _3$ may be performed using the following results,
\bea
 \int _\Sep \om_t \wedge \oom_1 (x) \gamma (x) & = & - { \tau_2 \over 2 \pi} \p_v g_2(v)
 \no \\
  \int _\Sep \om_1 \wedge \oom_1 (x) \gamma (x) & = & -2 i \tau_2 \left ( { \pi t \over 12} +{ 3 F_2 \over 8 \pi t} \right )
 \no \\
  \int _\Sep \om_t \wedge \oom_t (x) \gamma (x) & = & - {5 \pi i \over 6} t^2 - i t g(v) - { 3 i F_2(v) \over 4 \pi}
 \eea
Using the rearrangement formula (\ref{pvg2}),  we arrive at the following results,
\bea
\label{g1minsep}
\gamma_1 &=&\frac{\pi t}{4} + \frac14 \gone + \frac{3F_2}{8\pi t} +\cO(e^{-2\pi t})
\no \\
\gamma_2& =& \frac{41\pi^2 t^2}{360}+\frac{5\pi t}{24} \gone+\frac{1}{24}(5E_2-2g_2+3\gone^2)
\no \\ && 
+ { 1 \over 16 \pi t} \left ( D_3 -D_3^{(1)} + 2 gF_2 +{ \Delta _v F_4 \over 4 \pi} \right )
+\frac{F_4 +2F_2^2}{16\pi^2 t^2} +\cO(e^{-2\pi t})
\no \\
\gamma_3
&=& 
\frac{5 \pi ^2 t^2}{18}
+\frac{1}{3} \pi  \gone t
+\frac{3}{2}F_2
+\frac{2 \gone  F_2 }{ \pi  t} - { \Delta _v F_2^2 \over 16 \pi^2 t} 
+\frac{9   F_2^2}{8 \pi ^2 t^2} +\cO(e^{-2\pi t})
\eea
It is now a simple matter of algebraic substitution  to use formulas (\ref{BtocB}) with $Z_1$ given in  (\ref{B1minsep}),
$Z_2$ given in (\ref{B2minsep}), and $Z_3$ given in (\ref{B1minsep}) and the above expressions for $\gamma_1, \gamma _2, \gamma _3$ in order to obtain the expressions for $\cBZ_1, \cBZ_2$ and $\cBZ_3$ given in the body of the paper in (\ref{bark3}).

\newpage
\section{Tropical limits of modular graph functions}
\setcounter{equation}{0}
\label{sec:D}

In this appendix, we shall derive the limit as $\tau \to i \infty$ of the various modular graph functions which appear as coefficients in the non-separating degeneration of the genus-two string invariants $\f$, $\cZ_i$ and $\cB_{(2,0)}$. The results give the behavior of these functions in their tropical limit near the non-separating degeneration node of Section~\ref{sec_trop}. 

\sm

In the first two subsections we review, without derivation, the Bernoulli polynomials and  the limits of standard modular graph functions and elliptic polylogarithms. In the third subsection, we derive the limits of the functions $D_3^{(1)} $, $D_4 ^{(1)}$, $D_4^{(a)}$, $F_2$ and $F_4$. In the fourth subsection, we present the limit for the one-loop self-energy graph, and related graphs. In the four subsequent subsections, we obtain the limits of the modular graph functions $\cK_{abab}, \cK_{abba}, \cK_{aabb}^0, \cK_{aaab}^0$ and $\cK_{aaaa}^0$.

\subsection{Bernoulli polynomials}

The  Bernoulli polynomials $B_k(x)$ are defined for all integers $k \geq 0$ by the Taylor series,
\bea
\sum _{k=0}^\infty { z^k \over k!} \, B_k(x) = { z \, e^{xz} \over e^z-1}
\eea
for $x\in \CC$, and $z \in \CC \setminus 2\pi i \ZZ$. From this definition, we have the following relations, 
\bea
\label{Bref}
B_k(1-x) & = & (-)^k B_k(x)
\no \\
B_k' (x) & = & k B_{k-1}(x)
\eea
The Bernoulli polynomials sum up the following Fourier series when $0 \leq x \leq 1$ and $k \geq 2$,
\bea
\label{bern}
\sum _{n\not=0} {e^{2 \pi i n x} \over n^k} = -{ (2 \pi i)^k  \over k!} \,  B_k(x)
\eea
The function defined by $B_k(x)$ in the interval $x \in [0,1]$ for $k \geq 2$ takes the same values at $x=1$ and at $x=0$, and extends to a {\sl continuous periodic function} on $\RR$ by translation of the interval by $\ZZ$. Its successive derivatives, however, are not continuous. The validity of the formula may be extended to  all $x \in \RR$, 
\bea
\label{bern3}
\sum _{n\not=0} {e^{2 \pi i n x} \over n^k} = -{ (2 \pi i)^k  \over k!}   B_k( \{ x \})
\eea
where the  fractional part $\{ x\}$ of $x \in \RR$, defined so that $0 \leq \{ x \} <1$ and $x-\{ x\} \in \ZZ$. The Bernoulli polynomials with even index $k$ for $k \leq 8$ are given explicitly by $B_0(x)=1$ and 
\bea
\label{listBernoulli}
B_2(x) &=& \frac{1}{6} - x + x^2  
\nn\\
B_4(x) &=& -\frac{1}{30} + x^2 - 2 x^3 + x^4 
\nn\\
B_6(x)&=& \frac{1}{42} - \half x^2 + \frac{5}{2}  x^4 - 3 x^5 + x^6  
\nn\\
B_8(x) &=& -\frac{1}{30} + \frac{2}{3}\,x^2- \frac{7}{3}\,x^4 + \frac{14}{3} \,x^6 - 4 x^7 + x^8\,.
\label{berndef}
\eea
Note that the Bernoulli numbers $B_k$ are related to the Bernoulli polynomials by $B_{k}=B_{k}(0)$.

\subsection{Eisenstein series and standard modular graph functions}

The limit $\tau_2\to \infty$ of the non-holomorphic Eisenstein series $E_n(\tau)$ and of the elliptic polylogarithms
$g_n(v|\tau)$ and $D_{a,b}(v|\tau)$ for $u_2=v_2/\tau_2$ fixed and $0 < u_2 < \half$ is well known, 
\bea
E_n(\tau) &=& 
 { 2\zeta(2n) \over \pi^{2n}} y^n + 2\,  \frac{\Gamma(n-\frac12)}{\sqrt{\pi} \, \Gamma(n)} \zeta(2n+1) y^{1-n} 
+ \cO(e^{-2y})
\no \\
g_n(v|\tau) &=& -\frac{(-4y)^n}{(2n)!} \, B_{2n}(u_2) +\cO(e^{-2 y u_2} ) 
\no \\
D_{a,b}(v|\tau) &=& \frac{(4y)^{a+b-1}}{(a+b)!} B_{a+b}(u_2) +\cO(e^{-2yu_2})
\eea
where $y=\pi \tau_2$. The asymptotics of the modular graph functions $D_\ell$ defined in \eqref{defDk} was studied in \cite{D'Hoker:2015foa}. For the values $\ell=3,4$ relevant to this paper we have,
\bea
D_3 (\tau) &=&  \frac{2}{945} \, y^3+ \zeta(3) + \frac{3\zeta(5)}{4\, y^2} + \cO(e^{-2y}) 
\label{D3def}\\
D_4 (\tau)
&=&
\frac{y^4}{945}+\frac{2\zeta(3)}{3}\, y+\frac{10\zeta(5)}{y}-\frac{3\zeta(3)^2}{y^2}+\frac{9\zeta(7)}{4y^3}+ \cO(e^{-2y}) 
\label{D4def}
\eea
We note the relations $D_3=E_3+\zeta(3)$ and $D_4=24 C_{2,1,1}+ 3 E_2^2 - 18 E_4$ established in 
\cite{D'Hoker:2015foa}.

\subsection{Degeneration  of $D_\ell^{(1)}$, $D_4^{(2)} $, $D_4^{(a)}$, $F_2$ and $F_4$}

The method developed in \cite{DHoker:2015wxz} for computing the asymptotics of  $D_\ell(\tau)$ defined in \eqref{Dlk} applies just as well to the generalized modular graph function $D_\ell^{(1)}(v|\tau)$.  Consider the decomposition of the genus-one Green function into $\gone(z|\tau)=\mg_1(z|\tau )+\mg_2(z|\tau )+\mg_3(z|\tau )$, where we use  the parametrization $z = \alpha + \beta \tau$ for $\alpha \in \RR/\ZZ$ and $- 1 < \beta < 1$, 
\bea
\mg_1(z) &=& 2y \, B_2(|\beta| ) 
\no \\
\mg_2 (z) &=& \sum_{m\neq 0} \frac{1}{|m|} e^{2\pi\I m(-\alpha +\tau_1 \beta )-2y |m \beta |} 
\no \\
\mg_3 (z) &=&  \sum_{m\neq 0} \frac{1}{|m|} \sum_{k\neq 0} e^{2\pi\I m[-\alpha +\tau_1 (\beta +k)]-2y |m (\beta+k)|}
\eea
Given the range for $\beta$, with strict inequalities,  the term $\mg_3 (z|\tau )$ contributes to $D_\ell ^{(1)}$  terms which are exponentially suppressed by $\cO(e^{-2 \pi \tau_2 u_2})$ and  can therefore be omitted.  Similarly terms linear in $\mg_2(z)$ integrate to zero.  In this way we find, 
 \be
 \begin{split} 
D_\ell^{(1)}(v|\tau) = & (2y)^\ell \int_{-1/2}^{1/2} d \beta\,  B_2(|\beta|)  ^{\ell-1}\, B_2(|u_2-\beta|)
\\
&+\sum_{\ell-1=\ell_1+\ell_2 \atop \ell_1 \geq 0, \ell_2 \geq 2} 
\frac{(\ell_1+\ell_2)!}{\ell_1!\,\ell_2!} \sum_{n=1}^{2\ell_1+3}  P(\ell_1,n;u_2)\, 
S(\ell_2,n)\, (2y)^{\ell_1-n+1}
+\cO(e^{-2y u_2})
\end{split}
\label{del1def}
\ee
where $ P(\ell_1,n;u_2)$ are quadratic polynomials in $u_2$, defined by 
\be
\label{defPtilde}
\int_{-\infty}^{\infty} d \beta \, B_2(|\beta |)^{\ell_1} B_2(u_2-\beta ) \, e^{-y |\beta |} 
= \sum_{n=1}^{2\ell_1+3}  P(\ell_1,n;u_2)\, y^{-n}
\ee
and $S(\ell_2,n)$ is defined by the multiple sum,
\bea
S(\ell_2,n) = \sum _{m_1, \cdots, m_{\ell_2} \not=0} { \delta (\sum_{i=1}^{\ell_2} m_i) \over | m_1 \cdots m_{\ell_2} | ( |m_1| +  \cdots + |m_{\ell_2} |)} 
\eea
Here, we have replaced $B_2(|u_2-\beta|)$ by $B_2(u_2-\beta)$, since for $y\to \infty$ the integral is dominated by contributions from the region $|\beta|\ll 1$ so that effectively $\beta<u_2$, up to corrections of order  $\cO(e^{-2y u_2})$, which we neglect. In particular, from \eqref{del1def}, we have,
\bea
D_3^{(1)} (v|\tau)&=& - \frac{8y^3}{15} B_6 - \frac{4y^3}{9} B_4 + 
2 \zeta(3)\, B_2 + \frac{\zeta(5)}{4\, y^2}+ \cO(e^{-2y u_2})
\nn\\
\no \\
D_4^{(1)} (v|\tau)&=& y^4 \left (
\frac47 B_8 + \frac{16}{15}B_6 +\frac29 B_4 \right ) +2\zeta(3) B_2 \, y
\nn\\&& 
+ \frac{3\zeta(5)}{2\, y} \left (B_2 +\tfrac16 \right ) -\frac{3\zeta(3)^2}{4\, y^2}+\frac{9\zeta(7)}{8\, y^3}+ \cO(e^{-2y u_2})
\label{d41def}
\eea
where here and henceforth, we omit the argument in $B_{2n}(|u_2|)$. 
One may check that these results are consistent with the differential equation (\ref{delDk}).
We  note that, upon setting $u_2=0$, the polynomial part of $D_\ell^{(1)}(v|\tau )$ does not reduce to the 
polynomial part of  $D_\ell (\tau)$,  since terms of order $\cO(e^{-2y  u_2})$ cannot be neglected in this limit.

\subsubsection{Degeneration of $D_4 ^{(2)}(v |\tau)$}

The same method readily applies to the calculation of $D_4^{(2)}(v|\tau)$.  Replacing $g$ by $\mg_1+\mg_2$ and using the fact that terms linear in $\mg_2$ integrate to zero,  we get, 
\bea
D_4^{(2)} =   \int_{\Sigma_1} \kappa _1(z) \Big ( \mg_1^2(z)\, \mg_1^2(z-v)
+2 \mg_1^2(z-v) \, \mg_2^2(z) +  \mg_2(z)^2\, \mg_2(z-v)^2 \Big ) + \cO(e^{- 2 y u_2})
\quad
\label{D42def}
\eea
The first term can be evaluated directly and produces a linear combination of Bernoulli polynomials $B_{2k}(u_2)$ for $k\leq 4$. The last term is exponentially suppressed as $y\to \infty$. The second term in \eqref{D42def} gives,
\be
8\pi^2\tau_2^2 \, \sum_{m\neq 0} \frac{1}{m^2} \,
\int_{-1/2}^{1/2} d \beta \, B_2(|u_2-\beta |)^2\, e^{-4\pi\tau_2 |m \beta|}
\ee
which can be evaluated by extending the integral to the full real axis.  In total, we find,
\bea
\label{D42}
\begin{split}
D_4^{(2)}(v|\tau) =& - y^4 \left(  \frac{8}{35} B_{8}+\frac{32}{45} B_6+\frac{8}{27} B_4-\frac{1}{2025} \right)
 \\
&
+  y\, \zeta(3)\,\left( 8 B_4+\frac83 B_2+\frac{4}{45} \right) 
+ \frac{\zeta(5)}{y} \left( 6 B_2+\frac13\right) 
 + \frac{3\zeta(7)}{4\, y^3}  + \cO(e^{-2y u_2})
\end{split}
\eea

\subsubsection{Degeneration  of  $D_4^{(a)}(v|\tau)$}

Although the  function $D_4^{(a)}(v|\tau)$ does not enter into the final expressions for the degeneration of the genus-two string invariants considered in this paper, it will be useful to have at intermediate stages. It was defined in (\ref{D4a}) and  related to the function $D_4^{(2)}(v|\tau)$ in (\ref{D4adiff}). Thus, its degeneration may be obtained directly from that of $D_4^{(2)}$ by differentiation in $u_2$, 
\bea
D_4^{(a)} (v|\tau)= - { 1 \over 16 \pi \tau_2} \p_{u_2}^2 D_4^{(2)}(v|\tau)
\eea
which is readily computed using the differentiation rule for Bernoulli polynomials,
\be
D_4^{(a)} (v|\tau) = 
y^3 \left ( { 4 B_6 \over 5} +{ 4 B_4 \over 3} +{ 2B_2\over 9} \right )
-  \zeta (3) \left ( 6 B_2 + { 1 \over 3}    \right ) - { 3\zeta (5) \over  4\, y^2}  
+ \cO(e^{-2 y u_2})
\ee

\subsubsection{Degeneration of $F_2$, $F_2^2$ and $F_4$}

The degeneration of $F_2$, $F_2^2$, and $F_4$ are obtained from the definitions of $F_2$ and $F_4$  in (\ref{F4}), 
and the relation to the modular graph functions $D_4$, $D_4^{(1)}$ and $D_2^{(2)}$ which have already been calculated above, and we find, 
\bea
F_2 (v|\tau) & = & {  y^2 \over 45} + { \zeta (3) \over y} + {2 \over 3} y^2 B_4 +\cO(e^{-2 y u_2})
\no \\
F_2(v|\tau) ^2
& = & y^4 \left(\frac{2}{2835}+\frac{16}{27}B_6+\frac{4}{9} B_8\right)
+y \zeta(3) \left(\frac{2}{45} +\frac{4}{3}B_4\right)+\frac{\zeta(3)^2}{y^2} +\cO(e^{-2 y u_2})
\no \\
 F_4(v |\tau)
&=& { 2 y^4 \over 15} \left(\frac{1}{630} + \frac{4}{3} B_6+ B_8\right)
+ 2y\, \zeta(3)\,  \left(\frac{1}{30}+ B_4\right)
+\frac{1}{y}\zeta(5)\,  \left({ 5 \over 6} + B_2\right) +\cO(e^{-2 y u_2}) 
\nn\\
\label{tropAterm}
\eea
Note that every term in the combination $10F_4-3 F_2^2$  involves either $\zeta(3)$ or $\zeta(5)$.

\subsection{Self-energy and related graphs}

For the evaluation of the remaining modular graph functions, we shall make use of one-loop graphs with two Green function factors, possibly with various derivatives. Some of these graphs were already studied in  \cite{D'Hoker:2015foa} to which we refer for their complete derivation and degeneration, while other graphs appear for the first time, and for which we shall give a complete derivation. 

\sm

The Fourier transform $\cT(M,N)$ of the square of the Green function (which is often referred to as the self-energy graph in quantum field theory), is given by,
\bea
\label{FTg2}
 g(z)^2 = \sum _{M,N\in \ZZ} \cT(M,N) \, e^{2 \pi i (Mz_2-Nz_1)} 
\eea
where $z=z_1+\tau z_2$ with $z_1,z_2 \in \RR$, and the Fourier coefficients $\cT(M,N)$ are given as follows,
\bea
\label{cTMN}
\cT(M,N) = \sum _{(m,n) \not= (0,0), (M,N)} { \tau_2^2 \over \pi^2 \big |m+n \tau \big |^2 \,  \big |m-M + (n-N)\tau \big |^2}
\eea
We have $\cT(0,0) = E_2$.  Closely related is the following integral,
\bea
\label{cFMN}
\cF(M,N) = { \tau_2 \over \pi} \int _{\Sigma _1} \kappa _1 (z) |\p_z g(z) |^2 \Big ( e^{ 2 \pi i (Mz_2-Nz_1) } -1 \Big )
\eea
which clearly satisfies $\cF(0,0)=0$. For $(M,N) \not= (0,0)$, both $\cT(M,N)$ and $\cF(M,N)$ are invariant under $(M,N) \to (-M,-N)$. The function $\cF(M,N)$ may be evaluated by integrating by parts successively  in $z$ and $\bar z$, using the fact that  $\delta(z)$  cancels against the expression in the parentheses  and expressing the remaining integral in terms of $\cT(M,N)$, 
\bea
\label{cFcT}
\cF(M,N) = - { \tau_2 \over \pi |M+N\tau|^2} - { \pi \over 2 \tau_2} |M+N \tau|^2 \cT(M,N)
\eea
The degeneration of $\cT(M,N)$ for $(M,N) \not= (0,0)$ was evaluated in the Appendix of \cite{D'Hoker:2015foa}, while that of $\cF(M,N)$ may be deduced from the above relation between the two quantities. For $M \not= 0$, we have, 
\bea
\label{cTcF}
\cT(M,0) & = & { 2 \tau_2^2 \over 3 M^2} -{ 6 \tau_2^2 \over \pi^2 M^4} + { 8 \tau_2^2 \over M^2} \, \mJ(M) + \cO(e^{-2 \pi \tau_2})
\no \\
\cF(M,0) & = & { 2 \tau_2 \over \pi M^2} -{ \pi \tau_2 \over 3} -  4 \pi \tau_2 \, \mJ(M) + \cO(e^{-2 \pi \tau_2})
\eea
where the function $\mJ(M)$ is conveniently given by the following integral representation, 
\bea
\mI(M) = \int _0^\infty dt \,  { 2-e^{2 \pi i Mt} - e^{-2\pi i Mt} \over e^{ 4 y t} -1}
\eea
and we continue to use the notation $y = \pi \tau_2$. Evaluating the integral, we find,
\be
\mJ(M) = \frac{1}{2y} \left ( \gamma_E + \Psi \left(\frac{i \pi M}{2y} \right) \right )
\ee
where $\Psi(z)=d \ln \Gamma (z) $ and $\gamma_E$ is the Euler-Mascheroni constant.
Thus $\mJ(M)$ grows logarithmically with $M$.

\subsection{Sums involving powers of $\mJ(M)$}

We shall now present a general procedure to evaluate sums over $M$ involving powers of $\mJ(M)$, in an expansion wherewe omit exponential corrections $\cO(e^{-2yu_2})$.  As usual, we 
assume $0<u_2<\frac12$. The sums of  interest may be expressed as follows, 
\bea
\label{C31}
\sum _{M\not=0} e^{2 \pi i Mu_2} {\mI(M)^p \over \pi^n M^n}
= 
- { (2 i)^n \over n!}  \prod _{i=1}^p \int _0 ^\infty { dt_i \over e^{4 y t_i}-1} \, \mB_n^{(p)} (u_2; t)
\eea
where the double-index family of functions $\mB_n^{(p)}(u_2, t)$ is defined by,
\bea
\label{C32}
\mB_n^{(p)} (u_2;t) = - { n! \over (2 \pi i)^n} \sum _{M\not=0} { e^{2\pi i Mu_2} \over M^n} 
 \, \prod _{i=1} ^p ( 2 - e^{ 2 \pi i Mt_i} - e^{-2\pi i M t_i} )
\eea
and where $t$ stands for the array  $t = (t_1, \cdots, t_p)$. The function $\mB_n^{(p)} (u_2;t)$ is the sum of $3^p$ terms, obtained by summing the Bernoulli polynomial $B_n$ in the variable $u_2$ shifted by the various combinations of $\pm t_i$, $\pm t_i \pm t_j$ and so on for $i,j$  mutually 
distinct\footnote{We assume that the $t_i$'s are small enough so that the shifted $\tilde u_2$
remains in the interval $0<\tilde u_2<1/2$. This is indeed the region which dominates the 
integral in the limit $y\to\infty$. } The normalization factor has been included so that each term contributes a Bernoulli polynomial with its natural multiplicity. For example, we have,
\bea
\label{C33}
\mB_n^{(1)} (u_2;t) & = & 2 B_n(u_2) - B_n (u_2+t) - B_n (u_2-t)
\no \\
\mB_n^{(2)} (u_2;t) & = & 4 B_n(u_2) - 2 B_n (u_2+t_1) - 2 B_n (u_2-t_1)
\no \\ &&
- 2 B_n (u_2+t_2)- 2 B_n (u_2-t_2) + B_n(u_2+t_1+t_2) 
\no \\ &&
+ B_n(u_2+t_1-t_2)+ B_n(u_2-t_1+t_2)+ B_n(u_2-t_1-t_2)
\eea
The function $\mB_n^{(p)}(u_2;t)$ is a polynomial in $u_2$ and the variables $t_i$ of overall degree at most $n$; it is a symmetric function of the $t_i$; it is even in each $t_i$ separately and  vanishes whenever $t_i=0$ for at least one value of $i$. With those properties in mind, it becomes straightforward to compute and simplify these functions, and for $p=1$ we find, 
\bea
\label{C34}
\mB_n ^{(1)} (u_2;t) = - 2 \sum_{k=1} ^{[n/2]} \binom{n}{2k} t^{2k} B_{n-2k}(u_2)
\eea
To the orders needed here, we shall also make use of the following results, 
\begin{align} 
\label{C34a}
\mB_2^{(2)}(u_2, t)  & = 0  &  \mB_2^{(3)}(u_2, t)  & = 0 
\no \\
 \mB_3^{(2)}(u_2, t) & = 0 & 
\bp{\mB_4^{(2)}(u_2,t)} & = 24\, t_1^2\, t_2^2 
\end{align}
The  integrals for the remaining terms may be  evaluated using the following formula,
\bea
\label{intJ}
\int _0 ^ \infty dt \, { t^n \over e^{4yt} -1} = { n! \, \zeta (n+1) \over (4y)^{n+1}}
\eea
Applying the integrals to the formula in (\ref{C34}), we find, 
\bea
\label{C32}
\sum _{M\not=0} e^{2 \pi i Mu_2} \, {\mI(M) \over \pi^n M^n}= 2 (2 i)^n \sum_{k=1}^{[n/2]} { B_{n-2k}(u_2) \over (n-2k)!} \, { \zeta (2k+1) \over (4y)^{2k+1}}
\eea
Applying the integrals to the formula in (\ref{C34a}), we find, 
\bea
&&
\sum _{M\not=0} e^{2 \pi i Mu_2} \, {\mI(M)^2 \over \pi^n M^n}=0 \hskip 1in n=2,3
\no \\ &&
\sum _{M\not=0} e^{2 \pi i Mu_2} \, {\mI(M)^2 \over \pi^4 M^4}= - { \zeta(3)^2 \over 64 y^6}
\eea
When $u_2=0$, we use instead the identity, valid up to $\cO(e^{2y})$ terms 
\be
\sum _{M\not=0}  \, {\mI(M) \over \pi^n M^n}= \frac{2(2\I)^n}{n!} 
\int_0^\infty \de t \frac{B_n(t) - B_n(0)}{e^{4yt}-1}
\ee
and evaluate the integral using \eqref{intJ}.

\subsection{Degeneration of $\cK^0_{aabb}$ \label{sec_aabb}}

This function was defined in (\ref{Kzero}),  and we shall use translation invariance to shift $z$ and $w$ by $p_a$, so that the integral may be expressed directly in terms of $v$,
\bea
\cK _{aabb} ^0 & = & 
{\tau_2^2 \over \pi^2}  \int _{\Sigma_1} \! \kappa_1 ( z)  \int _{\Sigma_1}  \! \kappa_1 (w)   |\p_z g(z) |^2  
| \p_w g(w-v) |^2 
\no \\ && \hskip 1in \times
 \Big ( g(z-w)^2 - g(w)^2 -g(z-v)^2 + g(v)^2 \Big )   
 \label{Kaabb0}
 \eea
The Fourier transform of the combination in the parentheses of the integrand may be evaluated with the help of (\ref{FTg2}) and its dependence on $z$ and $w$ may be factored using the following formula, and its analogue for $w \to v$,
\bea
\label{gfac}
g(z-w)^2 - g(w)^2= \sum _{M,N \in \ZZ} \cT(M,N) \, e^{- 2 \pi i (Mw_2-Nw_1) } \Big  ( 
e^{2 \pi i (Mz_2-Nz_1) } - 1 \Big )
\eea
where we continue to use the notation $v=u_1 + \tau u_2$, $z=z_1+\tau z_2$ and $w=w_1+\tau w_2$. In terms of the Fourier coefficients $\cT$ and $\cF$, the function $\cK^0_{aabb}$ takes on the following form, 
\bea
\cK _{aabb} ^0  =  \sum _{M,N\in \ZZ} \cT(M,N)   |\cF(M,N)|^2 \, e^{2 \pi i (Mu_2-Nu_1)}
\eea
Retaining only the constant Fourier mode in $u_1$ in the limit where we omit exponential dependence on $u_2$, we are led to keep only the contribution from $N=0$, and we find, 
\bea
\cK _{aabb} ^0   =   \sum _{M\not=0} \cT(M,0)  |\cF(M,0)|^2 \, e^{2 \pi i Mu_2} + \cO(e^{ - 2 y u_2})
\eea
Note that $F(0,0)=0$ so that only $M \not =0$ contributes. Expressing both $\cT$ and $\cF$ in terms of the function $\mJ$ using (\ref{cTcF}), we find, 
\bea
\cK _{aabb} ^0  =  128 y^4 \sum _{M\not=0}  { e^{2 \pi i Mu_2} \over  \pi^2 M^2} \left (  {1  \over 12} - { 3  \over 4 \pi^2 M^2}  +  \mI(M) \right )  
\left ( { 1 \over 12}  -{ 1 \over 2 \pi^2 M^2}  + \mI(M) \right )^2
\eea
As a result of  the right most equation in (\ref{C34a}), the contribution from the term in $\mI^3$ sums to zero. The  integrals for the remaining terms may be  evaluated using the remaining formulas in (\ref{C34}), (\ref{C34a}), and (\ref{C32}) and we find, 
\bea
\cK^0_{aabb} & = & 
{ 4 y^4 \over 135}   \left ( 5 B_2 + 35 B_4 + 32 B_6 + { 36 \over 7} B_8 \right )
 \\ &&
- {y \over 3}   \zeta (3)   \Big ( 1 +28   B_2 + 32 B_4 \Big ) 
- { \zeta (5) \over 6 y} \Big (  7 +48  B_2  \Big )
+{ 7 \zeta (3)^2 \over 2 y^2} 
-{   \zeta (7) \over y^3}  + \cO(e^{-2 y u_2})
\no \eea
The leading term in fact agrees with the naive evaluation of the integral \eqref{Kaabb0} by replacing $g(z)$ by its polynomial approximation $\mg_1(z)$.

\subsection{Degeneration of $\cK_{abab}$ and $\cK_{abba}$\label{sec_abab}}

The starting point is  pair of  integrals  defined in (\ref{Kabab}). By translation invariance and reflection symmetry, we may shift $z,w$ by $p_a$ and express the result in terms of $v$, 
\bea
\cK _{abab} & = & {\tau_2^2 \over \pi^2} \int _{\Sigma_1} \! \kappa_1(z)  \int _{\Sigma_1} \! \kappa_1 (w)   \p_z g(z-v) \pbz g(z)   g(z-w)^2 \p_w g(w-v) \pbw g(w)  
\label{Kabab} \\
\cK _{abba} & = & {\tau_2^2 \over \pi^2}  \int _{\Sigma_1} \! \kappa_1( z)  \int _{\Sigma_1} \! \kappa_1 (w)   \p_z g(z-v) \pbz g(z)   g(z-w)^2 \p_w g(w) \pbw g(w-v)  
\label{Kabba}
\eea 
As we shall neglect exponential corrections we are interested only in the zero mode in $u_1$,
where $v=u_1+\tau u_2$. To extract it projection, we define the following integrals, 
\be
\begin{split}
L^+ (z,w;u_2) = & \int _0^1 du_1 \, \p_z g(z-v) \, \p_w g(w-v)
\\
L^- (z,w;u_2) = & \int _0^1 du_1 \, \p_z g(z-v) \, \pbw g(w-v)
\end{split}
\ee
The dependence of $L^\pm$ on the modulus $\tau$ is understood throughout. It will be convenient to parametrize $z=z_1+ \tau z_2$ and $w=w_1+\tau w_2$ with $z_1, z_2, w_1, w_2 \in \RR$ and with ranges $0 \leq z_1, w_1 \leq 1$ and $-\half \leq z_2 , w_2 \leq \half$. Since $L^\pm(z,w;u_2)$ depends only on the combination $z_1-w_1$, it is clear that the remaining integration over $w_1$ has the net effect of projecting onto the constant Fourier mode of the complex conjugate combinations as well, and we have, 
\be
\begin{split}
\cK _{abab}  &= 
{\tau_2^2 \over \pi^2}  \int _0^1 dz_1 \int _{-\half }^\half dz_2  \int _{-\half }^\half dw_2 \,  g(z-w)^2
L^+(z,w;u_2) \overline{L^+(z,w;0)} + \cO(e^{- 2 \pi \tau_2 u_2}) 
\\
\cK _{abba} &=
{\tau_2^2 \over \pi^2}  \int _0^1 dz_1 \int _{-\half }^\half dz_2  \int _{-\half }^\half dw_2  \, g(z-w)^2
 L^-(z,w;u_2) \overline{L^- (z,w;0)} + \cO(e^{- 2 \pi \tau_2 u_2}) 
\end{split}
\ee
We shall compute $L^\pm$ for $-\half  \leq u_2 < \half$ using the following formula,
\bea
\label{partg}
\p_z g(z) = - i \pi \, { e^{ 2 \pi i (z_1+\tau z_2)} + 1 \over e^{ 2 \pi i (z_1+\tau z_2)} - 1} - 2 \pi i z_2
+ \cO(e^{- \pi \tau_2}) 
\eea
It will be convenient to organize the result in terms of the Fourier components in $z_1-w_1$, 
 \bea
 L^\pm (z,w;u_2) = 4 \pi^2 \sum _{n=-\infty} ^\infty \ell^\pm _n (z_2,w_2;u_2) \, e^{ 2 \pi i n (z_1-w_1)}
 \eea
where the Fourier coefficients $\ell^\pm _n$ are given by,
\bea
\ell_0 ^\pm (z_2, w_2;u_2) & = & \mp \Big (z_2-u_2 - \half \ep(z_2-u_2) \Big ) \Big (w_2-u_2 - \half \ep(w_2-u_2) \Big ) 
\no \\
\ell^+_n (z_2, w_2;u_2) & = & e^{ 2 \pi i n \tau(z_2-w_2)} \, \tilde \ell^+_n (z_2, w_2;u_2)
\no \\
\ell^-_n (z_2, w_2;u_2) & = & e^{ 2 \pi i n \big (\tau(z_2-u_2) -\bar \tau (w_2-u_2)\big )} \, \tilde \ell^-_n (z_2, w_2;u_2)
\eea
and the functions $\tilde \ell_n ^\pm $ are given by (using the convention $\theta (0)=0$),
\bea
\tilde \ell ^+ _n (z_2,w_w;u_2) & = & 
\theta (n) \theta (z_2-u_2) \theta (u_2-w_2) + \theta (-n) \theta (u_2-z_2) \theta (w_2-u_2)
\no \\
\tilde \ell ^- _n (z_2,w_w;u_2) & = & 
\theta (n) \theta (z_2-u_2) \theta (w_2-u_2) + \theta (-n) \theta (u_2-z_2) \theta (u_2-w_2)
\eea
We shall also use the Fourier expansion of $g(z-w)^2$, given in (\ref{FTg2}).
The integrals over $z_1$ may now be  performed (where we abbreviate $z=z_2, w=w_2$),
\begin{align}
\cK _{abab} & =  
16 y^2 \sum _{M,N,n}  T(M,N) \int ^{\half } _{-\half } dz \int ^{\half } _{-\half } dw\, 
e^{2 \pi i M(z-w)} \, \ell^+ _n (z, w;u_2) \, \overline{\ell^+ _{n-N} (z,w;0) }
\no \\
\cK _{abba} & =  
16 y^2 \sum _{M,N,n}   T(M,N) \int ^{\half } _{-\half } dz \int ^{\half } _{-\half } dw\, 
e^{2 \pi i M(z-w)} \, \ell^- _n (z, w;u_2) \, \overline{\ell^- _{n-N} (z,w;0) }
\end{align}
up to exponential corrections. There are five cases  to be distinguished,
\bea
(1) & \hskip 0.4in & n \not = 0,  \hskip 0.3in  n \not= N
\no \\
(2) & \hskip 0.4in & n  = 0,  \hskip 0.3in  n \not= N \hskip 1.53in \hbox{requiring } N \not= 0
\no \\
(3) & \hskip 0.4in & n \not = 0,  \hskip 0.3in  n = N  \hskip 1.53in \hbox{requiring } N \not= 0
\no \\
(4) & \hskip 0.4in & n = 0,  \hskip 0.3in  n = N, M\not=0   \hskip 1in \hbox{requiring } N= 0
\no \\
(5) & \hskip 0.4in & n = 0,  \hskip 0.3in  n = N, M=0   \hskip 1in \hbox{requiring } N= 0
\eea
We designate the contributions to $\cK_{abab}$ and $\cK_{abba}$ of each sum range above respectively by $\cK_{abab}^{(i)}$ and $\cK_{abba}^{(i)}$ for $i=1,2,3,4,5$.  The following contributions are pairwise equal,
\bea
\cK_{abab}^{(1)}  =  \cK_{abba}^{(1)} & = & \cO  (e^{-2 y u_2} )
\no \\
\cK_{abab}^{(5)}  =  \cK_{abba}^{(5)} & = & 4 y^2 E_2 \, B_2 ^2 + \cO  (e^{-2 y u_2} )
\eea
as well as,
\bea
\cK_{abab}^{(4)} =  \cK_{abba}^{(4)} & = &
y^2 \sum _{M\not=0}  T(M,0) e^{2 \pi i M u_2} \left ( { 1 \over \pi^4 M^4}
-{ 4 B_2+ \frac13  \over \pi^2 M^2} - { 4 i B_1 \over \pi^3 M^3} \right ) + \hbox{c.c.}
\no \\ && + 
y^2 \sum _{M\not=0}  T(M,0) \left ( { 2 \over \pi^4 M^4}
+{ 8 B_2+ \frac23  \over \pi^2 M^2} \right ) 
 + \cO  (e^{-2 y u_2} )
 \label{eqKabab4}\\
\cK_{abab}^{(2)} = \overline{\cK_{abab}^{(3)}} & = &
y^2 \sum _M \sum _{N\not=0} T(M,N) \left ( { 1 \over \pi^4 (M+\bar \tau N)^4}
+{ 4 B_2+ \frac13  \over \pi^2 (M+ \bar \tau N)^2} \right ) 
\no \\
\cK_{abba}^{(2)} = \cK_{abba}^{(3)} & = & 
 y^2  \sum _M \sum _{N\not=0} T(M,N)  \left ( { 1+4y N B_1(u_2) 
\over\pi^4 |M+ \tau N|^4}
+{ 4 B_2+ \frac13  \over \pi^2 |M+  \tau N|^2} \right )\nn
\eea
We shall denote the contribution of the first and second line of \eqref{eqKabab4} by 
$\cK_{abab}^{(4')}$ and $\cK_{abab}^{(4'')}$, respectively.

\sm

The term proportional to $B_1(u_2)$ cancels since the summand that multiplies it is odd in $(M,N)\to (-M,-N)$ while $T(M,N)$ is even. The asymptotics of  the first line of \eqref{eqKabab4} can be computed using the techniques developed earlier,
\bea
\cK_{abab}^{(4')} =   \cK_{abba}^{(4')} &=& 
2 y^4 \left ( { 164 \over 105} B_8 +{ 416 \over 135} B_6 +{92 \over 135} B_4 
- { 8 \over 14175} \right )
 \\ &&
- 2 y \zeta (3) \left ( {10 \over 3}  B_4 +2 B_2 + { 4 \over 45} \right )
+ { \zeta (5) \over y} \left ( B_2 +{1 \over 6} \right )
 - { \zeta (7) \over 8 y^3} + \cO(e^{-2yu_2}) \nn
 \eea
For the remaining terms, we recognize the  contributions with $(M,N) \not=0$ as arising from 2-loop modular graph forms, whose  definition and normalization we recall here,
\bea
\cC \left [ \begin{matrix} a_1 ~ a_2 ~ a_3 \cr b_1 ~ b_2 ~ b_3 \cr \end{matrix} \right ]
=
\left ( { \tau_2 \over \pi} \right ) ^{\half \sum _i (a_i+b_i)} 
\sum _{p_1, p_2, p_3 \in \Lambda '} { \delta (p_1+p_2+p_3)
\over p_1^{a_1} p_2^{a_2} p_3^{a_3} 
\bar p_1^{b_1} \bar p_2 ^{b_2} \bar p_3 ^{b_3}}
\eea
for $\Lambda = \ZZ + \tau \ZZ$ and $\Lambda ' = \Lambda \setminus \{ 0 \}$. When $b_i=a_i$ for $i=1,2,3$, we shall use the simplified notation $\cC_{a_1, a_2, a_3}$ instead. In terms of these functions, we have
\bea
\cK_{abab}^{(3)}  + \frac12 \cK_{abab}^{(4'')} & = & 
 \cC \left [ \begin{matrix} 4 ~ 1 ~ 1 \cr 0 ~ 1 ~ 1 \cr \end{matrix} \right ]
+ {4 y  \left( B_2 +\frac{1}{12} \right)} \, 
\, \cC \left [ \begin{matrix} 2 ~ 1 ~ 1 \cr 0 ~ 1 ~ 1 \cr \end{matrix} \right ]
\no \\
\cK_{abba}^{(2)}  +\frac12 \cK_{abba}^{(4'')} & = & 
\cC _{2,1,1}
+ 4 y \left( B_2 +\frac{1}{12} \right) \,  \cC _{1,1,1}
\eea
We shall need the asymptotics of $\cC_{1,1,1}$ and $\cC_{2,1,1}$, as well as of $D_4$, 
\bea
\label{C111asymp}
\cC _{1,1,1}  & = & 
{ 2 y^3 \over 945} + \zeta (3) + { 3 \zeta (5) \over 4 y^2} + \cO(e^{-2 y})
\\
\label{C211asymp}
\cC _{2,1,1}  & = & 
{ 2 y^4 \over 14175} + { y \zeta (3) \over 45}  + { 5 \zeta (5) \over 12 y}
- {  \zeta (3)^2 \over 4 y^2} + { 9 \zeta (7) \over 16 y^3} + \cO(e^{-2 y})
\no \\
D_4 & = & {y^4 \over 945 } +{2 \over 3} y \zeta (3) +{10 \zeta(5) \over y} -{ 3 \zeta (3)^2 \over y^2} +{9 \zeta (7 ) \over 4 y^3} + \cO(e^{-2 y})
\eea
The asymptotics of $\cC{\scriptsize \left [ \begin{matrix} 2 ~ 1 ~ 1 \cr 0 ~ 1 ~ 1 \cr \end{matrix} \right ]}$ and
$\cC {\scriptsize \left [ \begin{matrix} 4 ~ 1 ~ 1 \cr 0 ~ 1 ~ 1 \cr \end{matrix} \right ]}$ can be obtained 
by taking successive derivatives with respect to $\nabla = 2 i \tau_2^2 \p_{\tau}$ (which reduces to 
$\tau_2^2 \p_{\tau_2}$ when acting on functions independent of $\tau_1$):
\bea
\label{C211011}
\cC \left [ \begin{matrix} 2 ~ 1 ~ 1 \cr 0 ~ 1 ~ 1 \cr \end{matrix} \right ] =   { 1 \over 3\tau_2} \nabla E_3
={ 2 y^3 \over 945} -{ \zeta (5) \over 2 y^2} + \cO(e^{-2 y})
\eea
while we also have (see eq (4.26) of \cite{DHoker:2016mwo}),
\bea
\label{C411011}
\cC \left [ \begin{matrix} 4 ~ 1 ~ 1 \cr 0 ~ 1 ~ 1 \cr \end{matrix} \right ] 
= {1 \over 24 \tau_2^2} \nabla ^2 D_4  - {1 \over 4 \tau_2^2}  E_2\,\nabla ^2 E_2
\eea
As a result we obtain the following asymptotics, 
\bea
\cC \left [ \begin{matrix} 4 ~ 1 ~ 1 \cr 0 ~ 1 ~ 1 \cr \end{matrix} \right ] =  
{ 2 y^4 \over 14175} +{ y \zeta (3) \over 45} -{ \zeta(3)^2 \over 4 y^2} +{ 9 \zeta (7) \over 16 y^3} + \cO(e^{-2 y})
\eea
Combining these results, we find 
\bea
\cK_{abab}  &=& \frac{y^4}{945}
\left( 1+44 B_2 + 1372 B_4+ 5824 B_6 +  2952 B_8 \right)  \\
&& -\frac{y\zeta(3)}{3} \left( 8 B_4+8 B_2+\frac{1}{3}\right) -\frac{\zeta(5)}{y} \left( 3B_2+\frac16 \right)
- \frac{\zeta(3)^2}{2y^2} 
+\frac{\zeta(7)}{y^3}
+ \cO(e^{-2yu_2}) \nn
\\
\cK_{abba}  &=& \frac{y^4}{945}
\left( 1+44 B_2 + 1372 B_4+ 5824 B_6 +  2952 B_8 \right) \\
&& -\frac{y\zeta(3)}{3} \left( 8 B_4-16 B_2-\frac53\right) 
+\frac{\zeta(5)}{y} \left( 7B_2+\frac32 \right)
-\frac{\zeta(3)^2}{2y^2} 
+ \frac{\zeta(7)}{y^3} 
+ \cO(e^{-2yu_2}) \nn
\eea
The leading term in each expression agrees with the result obtained by evaluating
the integrals \eqref{Kabab}, \eqref{Kabba} after replacing $g(z)$ by its polynomial approximation
$\mg_1(z)$.

\subsection{Degeneration of $\cK_{aaab}^0$ \label{sec_aaab}}

The modular graph function $\cK_{aaab}^0$ was defined in (\ref{Kzero}). Using translation invariance, we may shift $z$ and $w$ by $p_a$ and express the result solely in terms of $v$,
\bea
 \cK _{aaab} ^0 = 
{\tau_2^2 \over \pi^2}  \int _{\Sigma_1} \! \kappa_1 ( z)  \int _{\Sigma _1} \! \kappa_1 (w)   |\p_z g(z) |^2  
 \p_w g(w) \pbw g(w-v) \Big ( g(z,w)^2 - g(w)^2 \Big ) 
 \label{Kaaab0}
 \eea
 We use the relation (\ref{gfac}) to factorize the dependence of the integrand on $z$ and $w$. The result may be expressed as follows,
 \bea
 \label{C57}
\cK _{aaab} ^0 =    \sum_{(M,N)\not= (0,0)}   \cT(M,N) \, \cF(M,N) \, \cG(M,N)
\eea
where $\cT$ was defined in (\ref{cTMN})   and $\cF$ was calculated in (\ref{cFMN}) in terms of $\cT$. The coefficients $\cG(M,N)$ are defined as follows,
\bea
\cG(M,N) = { \tau_2 \over \pi} \int _{\Sigma _1} \kappa _1(w) \p_w g(w) \pbw g(w-v) \, e^{- 2 \pi i (Mw_2-Nw_1)}
\eea
Since we neglect exponentially suppressed contributions, we retain only the zero mode in $u_1$, which is calculated by integrating over $u_1$, 
\bea
\int _0^1 d u_1 g(w-v) = 2 y B_2 (|w_2-u_2|)
\eea 
Substituting this result into the definition of $\cG(M,N)$, we find, 
\bea
\cG(M,N) = i \tau_2 \int _{-\half} ^\half dw_2 \, \p_{u_2}  B_2 (|w_2-u_2|) e^{- 2 \pi i Mw_2} \int _0 ^1 dw_1 \p_w g(w) \, e^{2 \pi i Nw_1} + \cO(e^{-2yu_2})
\quad
\eea 
Within this approximation, the $w_1$-integral  may be carried out using (\ref{partg}), and we find, 
\bea
\cG(M,0) & = & 
-  { 2 i y B_1(u_2) \over \pi M} + { y \over \pi^2 M^2} 
+ \left ( { 2 i y B_1(u_2) \over \pi M} + { y \over \pi^2 M^2} \right )  e^{-2 \pi i Mu_2} 
  \no \\
\cG(M,N) & = & - { 2 i y B_1(u_2) \over \pi (M+N \tau) }  + { y \over \pi^2 (M+ N \tau) ^2}  
 \eea
where on the first line $M \not=0$ and on the second line $N \not=0$. Since we have retained only the zero mode the above formulas are valid up to exponentially suppressed contributions. 
 
\sm

We split the sum in (\ref{C57}) according to whether $N=0$ or not, and further split the $N=0$ part into parts with and without exponential $u_2$-dependence  in $\cG(M,0)$, 
\bea
\cK_{aaab}^0 = \cK_A+\cK_B+\cK_C
\eea
where 
\bea
\cK_A & = & y \sum _{M\not=0} \cT(M,0) \, \cF(M,0) \, \left ( { 2 i  B_1(u_2) \over \pi M} + { 1 \over \pi^2 M^2} \right )\, e^{- 2 \pi i M u_2}
\no \\
\cK_B & = &  y \sum _{M\not=0} \cT(M,0) \, \cF(M,0) \,  { 1 \over \pi^2 M^2} 
\no \\
\cK_C & = & y \sum _{N\not=0} \sum_{M} \cT(M,N) \, \cF(M,N) \, { 1 \over \pi^2 (M+ N \tau) ^2} 
\eea
We have simplified these expressions by using the fact that the terms which are proportional to $B_1(u_2) $ in the non-exponential terms in $\cG$ are  odd under $(M,N) \to (-M, -N)$ and sum to zero since both $\cT$ and $\cF$ are even. 

\subsubsection{Calculating $\cK_B+\cK_C$}

As a result, $\cK_B$ and $\cK_C$ are independent of $u_2$. Their sum is an ordinary genus-one modular graph function, 
\bea
\cK_B+ \cK_C =   \sum _{(M,N)\not=(0,0)} \cT(M,N) \, \cF(M,N) \, { \tau_2 \over \pi (M+ N \tau) ^2} 
\eea
Using the explicit expression for $\cF$ of (\ref{cFcT}), we find more explicitly,
\bea
\cK_B + \cK_C =
-  \sum _{(M,N)\not=(0,0)} \left ( { \tau _2 ^2 \,  \cT(M,N)  \over \pi^2 (M+N\tau)^3 (M+ N \bar \tau)}
+\half  \cT(M,N)^2 \, { M+N \bar \tau  \over  M+N\tau } \right )
\eea
This is recognized as the sum of the following modular graph functions \cite{DHoker:2016mwo},
\bea
\cK_B + \cK_C = 
-\cC \left [ \begin{matrix} 3 ~ 1 ~ 1 \cr 1 ~ 1 ~ 1 \cr \end{matrix} \right ]
-\frac12\, \cC \left [ 
\begin{matrix} 1 ~ 1 \cr 1 ~ 1 \end{matrix} \bigg | 
\begin{matrix} 1 ~ 1 \cr 1 ~ 1 \end{matrix} \bigg |
\begin{matrix} 1  \cr -1 \end{matrix} \right ]
\eea
We may simplify the trihedral modular graph function using the rules of \cite[\S 7]{DHoker:2016mwo}, 
\bea
\cC \left [ 
\begin{matrix} 1 ~ 1 \cr 1 ~ 1 \end{matrix} \bigg | 
\begin{matrix} 1 ~ 1 \cr 1 ~ 1 \end{matrix} \bigg |
\begin{matrix} 1  \cr -1 \end{matrix} \right ]
&=&
2\, \cC \left [ 
\begin{matrix} 1 ~ 1 \cr 1 ~ 1 \end{matrix} \bigg | 
\begin{matrix} 1 ~ 1 \cr 1 ~ 0 \end{matrix} \bigg |
\begin{matrix} 1  \cr 0 \end{matrix} \right ]
=
2\, \cC \left [ 
\begin{matrix} 1 ~ 1 \cr 1 ~ 1 \end{matrix} \bigg | 
\begin{matrix} 2 ~ 1 \cr 1 ~ 0 \end{matrix} \bigg |
\begin{matrix} 0  \cr 0 \end{matrix} \right ]
-
2\, \cC \left [ 
\begin{matrix} 1 ~ 1 \cr 1 ~ 1 \end{matrix} \bigg | 
\begin{matrix} 2 ~ 0 \cr 1 ~ 0 \end{matrix} \bigg |
\begin{matrix} 1  \cr 0 \end{matrix} \right ]
\eea
Using now also the algebraic reduction formulas of \cite{DHoker:2016mwo}, we find, 
\bea
\cC \left [ 
\begin{matrix} 1 ~ 1 \cr 1 ~ 1 \end{matrix} \bigg | 
\begin{matrix} 1 ~ 1 \cr 1 ~ 1 \end{matrix} \bigg |
\begin{matrix} 1  \cr -1 \end{matrix} \right ]
&=&
 { 1 \over 2\tau_2}  \nabla E_2^2 -{1 \over 6 \tau_2 } \nabla D_4 
+ 2 \, \cC \left [ \begin{matrix} 3 ~ 1 ~ 1 \cr 1 ~ 1 ~ 1 \cr \end{matrix} \right ]
\eea
Hence we have, 
\bea
\cK_B+\cK_C = -{ 1 \over 4\tau_2}  \nabla E_2^2 +{1 \over 12 \tau_2 } \nabla D_4  
-2 \, \cC \left [ \begin{matrix} 3 ~ 1 ~ 1 \cr 1 ~ 1 ~ 1 \cr \end{matrix} \right ]
\eea
To evaluate the  last term is more involved. We begin with the observation that this modular graph form satisfies the following differential equation (for the rules of differentiation and further manipulation of modular graph forms, see \cite{DHoker:2016mwo}),
\bea
\nabla \left( \tau_2\, \cC \left [ \begin{matrix} 3 ~ 1 ~ 1 \cr 1 ~ 1 ~ 1 \cr \end{matrix} \right ] \right)
= 3 \tau_2^2 \cC \left [ \begin{matrix} 4 ~ 1 ~ 1 \cr 0 ~ 1 ~ 1 \cr \end{matrix} \right ] 
+\frac14 (\nabla E_2)^2 -\frac{1}{20} \nabla^2 E_4
\eea
Since the Laurent expansion on the right side is known, we obtain the following Laurent expansion by integrating the above differential equation
\bea
\cC \left [ \begin{matrix} 3 ~ 1 ~ 1 \cr 1 ~ 1 ~ 1 \cr \end{matrix} \right ] 
= { 2 y^4 \over 14175} +{ \zeta (3) y \over 45} + { c \over y} +{ \zeta (3)^2 \over 2 y^2} -{ 3 \zeta (7) \over 4 y^3} + \cO(e^{-2y})
 \eea
 where $c$ is the integration constant which is left undetermined by the above calculation. It may be evaluated by direct summation, and one finds $c= - 5 \zeta (5)/12$. 
Using these asymptotics  we obtain,
\be
\cK_B+\cK_C = -{2 y^4 \over 4725}+  {15\zeta(7)\over 16 y^3}  +\cO(e^{-2y}) 
\ee

\subsubsection{Calculating $\cK_A$}

It remains to evaluate $\cK_A$, which we write out explicitly as follows,
\bea
\cK_A & = & -32 y^4 \sum _{M\not=0} 
\left (  {1 \over 12}   -{ 3  \over 4 \pi^2 M^2} +   \mJ(M) \right )  
\left ( { 1 \over 12} - { 1  \over 2 \pi^2 M^2} +  \mJ(M) \right )  
\no \\ && \hskip 0.7in \times 
\left ( { 2 i  B_1 \over \pi^3 M^3} + { 1 \over \pi^4 M^4} \right )\, e^{- 2 \pi i M u_2}
\eea
The contributions to $\cK_A= \cK_A^+ + \cK_A^-$ respectively with an even or odd power of $M$ multiplying the exponential, are given by,
\bea
\cK_A ^+ & = &  -y^4 \sum _{M\not=0}  { e^{2 \pi i M u_2} \over \pi^4 M^4} 
\left ( 32 \mJ(M)^2 -{ 40 \mJ(M) \over \pi^2 M^2} 
+ { 16 \mJ(M) \over 3}  +{ 12 \over \pi^4 M^4} -{ 10 \over 3 \pi^2 M^2} +{ 2 \over 9}  \right )   
 \\
\cK_A^-  & = & -2 i  B_1 y^4 \sum _{M\not=0} { e^{2 \pi i M u_2} \over \pi^3 M^3} 
\left ( 32 \mJ(M)^2 -{ 40 \mJ(M) \over \pi^2 M^2} 
+ { 16 \mJ(M) \over 3}  +{ 12 \over \pi^4 M^4} -{ 10 \over 3 \pi^2 M^2} +{ 2 \over 9}  \right )   
\no
\eea 
We use (\ref{C31}), (\ref{C32}), (\ref{C33}), (\ref{C34}) to evaluate the remaining integrals, and we find, 
\bea
\cK^+ _A & = &y^4 \left(  { 8  \over 105} B_8   +{ 8  \over 27} B_6 
 +{ 4  \over 27} B_4  \right) - {y \over 3} \zeta (3) (10 B_4 + 4 B_2 ) 
 -{ \zeta (5) \over 6 y} (15 B_2 + 1) \no \\ &&
 + {\zeta(3)^2 \over 2 y^2} -{5 \zeta (7) \over 16 y^3} +\cO(e^{-2yu_2}) 
 \no \\
\cK^- _A & = & y^4 \left( -{ 64  \over 105} B_8   -{ 32  \over 15} B_6 
 -{ 52  \over 45} B_4 -{ 16 \over 315} B_2 +{ 2  \over 4725} \right)
 \no \\ &&
 + {y \over 3} \zeta (3) \left (40 B_4 + 18  B_2 + { 1 \over 3} \right ) 
 + { 5 \zeta (5) \over  y} \left ( B_2 + {1 \over 12} \right ) +\cO(e^{-2yu_2}) 
 \eea
Collecting all terms together we get 
\bea
\cK^0_{aaab} & =& -\frac{y^4}{945} \left( 48 B_2+ 952 B_4 + 1736 B_6 
+ 504 B_8  \right)  \\ &&
+ y  \zeta (3) \left(10 B_4 + \frac{14}{3} B_2 +\frac19 \right) 
 +{ \zeta (5) \over 4 y} (10 B_2 + 1) + {\zeta(3)^2 \over 2 y^2} +{5 \zeta (7) \over 8 y^3}
+\cO(e^{-2yu_2}) \nn
\eea
The leading term agrees with the result obtained by evaluating the integral \eqref{Kaaab0}
after replacing $g(z)$ by its polynomial approximation $\mg_1(z)$.

%

\subsection{Summary of the tropical degeneration}

Collecting the various contributions computed in the previous subsections we can now state 
the tropical limits of the string invariants considered in Section~\ref{sec_tropinv}. For the 
Kawazumi-Zhang invariant, we recover the result obtained earlier in \cite{Pioline:2015qha}:
 \be
\label{tropkz}
\varphi^{(t)} = \frac{\pi t}{6} +y \, B_2 + \frac{5 y^2 }{6\pi t} \left( B_4 + \frac{1}{30} \right) + \frac{5\zeta(3)}{4 \pi y t} 
\ee 
where we recall that we denote $B_{2n}=B_{2n}(|u_2|)$, and assume that $|u_2|<1$.

For the invariants $\cZ_2$ and $\cZ_3$ defined in \eqref{bark3}, we find

\bea
 \cBZ^{(t)}_2 &=&  -\frac{7(\pi t)^2}{90} - \frac{2 \pi t y}{3} B_2
 -y^2 \left( \frac{5}{3}B_4+\frac{2}{3}B_2+\frac{1}{45} \right)
 -\frac{y^3}{\pi t} \left( \frac{74}{45}B_6+\frac{4}{3} B_4+\frac{1}{189} \right) \nn\\
 && -17 \frac{y^4}{(\pi t) ^2} \left( \frac{1}{30}B_8+\frac{2}{45}B_6+\frac{1}{18900}\right)
- \frac{\zeta(3)}{2}\left( \frac{1}{y}+ \frac{6B_2}{\pi t} + \frac{(5B_4+\frac16)y }{(\pi t)^2}\right) \nn\\
  &&
  -\frac{7\zeta(5)}{4} \left( \frac{1}{y^2 \pi t} + \frac{(B_2+\frac56)y}{(\pi t)^2} \right)
    \label{cB2trop}
\eea

\bea
 \cBZ^{(t)}_3 &=&   \frac{ (\pi t)^2}{18} +\frac{2 \pi t y}{3} B_2
 +y \left( \frac{19}{9}B_4+\frac{2}{3}B_2+\frac{2}{135} \right)
 +\frac{y^3}{\pi t} \left( \frac{22}{9}B_6+\frac{16}{9} B_4+\frac{1}{945} \right) \nn\\
 && +17 \frac{y^4}{(\pi t)^2} \left( \frac{1}{18}B_8+\frac{2}{27}B_6+\frac{1}{11340}\right)
 \nn \\
 &&+ \frac{\zeta(3)}{6}\left( \frac{1}{y} + \frac{6B_2}{\pi t} + \frac{(5B_4+\frac16 )y }{(\pi t)^2}\right)
 +\frac{11}{8 (\pi y t)^2} \zeta(3)^2 
    \label{cB3trop}
\eea

For the invariant $\cZ_1$ defined in \eqref{bark3}, we require the tropical limit of the term $\cK^c$ 
defined 
in \eqref{defKc}. Using the results of Sections \ref{sec_aabb}-\ref{sec_aaab} we find
\bea
\label{Kctrop}
\cK^c &=&  y^4 \left( \frac{16}{5} B_8 + \frac{64}{15}B_6+ \frac{2}{675}\right)
-32y\zeta(3) \left( B_4 + \frac1{60}\right) - \frac{\zeta(5)}{y} \left( 16 B_2 -\frac{65}{6} \right) \nn
\\&& -\frac{9\,\zeta(3)^2}{2y^2}+ \frac{3\zeta(7)}{4y^3} +2\cK_{aaaa}^0 + \cO(e^{-2yu_2}) 
\eea
where $\cK_{aaaa}^0$ is the integral \eqref{Kaaaa}. We have not analyzed the tropical limit
of this integral (which is a function of {$\tau$} but not of $v$), but we shall be able to infer it indirectly, up to an undetermined term proportional to $1/y^2$ (see \eqref{Kaaaa0predict}).   It follows from 
\eqref{Kctrop} and earlier results in this section that the tropical
limit of $ \cBZ_1$ is given by 
\bea
 \cBZ^{(t)}_1&=&   \frac{13(\pi t)^2}{90} + \frac{2\pi t y}{3} B_2 
+y^2 \left( \frac{5}{3} B_4+\frac{2}{3} B_2 + \frac{4}{45}\right) 
+ \frac{y^3}{ \pi t} \left(
\frac{86}{45}B_6+\frac{4}{3}B_4-\frac{1}{945}\right) \nn  \\
&& + \frac{y^4}{(\pi t)^2} \left( \frac{23}{30} B_8 + \frac{46}{45} B_6 + \frac{1}{1050}\right)  
+ \zeta(3) \left(\frac{7}{2y} + \frac{3B_2+ 3}{\pi t} -\frac{y(B_4(u_2)-\frac{11}{30})}{2(\pi t)^2}   
  \right) 
 \nn \\
&&  
 +\frac{\zeta(5)}{2}\left(-\frac{1}{ 2\pi t\, y^2}  -\frac{B_2(u_2)-\frac{125}{24}}{2(\pi t)^2\, y}    \right)   
 +\frac{3\zeta(7)}{32\pi^2y^3 t^2}  
-\frac{3\zeta(3)^2 }{(4\pi y  t)^2}  
+\frac{\cK_{aaaa}^0}{4\pi^2 t^2} 
\label{cB1trop}
\eea
The tropical limit of the complete string invariant 
$\cB_{(2,0)}^{(t)}=\frac12\left( \cBZ^{(t)}_1-2 \cBZ^{(t)}_2 + \cBZ^{(t)}_3 \right) $ is then given by 
\bea
\label{B20trop}
\cB_{(2,0)}^{(t)}
&=& \frac{8 (\pi t)^2}{45} + \frac{4 \pi ty}{3} B_2+ 
 y^2 \left( \frac{32}{9}  B_4+\frac{4}{3} B_2+\frac{2}{27} \right) 
 +\frac{y^3}{\pi t}\left( \frac{172}{45} B_6+\frac{26}{9} B_4+\frac{1}{189} \right)
 \nn\\
 && + \frac{y^4}{(\pi t)^2} \left( \frac{64}{45} B_8 + \frac{256}{135} B_6 + \frac{241}{113400}\right)  
+\zeta(3)\left( \frac{7}{3 y} +\frac{5 B_2+\frac32 }{ \pi t}+\frac{ (8B_4+\frac{17}{30}) \,y}{3(\pi t)^2} \right) 
 \nn\\&&
 + \,\zeta(5) \left(\frac{3}{2y^2 \pi t } + \frac{(3B_2+\frac{265}{48})}{2y\, (\pi t)^2}\right)
+\frac{3\zeta(7)}{64\pi^2y^3 t^2}  
+\frac{19\, \zeta(3)^2 }{32(\pi y  t)^2}  
+\frac{\cK_{aaaa}^0}{8\pi^2 t^2} 
 \eea
 Upon changing variables from $(t,y=\pi\tau_2,u_2)$ to $(V,S_1,S_2)$ using \eqref{defS12},
 and making use of the functions $A_{i,j}$ defined in \eqref{defA10}, \eqref{arewrite}, we recover
 the results announced in Section \ref{sec_tropinv}.

\newpage


\providecommand{\href}[2]{#2}\begingroup\raggedright\endgroup

\end{document}